\documentclass[usenatbib]{mnras}

\usepackage{newtxtext,newtxmath}

\usepackage[T1]{fontenc}

\DeclareRobustCommand{\VAN}[3]{#2}
\let\VANthebibliography\thebibliography
\def\thebibliography{\DeclareRobustCommand{\VAN}[3]{##3}\VANthebibliography}
\usepackage{xcolor}

\usepackage{graphicx}	

\usepackage{amsmath}	
\usepackage{mathtools}
\usepackage{hyperref}
\hypersetup{colorlinks=true,
            citecolor=blue,
            linkcolor=blue}
\usepackage[export]{adjustbox}

\usepackage{chngcntr}
\usepackage{newtxtext,newtxmath}

\newcommand{\Msun}{\mathrm{M}_{\odot}}	

\begin{document}


\title[Black Hole Binary Capture in AGN Accretion Discs]{Black Hole Binary Formation in AGN Discs: From Isolation to Merger}

\author[C. Rowan et al.]{
Connar Rowan$^{1}$\thanks{E-mail: connar.rowan@physics.ox.ac.uk}
,Tjarda Boekholt$^{1}$, Bence Kocsis$^{1,2}$ and Zoltán Haiman$^{3}$
\\
$^{1}$Rudolf Peierls Centre for Theoretical Physics, Clarendon Laboratory, Parks Road, Oxford, OX1 3PU, UK
\\
$^{2}$St Hugh's College, St Margaret's Rd, Oxford, OX2 6LE, UK
\\
$^{3}$Department of Astronomy, Columbia University, New York, NY 10027, USA\\
}


\date{\today}

\maketitle

\begin{abstract}
    Motivated by the increasing number of detections of merging black holes by LIGO-VIRGO-KAGRA, black hole (BH) binary mergers in the discs of active galactic nuclei (AGN) is investigated as a possible merger channel. In this pathway, BH encounters in the gas disc form mutually bound black hole binary systems through interaction with the gas in the disc and subsequently inspiral through gravitational torques induced by the local gas. To determine the feasibility of this merger pathway, we present the first 3D global hydrodynamic simulations of the formation and evolution of a stellar-mass BH binaries AGN discs with three different AGN disc masses and five different initial radial separations. These 15 simulations show binary capture of prograde and retrograde binaries can be successful in a range of disc densities including cases well below that of a standard radiatively efficient alpha disc, identifying that the majority of these captured binaries are then subsequently hardened by the surrounding gas. The eccentricity evolution depends strongly on the orbital rotation where prograde binaries are governed by gravitational torques form their circumbinary mini-disc, with eccentricities being damped, while for retrograde binaries the eccentricities are excited to $>\sim$ 0.9 by accretion torques. In two cases, retrograde binaries ultimately undergo a close periapsis passage which results in a merger via gravitational waves after only a few thousand binary orbits. Thus, the merger timescale can be far shorter than the AGN disc lifetime. These simulations support an efficient AGN disc merger pathway for BHs.
\end{abstract}




\section{Introduction}
\label{sec:intro}
Since the first detection of a black hole-black hole (BH-BH) merger in 2015 \citep{LIGO2016} there have been numerous additional observations of compact object mergers (e.g              \citealt{LIGO2019,LIGO2020a,LIGO2020b,LIGO2020c,LIGO2020d}). Of particular interest are merging BHs with masses in the ``mass gap'' \citep{LIGO2020e} associated with the mass loss predicted by pair-instability supernovae in the later stages of high mass stars' evolution \citep{Rakavy1967,Belczynski2016}. Since BHs of this mass cannot originate as stellar remnants, this implies the objects merging in this particular case have themselves been the result of prior mergers \citep{Gerosa2019,Gerosa2021} or accreted additional mass \citep{Safarzadeh2020}. The astrophysical mechanisms driving these specific merger events remain poorly understood, though several processes and environments have been proposed to form these systems (see \citealt{Gerosa2021} for review). These include many-body encounters in dense star clusters (e.g \citealt{Mouri2002,Miller2002,Liu2021,DiCarlo2020}) which demonstrate that a small percentage of BHs in a cluster can merge to form BHs within the aforementioned mass gap via binary-single interactions. 

Alternatively, BH mergers have been suggested to occur in the gaseous accretion discs of active galactic nuclei (AGN) where both theoretical arguments \citep{Bahcall1976,Bahcall1977,Miralda2000} and observational evidence (e.g \citealt{Hailey2018}) indicate that approximately $2\times10^4$ BHs reside within $\sim1$pc. Compared to the central low-mass stellar population the distribution of these objects in the centres of galaxies is expected to be less spherically symmetric and more disc-like due to a process known as vector resonant relaxation \citep{Rauch1996,Szolgyen2018,Gruzinov2020,Magnan+2022,Mathe+2022}. If a gas disc is present, gas dynamical friction \citep{Ostriker1999} with the disc can reduce the inclination of the orbits crossing the disc (e.g \citealt{Bartos2017,Panamarev2018}) and thereby increase the chance of encounters. A flattened distribution is observed in the distribution of the X-ray BH binary population of the Milky Way (e.g \citealt{Mori2021} as well as in the high-mass stellar population close to its SMBH, Sgr A$^{*}$ (e.g \citealt{Bartko2009,Ali2020}). A similarly disc-like population of the higher-mass objects is predicted by N-body simulations 
in rotating star clusters (e.g \citealt{Szolgyen2019,Szolgyen2021}). The compression of the BH distribution in the vertical axis is expected to increase the likelihood of close encounters and hence the chance for binary captures and mergers \citep[e.g.][]{Vergara21}. 
\newline\indent The capture process itself is a complex phenomenon and shows similarities to the problem of satellite capture in the Solar System, such as the capture of moons \citep[e.g.][]{Johnson2005,Agnor2006,nesvorny2007}. The interaction between two black holes most often results in a single flyby, during which a sufficient amount of orbital energy has to be dissipated by gravitational wave (GW) emission for a capture. Alternatively, the two black holes can engage in a Jacobi capture interaction, during which they encounter each other multiple times within their mutual Hill sphere \citep{Boekholt_2022}.
This increases the probability of an eventual permanent capture, and thus the formation of a binary black hole. The resultant binary systems are highly eccentric, i.e. approximately super-thermal. Additional gas effects, through dynamical friction and accretion, can potentially significantly increase the capture cross section, and also facilitate the subsequent binary hardening and eventual merger \citep{Tagawa2018}. However, the phase space related to gasless Jacobi captures is fractal, i.e. the outcome of an interaction depends sensitively on the initial impact parameter and speed of the encounter. The influence of the SMBH on the capture process might thus be non-trivial even when including gas. 
\newline\indent Numerous studies have looked at binary black hole (BBH) formation and/or evolution in an AGN environment using semianalytic methods (e.g \citealt{Tagawa2020,Grobner2020,Secunda2019}) for handling small scale BH-gas interactions such as accretion, dynamical friction, capture and crucially the subsequent post-capture torques on the binary. The majority of these processes indicate that the presence of a dense gaseous medium can aid binary formation and merger by inducing a net negative torque on the system.

Recent high-resolution hydrodynamical simulations of isolated binary evolution, where there are only two objects at the centre of a circumbinary gas disk, rather than embedded in the local shear flow of far larger circum-SMBH disc, find that disk torques can be either negative (causing the binary to inspiral) or positive (causing outspiral). 
Several 2D studies of circular, equal-mass binaries embedded in relatively thick, locally isothermal diskc with a fixed scale height to radius ratio $H/R=0.1$ have, in particular, converged on positive torques \citep{Tang2017,Munoz2019,Moody2019,Tiede2020}, which, in the case of \citet{Moody2019} was confirmed in a 3D simulation.  Subsequent work has, however, revealed that the disc torques become negative when the above assumptions are relaxed.

Binary inspiral is found for cooler and thinner discs with $H/R\lesssim 0.1$~\citep{Tiede2020, Heath2020}, with the precise critical $H/R$ value dependent on viscosity~\citep{Dittman2022}.  These cases may be more realistic for radiatively efficient discs that cool rapidly. Likewise, torques switch to a negative sign for eccentric binaries with $e\gtrsim 0.4$~\citep{DOrazio2021}, which may again be the more typical case, since the disc tends to drive binaries to eccentricities above this value, unless they start very close to circular orbits ~\citep{Zrake2021,DOrazio2021}.
Finally, unequal-mass binaries with $q\equiv M_2/M_1\lesssim 0.05$ have also been found to inspiral~\citep{Duffell2020} (although these may be atypical for stellar mass BH mergers, see \citealt{Tagawa2021_massgap}).
In summary, it appears at least plausible for most binaries with a circumbinary gas disc to be driven towards merger. Caution must still be exercised as the above simulations remain idealised in several ways, and would, in any case, only apply for a stellar-mass binary deeply inside the Hill radius in our case.

There also remain some numerical issues, with some studies finding a sensitivity to sink prescriptions~\citep{Tang2017} and softening lengths~\citep{Li2021}, although more recent work find torques to have converged with respect to most numerical choices, at least for circular binaries~\citep{Moody2019,Munoz2020,Duffell2020,Westernacher-Schneider2022}.

There have been only a few direct hydrodynamical simulations of binaries embedded in a SMBH accretion disc.  \citet{Baruteau2011} report that gas dynamical friction hardens a pre-existing binary in a 2D gas disc, regardless of whether it opens up a gap, where the binary is sufficiently massive to expel gas from its orbit around the SMBH faster than gas can refill it via viscosity driven diffusion  \citep[e.g.][]{Goldreich1980}. This result has been recently reaffirmed by \citet{Li2021} in the cases where the binary system orbit is retrograde with respect to its orbit around the disc/SMBH and is attributed to an increased velocity difference between the individual BHs in the binary and the local gas, leading to the destruction of positive torque sources near the BHs. \citet{Kaaz2021} consider a binary in a thick disc where the binary does not open a gap in the disc using 3D wind tunnel simulations and find binary hardening in all their models. At present, there is no clear consensus on how the differences in environment between isolated binaries, disc-embedded binaries, and wind tunnel simulations \citep{Kaaz2021} affect binary evolution. 

While the problem of binary evolution is not yet settled, even less is known about the correlation between the rates of successful gas-assisted binary captures and the nature of the AGN host environment they take place in, which may vary considerably depending on the evolution of its host galaxy. An environmental dependence would therefore have implications for the rates of BH mergers in AGN over cosmic time, based on the redshift dependence of galaxy morphology, merger rates etc. that are tied to the number and evolution of AGN  \citep[e.g.][]{Fanidakis2012,Conselice2014,Tagawa2020}.

A primary difficulty faced by hydrodynamic simulations of disc-embedded BBHs is achieving sufficient resolution to resolve the circumbinary "mini" discs (CMBDs) and the streamers between the individual components of the binary, whilst also modelling the larger global AGN gas disc or, more usually, annulus which the binary orbits within. This problem is commonly avoided by artificially increasing the BBH/SMBH mass ratio so the scale of the region of interest around the BBH increases relative to the disc, reducing the dynamical range to be covered. While this makes achieving an appropriate resolution around the binary less expensive, it describes a far more exotic scenario of two intermediate mass BHs (IMBHs) embedded in a AGN disc of a SMBH (e.g \citealt{Li2021}), and it is unclear if these systems may be simply extrapolated to the case of stellar mass BBHs. 

In this paper we investigate the efficiency and efficacy of BBH formation via gas dissipation during two body scatterings in an AGN disc. We also examine how the AGN disc mass and influence of the SMBH affect the encounter. A total of 15 three-dimensional simulations of the approach of two isolated BHs are performed, in the midplane of an AGN disc for simplicity. This work begins with a description of the literature related to the gas-driven binary capture mechanism in Sec.~\ref{sec:gas_driven_capture}. Sec.~\ref{sec:CompMethods} details the computational method used to model this system, followed by a description of the initial conditions in Sec.~\ref{sec:initcond}. We present our fiducial simulation results showing a successful gas-assisted binary forming event in Sec. \ref{sec:results_fiducial} and discuss the sources of torques and energy dissipation. Sec \ref{sec:results_all} presents the outcome of all other simulations showing that gas assisted captures may occur for a broad range of initial conditions for sufficiently high initial gas densities. We discuss our findings and methods in Sec. \ref{sec:discussion} before summarising our conclusions in Sec.~\ref{sec:conclusions}. Due to the number of models, we spare the main text of several large plots and instead refer to their position in the appendix. 

\begin{figure*}
\begin{tabular}{ccccc}
    \includegraphics[width=0.195\textwidth]{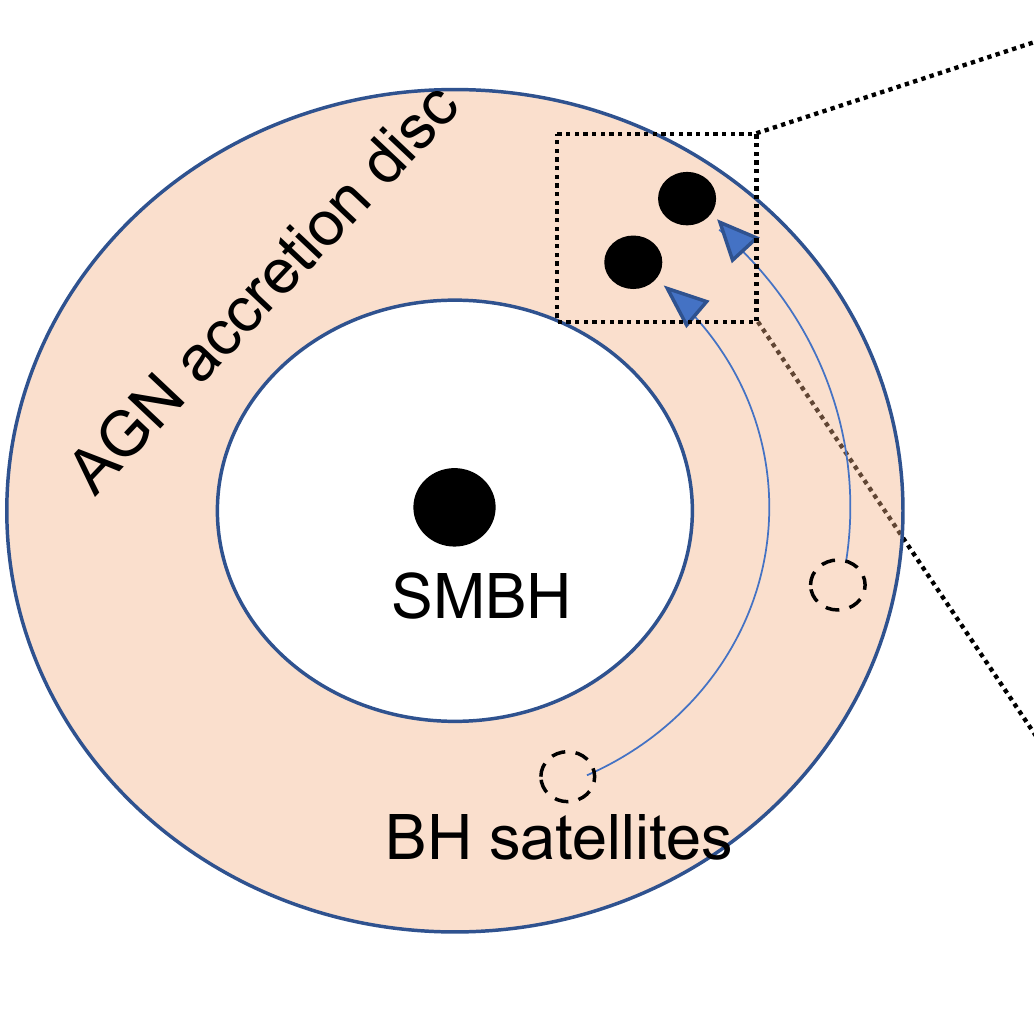}
    \includegraphics[width=0.195\textwidth]{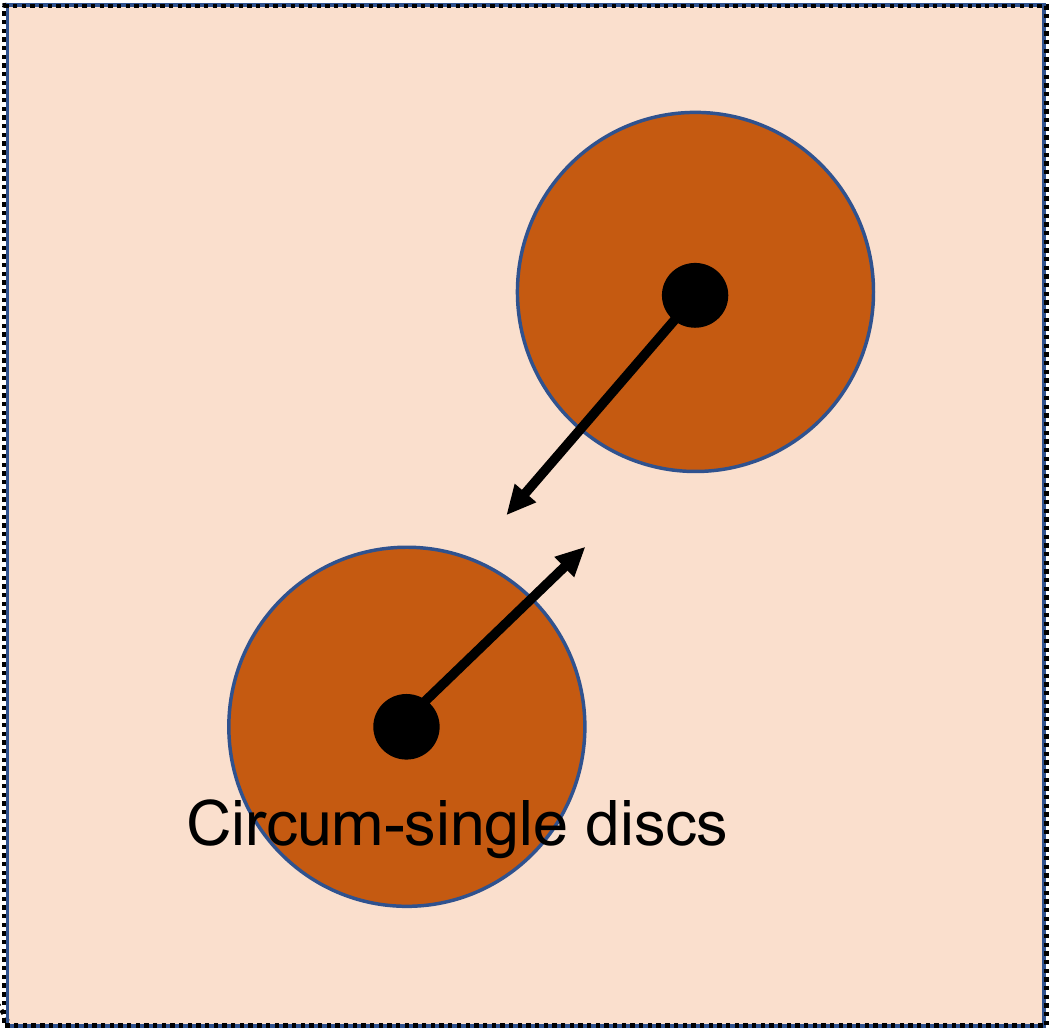} 
    \includegraphics[width=0.195\textwidth]{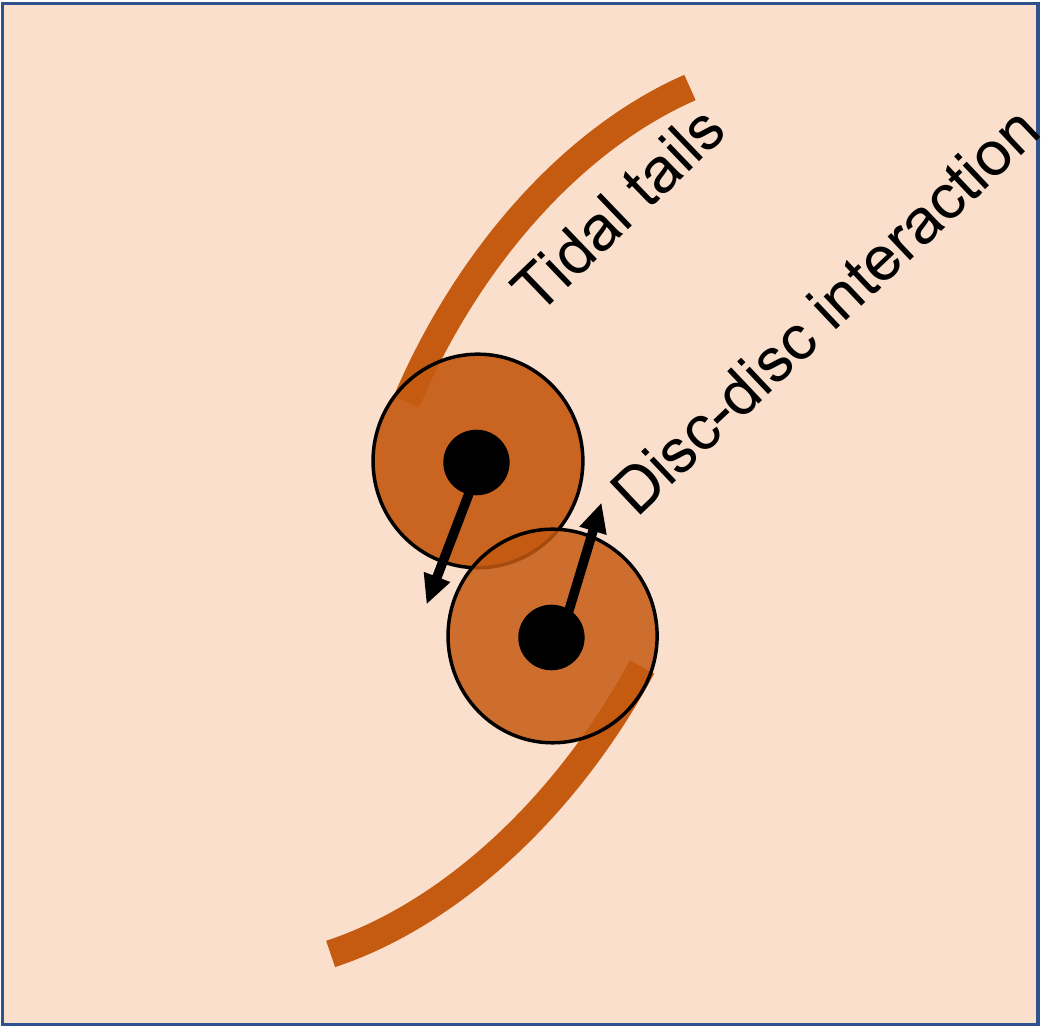}
    \includegraphics[width=0.195\textwidth]{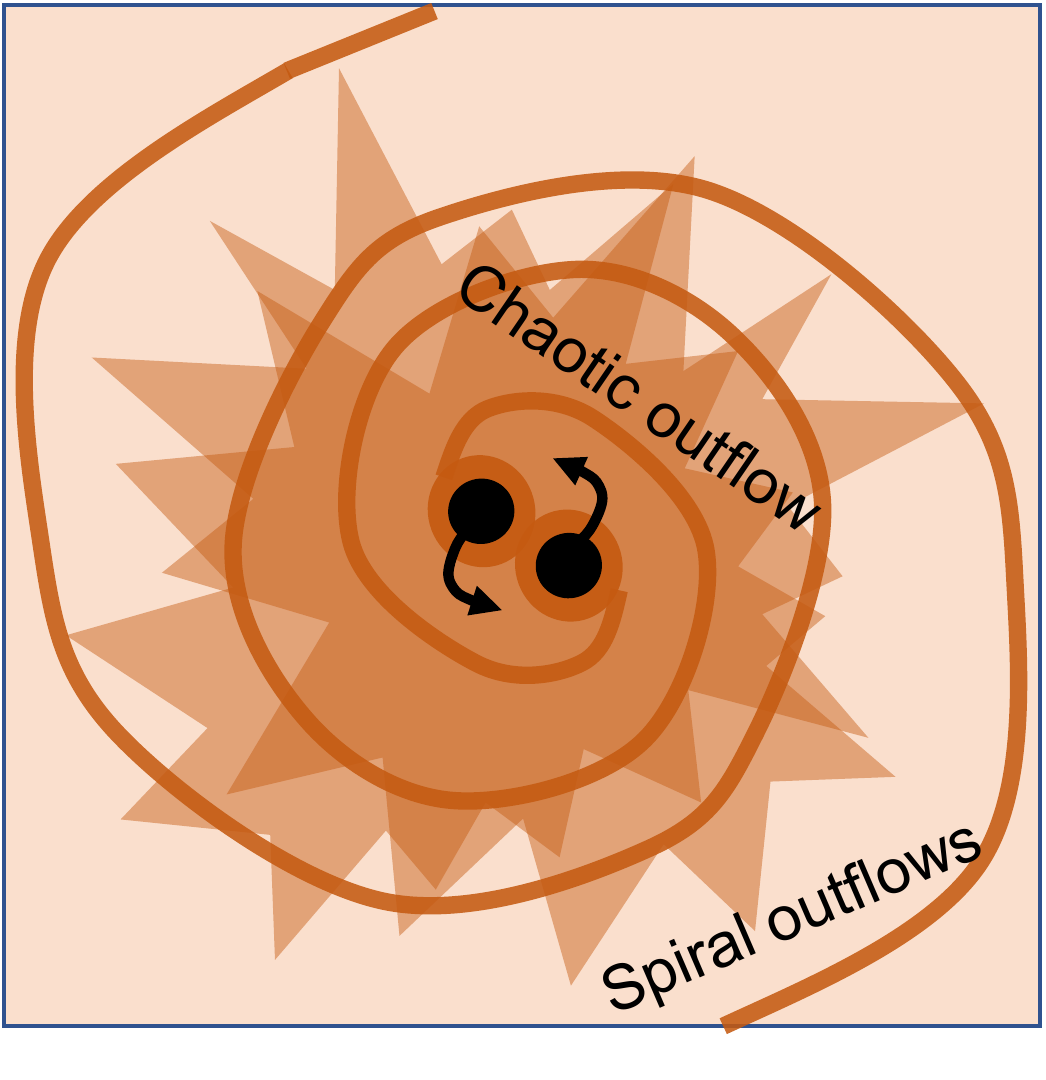}
    \includegraphics[width=0.195\textwidth]{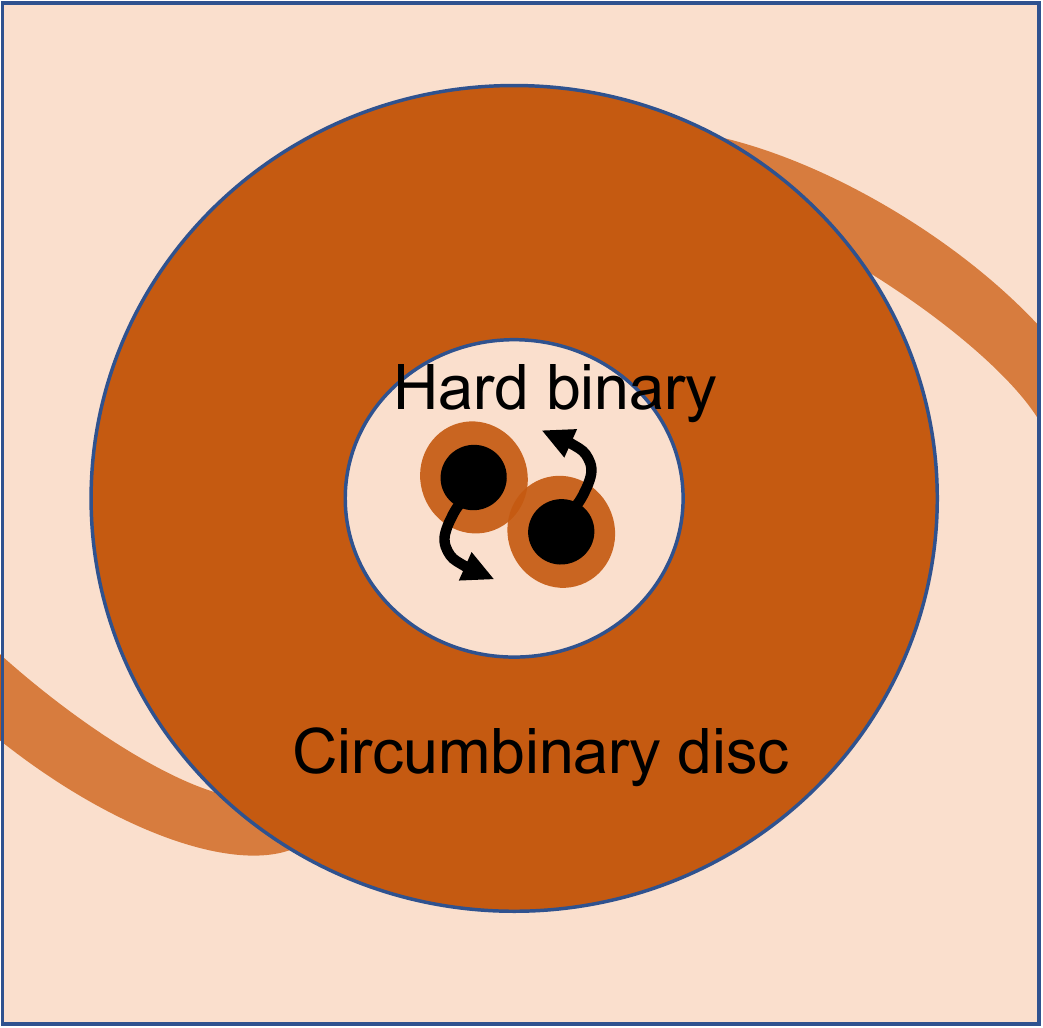}
\end{tabular}
\caption{Cartoon of the gas-driven binary capture mechanism, where the darkness of the orange shading represents the density of the medium. From left to right, i) Zoom out of scenario showing the SMBH, AGN disc and BH satellites ii) Zoom in of BHs just prior to encounter with their circum-single mini-discs iii) beginning of encounter showing tidal disruption of BH discs and disc collision iv) gas outflows and density spirals generated by wakes behind BHs v) late evolution of now hardened binary with circum-binary mini-disc. }
\label{fig:cartoon}
\end{figure*}

\section{Binary formation via gas dissipation}
\label{sec:gas_driven_capture}
We perform smoothed particle hydrodynamic (SPH) simulations of two stellar-mass BHs orbiting in an AGN disc, which undergo a mutual close encounter. Our setup differs from previous hydrodynamic simulations of binaries in AGN, since the binary has not yet formed at the start of our simulation. This setup allows us to investigate the gas-assisted binary formation process which \citet{Tagawa2020} predicted to be the formation pathway for >90\% of binaries that merge in AGN. 

Literature on binary formation through this pathway is scarce and currently only uses analytic methods. In \citet{Tagawa2020}, binary formation is considered successful when the timescale in which dynamical friction can dampen the relative velocity of the scattering objects is shorter than the crossing time of one object across the Hill sphere (defined with respect to the SMBH) of the other. However, this prescription assumes a constant density and unchanging uniform distribution of gas. Dynamical friction on a satellite in a uniform gaseous disc \citep{Kim2007} has been shown to behave nearly identically to that in an infinite uniform medium \citep{Ostriker1999} with the same limitations. \citet{Kim2008} analytically explored dynamical friction on two co-rotating perturbers in a uniform gaseous medium. They find dynamical friction still leads to inspiral though the dynamical friction force on each object is reduced due to the wake following the other. Similarly to the pre-existing binary simulations, this is only shown for circular, already bound binaries and ignores density gradients which can be expected to be large following the initial encounter since the objects will have filled their Hill spheres with gas. Therefore the interaction between two satellites and the gas during their first encounter is largely unknown. Understanding this process and its efficiency is crucial for estimates of binary fractions and merger rates in AGN.

Figure \ref{fig:cartoon} depicts a cartoon of the gas-driven binary formation process examined in this work. The left panel shows the large scale configuration of the system where two BH satellites moving initially on Keplerian orbits around the SMBH, embedded in the AGN disc, approach each other due to the difference in their angular velocities. The four panels on the right show a zoom-in on the BH satellites in different evolutionary stages in time, respectively, from left to right. Initially the BHs have distinct circum-single mini-discs (CSMD) just prior to the encounter (1st zoom-in panel). At the beginning of their encounter, these discs begin to be tidally stripped by one another (2nd panel) and there is also the opportunity for the discs to intersect which leads to strong accretion and gravitational drag forces on the BHs due to their interaction with the gas. This interaction is incredibly violent and leads to mass loss of gas from within the binary Hill sphere (3rd panel). It is this drag that decreases the energy and eccentricity of the BHs to the point that they can remain bound as a binary. 
The binary is further hardened by spiral gas outflows, originating from the CSMDs of the BHs that are continually being stripped of material. This is akin to the predictions of \citet{Kim2008}. Once the binary is sufficiently hardened that their separation is much smaller than their mutual Hill sphere with respect to the SMBH, a CBMD forms around the binary (4th panel) and the binary transitions to the problem of a pre-existing disc embedded binary with a high eccentricity, where its evolution is predominantly governed by the CBMD.

A close analogy to this scenario is work on protostellar disc collisions. The SPH simulations of \citet{Watkins1998a} demonstrate successful formation of stellar binaries with coplanar protostellar discs, where the stars are initially energetically unbound. This is extended to non-coplanar configurations in their second paper \citep{Watkins1998b} where they predict that for these systems 15\% of disc-disc encounters lead to a protostellar binary. This is encouraging for AGN, however the velocity dispersion is far larger and encounters will be more energetic on average. Curiously, for star-disc encounters they find energy can be dissipated from \textit{or} added to the stellar binary orbits (e.g \citealt{Boffin1998}). More recently, at higher resolutions, \citet{Shen2010} corroborate (albeit as a rare case) that gas can aid stellar binary formation, though their paper primarily focuses on brown dwarf formation due to the disc-disc collision. Most recently, \citet{Munoz2015} use the adaptive grid code AREPO to investigate the binary evolution specifically following such encounters and its dependence on initial periapsis distance. They again verify that capture is possible and that this becomes more likely the closer the first periapsis pass is. With the exception of \citet{Munoz2015}, all other disc collision papers that we are aware of utilise SPH for their encounters, which is well suited given the strongly non-axisymmetric gas morphology compared to grid based codes.

We note that resolving all three dimensions may be essential to obtain the correct outcome of the scattering process. As argued in \citet{Dempsey2022}, assuming the 2D symmetry adopted in most studies of disc embedded binaries can lead to different torques on the binary by ignoring the vertical gas behaviour. In particular, its omission leads to larger positive torques in the simulation driving the binary to outspiral. This discrepancy was found for simulations with an isothermal equation of state (EOS). 

SPH is well suited to investigate our scattering problem for a hierarchical triple in gas, where the encounter occurs far from the center of mass (COM) of the system, near the SMBH. By simulating an annulus of gas around the SMBH, we have the required high resolution in the vicinity of the binary to examine the process in 3D. 

While finalising this paper, three papers appeared on a similar subject. \citet{Li_Dempsey_Lai+2022} presented 2D grid-based hydrodynamics simulations of binary formation in a gaseous disk showing that bound systems may form in gas, similar to the scenario investigated in this paper independently with a 3D SPH method. \citet{Rozner+2022} and \citet{DeLaurentiis2022} ran semianalytic simulations with dynamical friction to indentify analytic conditions for binary formation beyond previous prescriptions by \citet{Goldreich2002} and \citet{Tagawa2020}.

\section{Computational Methods}
\label{sec:CompMethods}
We use the smoothed particle hydrodynamics (SPH) code PHANTOM (see \citealt{Price2018}) to solve the hydrodynamic equations in Lagrangian form. Radiative and magnetic effects, which for the AGN disc are more relevant closer to the SMBH than our orbital radius of interest \citep{Jiang2019,Davis2020} are neglected. PHANTOM uses a fixed particle mass treatment of SPH, where the particle density is directly proportional to the mass density. These particles are placed in a circular annulus around a SMBH between an inner radius $R_{\rm in}$ and outer radius $R_{\rm out}$ using a Monte Carlo scheme. For all our models, we consider a thin accretion disc of $2.5\times10^{7}$ gas particles around a SMBH of mass $M_{\rm SMBH}=4\times10^{6}\Msun$, representing an object in the more numerous population of low redshift AGN as predicted by SMBH mass functions (e.g \citealt{Shankar2004,Greene_Ho2007,Li2011}). 

The the orbiting stellar-mass BHs are represented by sink particles which may accrete mass, momentum and angular momentum from infalling gas particles. The corresponding accretion radii $r_{\rm acc}$ of these sinks are defined as a fraction of the Hill radius $r_{\rm H}$ of the associated object 
\begin{equation}
    \centering
    r_{\rm H} = R_{\rm BH}\left(\frac{M_{\rm BH}}{3M_{\rm SMBH}}\right)^{1/3},
    \label{eq:hill}
\end{equation}
where $M_{\rm BH}$ denotes the mass of the BH satellite with $R_{\rm BH}$ its radial distance from the SMBH. This fraction is set to $r_{\rm acc}=0.01r_{\rm H}$. Particles entering the accretion radius are checked to see if they meet multiple criteria before being accreted. Upon entering the sink accretion radius, a gas particle is accreted unconditionally inside $0.8r_{\rm acc}$. In the region $0.8r_{\rm acc} < r < r_{\rm acc}$, accretion occurs if three conditions are met: i) its specific angular momentum is less than that of a circular Keplerian orbit at it's current distance from the accreting BH, ii) it is energetically bound to the particular sink iii) it is more bound to that sink than any others. {Particles are softened near the sinks based on PHANTOM's cubic spline kernel softening with a characteristic softening length equal to the inner boundary of 0.8$r_{\rm acc}$. These are the default sink prescriptions in PHANTOM, which have been shown to reduce the pressure imbalance at the hard accretion boundary of 0.8$r_{\rm acc}$ (see \citealt{Price2018}).} 

\section{Initial Conditions}

\label{sec:initcond}
\subsection{AGN Disc}
The disc which our satellite BHs are embedded in is represented by an annulus of radius $R_{\rm mid}=0.0075\,$pc and radial width of $\Delta R_{\rm disk}=20\,r_{\rm H}$ so that any gap opening in the annulus is correctly captured by the simulation.
This allows density spirals in the disc from the BHs to dissipate sufficiently upon reaching its edges, whilst not wasting resources solving for gas motions inside or outside this annulus. All models use a purely Shakura-Sunyaev disc \citep{Shakura1973} with a constant alpha viscosity of $\alpha_{SS}= 0.1$. This disc viscosity is applied to both approaching and receding particles, implemented artificially into the momentum equation within PHANTOM. The disc viscosity behaviour in PHANTOM has been tested via measuring the diffusion rate of the surface density against 1D codes (see Figure 4 in \citealt{Lodato2010}). In our simulations we neglect viscous and shock heating. The adopted values of our model parameters are listed in Table~\ref{tab:initial_conditions}.

The surface density follows a power-law profile of the form
\begin{equation}
    \centering
    \Sigma(R) = \Sigma_{0}\bigg(\frac{R}{R_{\rm in}}\bigg)^{-p}\,,
    \label{eq:rho_R}
\end{equation}
\noindent where $R_{\rm in}=R_{\rm mid}-\frac12\Delta R_{\rm disk}$ is the inner radius of the annulus, $p=0.6$ is the power-law exponent and $\Sigma_{0}$ is the density at $R_{\rm in}$. The sound speed, $c_{\rm s}$, assumes a locally isothermal equation of state and is described in a similar form to the density:
\begin{align}
    c_{\rm s}(R) &= c_{s,\rm in}\left(\frac{R}{R_{\rm in}}\right)^{-q}\,,\label{eq:cs_prof}\\
    c_{s,\rm in} &= \left(\frac{H}{R}\right)_{R_{\rm in}}
    \sqrt{\frac{GM_{\rm SMBH}}{R_{\rm in}}}
    \,.
    \label{eq:cs0}
\end{align}
\noindent Here $q=0.45$ is a power-law index, $H=c_{\rm s}/\Omega$ is the disc scale height and $c_{\rm s,in}$ is the sound speed at $R_{\rm in}$, defined in Eq.~\eqref{eq:cs0} to satisfy the target $H/R$ at $R_{\rm in}$ and $G$ is the gravitational constant. Inline with a geometrically thin disc (e.g \citealt{Szuszkiewicz1996}), the value of $(H/R)_{\rm in}=0.005$ is used for all models. Under the assumption of local thermal equilibrium the vertical density is given by the usual relation:
\begin{equation}
    \centering
    \rho(R,z) = \frac{\Sigma(R)}{\sqrt{2\pi}H}\exp \bigg(\frac{-z^{2}}{2H^{2}}\bigg).
    \label{eq:rho_Rz}
\end{equation}
\noindent In alignment with the steady Shakura-Sunyaev alpha disc prescription of \citet{Goodman2004} the inner density $\Sigma_{0}$ may be expressed through the enclosed disc mass $M_{\rm d,0}$ and the SMBH mass $M_{\rm SMBH}$ as
\begin{multline} 
M_{\rm d,0}(<R)\approx 4.82\times10^{5}\alpha_{SS}^{-\frac{4}{5}}\hat{\kappa}^{-\frac{1}{5}}\mu^{-\frac{4}{5}} \\ 
\times\bigg(\frac{L_{E}}{\epsilon}\bigg)^{\frac{3}{5}}\bigg(\frac{M_{\rm SMBH}}{10^{8}\Msun}\bigg)^\frac{11}{5}\bigg(\frac{R}{1000r_{s}}\bigg)^{\frac{7}{5}}\Msun,
\label{eq:discmass}
\end{multline}
\noindent where $r_{s}$ is the Schwarzschild radius of the SMBH. The zero subscript denotes this as the fiducial disc mass which we later scale in our other models. This relation provides the disc mass enclosed within some radius $R$ given four parameters, the disc luminosity relative to the Eddington limit $L_{E}$, the mean molecular mass $\mu$ , opacity $\hat{\kappa}$ and radiative efficiency $\epsilon$. The adopted values listed in Table~\ref{tab:initial_conditions} correspond to an optically thick, geometrically thin, radiatively efficient disc. From the enclosed mass, $\Sigma_{0}$ is calculated via normalisation of the density function in Eq.~\eqref{eq:rho_Rz} between $R_{\rm in}$ and $R_{\rm out}$ such that 

\begin{equation}
    \centering
    2\pi\int_{R_{\rm{in}}}^{R_{\rm{out}}}  \Sigma(R)R dR = M_{\rm{d},0}(R_{\rm{out}}) - M_{\rm{d},0}(R_{\rm{in}})\, .
    \label{eq:sig_norm}
\end{equation}

\begin{table*}
    \centering
    
    \begin{tabular}{ccccccccccccccc}
    \hline\hline
$\dfrac{\Delta R_{\rm i}}{r_{\rm H}}$ & $\Delta \phi$
& $\dfrac{M_{\rm SMBH}}{\Msun}$ 
& $\dfrac{M_{\rm BH}}{\Msun}$ 
& $\dfrac{M_{\rm d}}{10^{-3}M_{\rm SMBH}}$ 
&  $\dfrac{R_{\rm mid}}{\rm mpc}$ & $\dfrac{\Delta R_{\rm disk}}{r_{\rm H}}$ 
&  $p$ & $q$ & $\dfrac{H}{R_{\rm in}}$ & $\alpha_{\rm SS}$  
&    $L_{\rm E}$ & $\epsilon$ & $\mu$ & $\hat{\kappa}$ 
\\ \hline 
    2.5--3.5
    & $20^{\circ}$ 
    & $4\times10^{6}$
    & 25 
    & $\{0.32,1.6,8\}$
    & 7.5 & 20 
    & 0.6 & 0.45 & 0.005 & 0.1 
    &  0.1 & 0.1 & 0.6 & 1.0 
    \\ \hline
    \end{tabular}
    \caption{Fiducial model parameters. Here $(\Delta R_i, \Delta \phi)$ are the initial radial offset and orbital phase between the two stellar BHs in their initially Keplerian orbits around the SMBH, $r_{\rm H}$ is the Hill radius, $(M_{\rm SMBH}, M_{\rm BH}, M_{\rm d})$ are respectively the SMBH mass, the individual stellar BH masses, and the total enclosed gaseous disk mass (Eq.~\ref{eq:discmass}), hence the gas mass per $r_{\rm H}$ radial width is $\frac75 (r_{\rm H}/R_{\rm mid})M_{\rm d}=0.026 M_{\rm d}=\{23\,\Msun,110\,\Msun,570\,\Msun\}$, 
    $(R_{\rm mid},\Delta R_{\rm disk})$ are the mean radius and the width of the simulated gaseous annulus, $p$ and $q$ set the radial dependence in the surface density and sound speed across the annulus (Eqs.~\ref{eq:rho_R}--\ref{eq:cs0}), $H$ is the scaleheight ($H=0.4\,r_{\rm H}$) which also sets the pressure and temperature in the disk via Eq.~\eqref{eq:cs0}, $L_{\rm E}$ is the Eddington ratio, $\epsilon$ is the radiative efficiency, $\mu$ is the mean molecular mass, $\hat{\kappa}$ is the opacity relative to the electron scattering opacity ($0.4\,{\rm cm}^{2}{\rm g}^{-1}$).
}
    \label{tab:initial_conditions}
\end{table*}

\subsection{BH Satellites}
Two equal-mass BHs with $m_1,m_2=25\Msun$ are inserted with circular Keplerian orbits with zero velocity dispersion around the SMBH, accounting for the radially enclosed disc mass, assuming cylindrical symmetry, with zero inclination. To simulate a possible capture within the limit of the observed density cusp at $\sim$0.01pc, the BHs are initialised symmetrically about 0.0075pc at an azimuthal separation of 20 degrees with the outer object ahead of the inner along its orbit. The orbital period of the BHs around the SMBH is $\sim30 yr$. The BHs are inserted without their own accretion discs, but prior to closing their angular separation they quickly accumulate significant disc mass until after about 10 years where the buildup becomes more gradual. This way their discs are formed entirely self-consistently within the simulation and do not rely on their own separate set of initial conditions. The satellite BHs are given a small sink accretion radius to minimise the chance the BHs cross each other's accretion radii where gas dissipation cannot take effect. The accretion radii for the BHs is set to $r_{\rm acc}=0.01r_{\rm H}$ to solve for the smallest practical accretion radius ensuring dynamical friction is incorporated directly and as accurately as possible from the local SPH particles. 
 
\subsection{Different Models} \label{sec:models} 
\begin{figure}
    \centering
    \includegraphics[width=8cm]{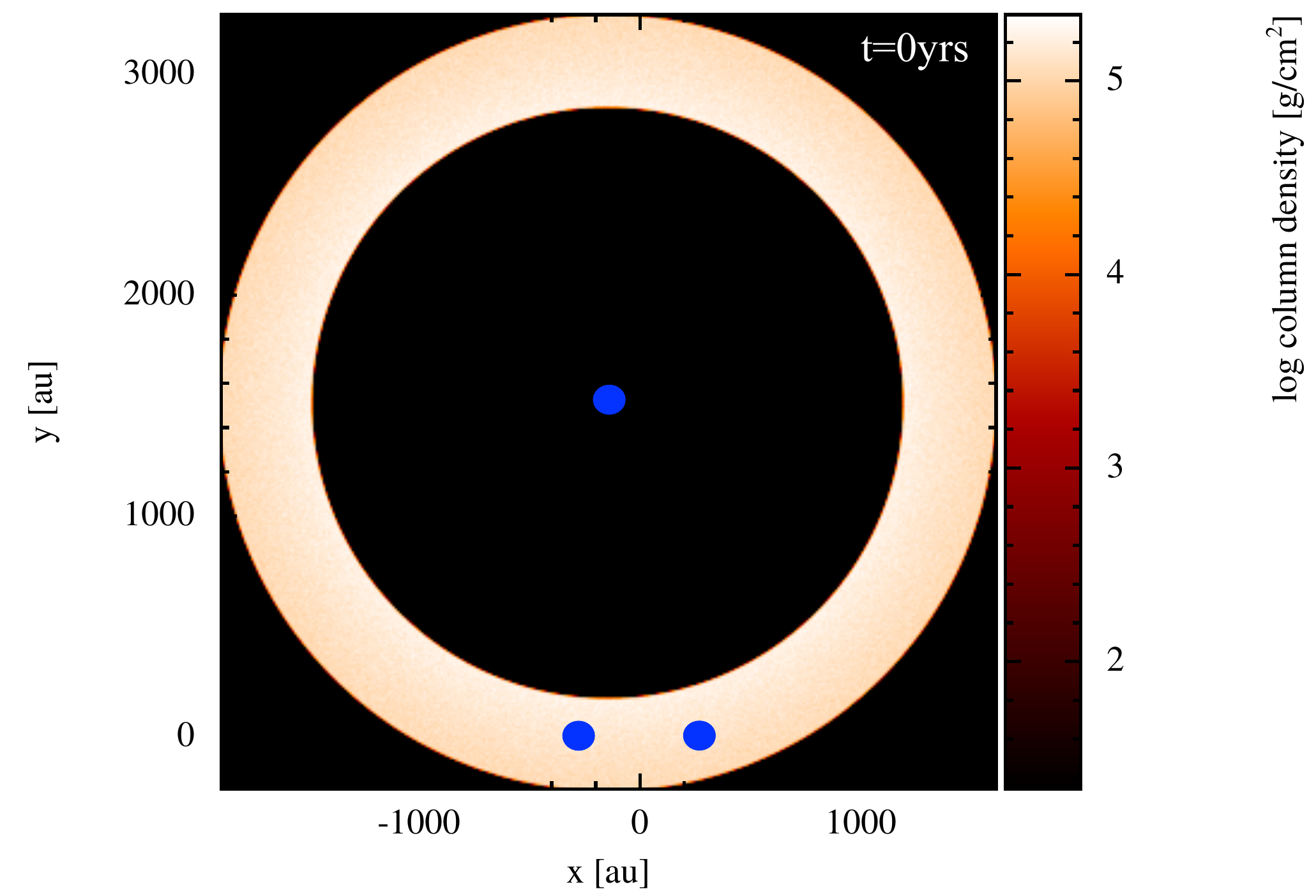}
    \caption{Initial gas surface density profile of  $\mathit{Cap_{5,3.5}}$ with the SMBH and satellite BHs represented by the blue markers.}
    \label{fig:incond}
\end{figure}

A total of 15 simulations are run, labelled $\mathit{Cap_{X,Y}}$. The effect of varying the initial disc mass is examined using three alternate enclosed disc masses denoted in the first subscript $X={0.2,1,5}$ corresponding to models with 1/5, 1 and 5 times the disc mass of the standard Shakura-Sunyaev disk ($M_{\rm d,0}$) according to Eq.~\eqref{eq:discmass} respectively. For each of these 3 cases, 5 models are run with different initial radial separations, $\Delta R_{i}$. This allows us to observe any stochastic behaviour of encounters with the same AGN disc mass as well as sample slightly different approaches for the encounter. In units of the BHs' initial Hill radii these have separations 2.5, 2.75, 3, 3.25 and 3.5. These are similarly denoted in the second subscript as  $Y={2.5, ..., 3.5}$, therefore the model with five times the expected disc mass and an initial radial separation of 3.5 $r_{\rm H}$ is labelled $\mathit{Cap_{5,3.5}}$. It may seem that our choice for the range of radial separations is large and that we should expect only weak encounters between the BHs. However, as shown in our previous paper that focuses on the pure N-body case \citep{Boekholt_2022}, as well as by others \citep[e.g.][]{higuchi2016}, gravitational focusing can still lead to close encounters well within the Hill sphere even where the initial radial separation is beyond this distance. Since we are including gas this range is increased further to allow for the fact the mass within the Hill radii of the objects will grow as they accrete mass and form CSMDs from the surrounding gas, which will lead to increased focusing in the lead up to the encounter. This was tested based on initial lower resolution test runs. All models with varying radial separations maintain 0.0075pc as the radial midpoint of the BHs as well as the annulus. As a visual example, Figure \ref{fig:incond} shows the initial gas annulus, its surface density, as well as the positions of the BHs in the disc. This figure corresponds to the first panel of the cartoon in Fig.~\ref{fig:cartoon}. 
\noindent We summarise all parameterised initial conditions in Table \ref{tab:initial_conditions}.

\section{RESULTS (FIDUCIAL)}
\label{sec:results_fiducial}
For clarity, we begin our discussion considering only a single model due to the many components of these simulations and their analysis that must be defined. Our fiducial model is $\mathit{Cap_{1,2.5}}$ with initial radial separation $\Delta R_{i} = 2.5r_{\rm H}$ and $M_{\rm d} = M_{\rm d,0}$ , i.e. matching a standard Shakura-Sunyaev accretion disc, see Sec.~\ref{sec:models}.

\subsection{Capture Overview}
\subsubsection{Gas Morphology}
\label{sec:capture_overview}

\begin{figure*}
    \centering
    \includegraphics[width=17cm]{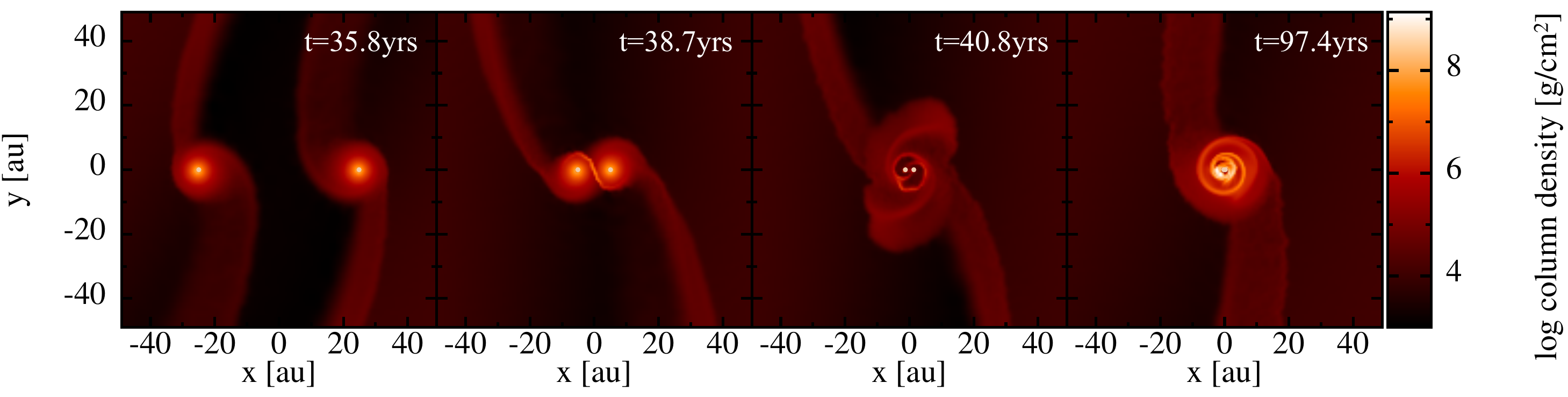}

    \caption{Time series of capture process for our fiducial model, snapshots taken at $35.81$,  $38.63$, $40.79$ and $97.33$yr. The BHs are moving clockwise around the SMBH with the SMBH to the left. Each panel is orientated to the line connecting the two BHs. From left to right in the figure, i) the two BHs and their accretion discs shortly before encounter ii) initial intersection of accretion discs leading to overdensity at the point of contact iii) violent gas outflows and spiral structure as material is stripped from their circum-single discs iv) final snapshot of simulation showing a well defined circumbinary disc around a well hardened binary.}
    \label{fig:capture_visualisation}
\end{figure*}

Figure \ref{fig:capture_visualisation} shows the surface density of the fiducial simulation just prior to, during and long after the first encounter, in direct correspondence to the last 4 panels in the cartoon of Figure \ref{fig:cartoon}. This gives a visual example of the simulated formation of a binary via gas dissipation. 

In the first panel, gas is observed to accumulate around each BH and forms CSMDs (1st panel), reaching out to $\sim r_{\rm H}/2$. At the same time, the presence of the objects leads to an underdensity of about one order of magnitude compared to the AGN disc between them as the gas on initially approximately Keplerian orbits is deflected towards each BH, forming dense gas streams feeding their accretion minidiscs. This is well understood in literature surrounding planets in protostellar discs (e.g \citealt{Lubow1999}, \citealt{Ogilvie2002}, \citealt{Kley2012}). 

The self-consistently formed CSMDs in our simulations orbit prograde (anticlockwise) with respect to the SMBH disc. Gas on horseshoe orbits outside the Hill sphere of the BHs dominate over the Keplerian velocity shear across the BHs' Hill radii due to the SMBH, leading to a prograde CSMD  \citep[see also][]{Lubow1999}. If the initial conditions for our satellite BH orbits had some eccentricity with respect to the orbit around the AGN, this could lead to retrograde CSMDs \citep[see][]{Li2022_retroflows}, which warrants further study in this context. Distinguishing the sense of the disc rotation is important as it implies that when the BH accretion discs intersect their flows meet head on rather than in parallel if their disc rotations were opposite. This is depicted in the second panel of the figure as the density increases by three orders of magnitude above the AGN disc at the point where the two discs intersect. This pileup is also observed in simulations of protostellar disc collisions, see Figure 2 in \citet{Shen2010} for example.

Strong spiral gas outflows are generated during the close encounter (3rd panel) due to strong tidal stripping of the initially well defined circumsingle accretion discs of the BHs as they orbit the COM in prograde with respect to the AGN disc. These outflows remove a portion of the mass initially retained in the individual Hill spheres of the BHs. Another portion is accreted onto the BHs. After a relatively short period of time (few tens of orbits of the newborn binary), the spiral structure dissipates and the binary is sufficiently hardened so that a circum-binary mini-disc (CBMD) may form around it (4th panel). At this point the evolution of the binary is governed by more complex and secular interactions with the CBMD and inner gas dynamics. This marks the transition from our novel initial conditions to the much more widely studied
problem of a pre-existing binary satellite (see references in Section~\ref{sec:intro}) which we can compare with.

\subsubsection{Binary Dynamics}
\begin{figure}
    \centering
    \includegraphics[width=0.49\textwidth]{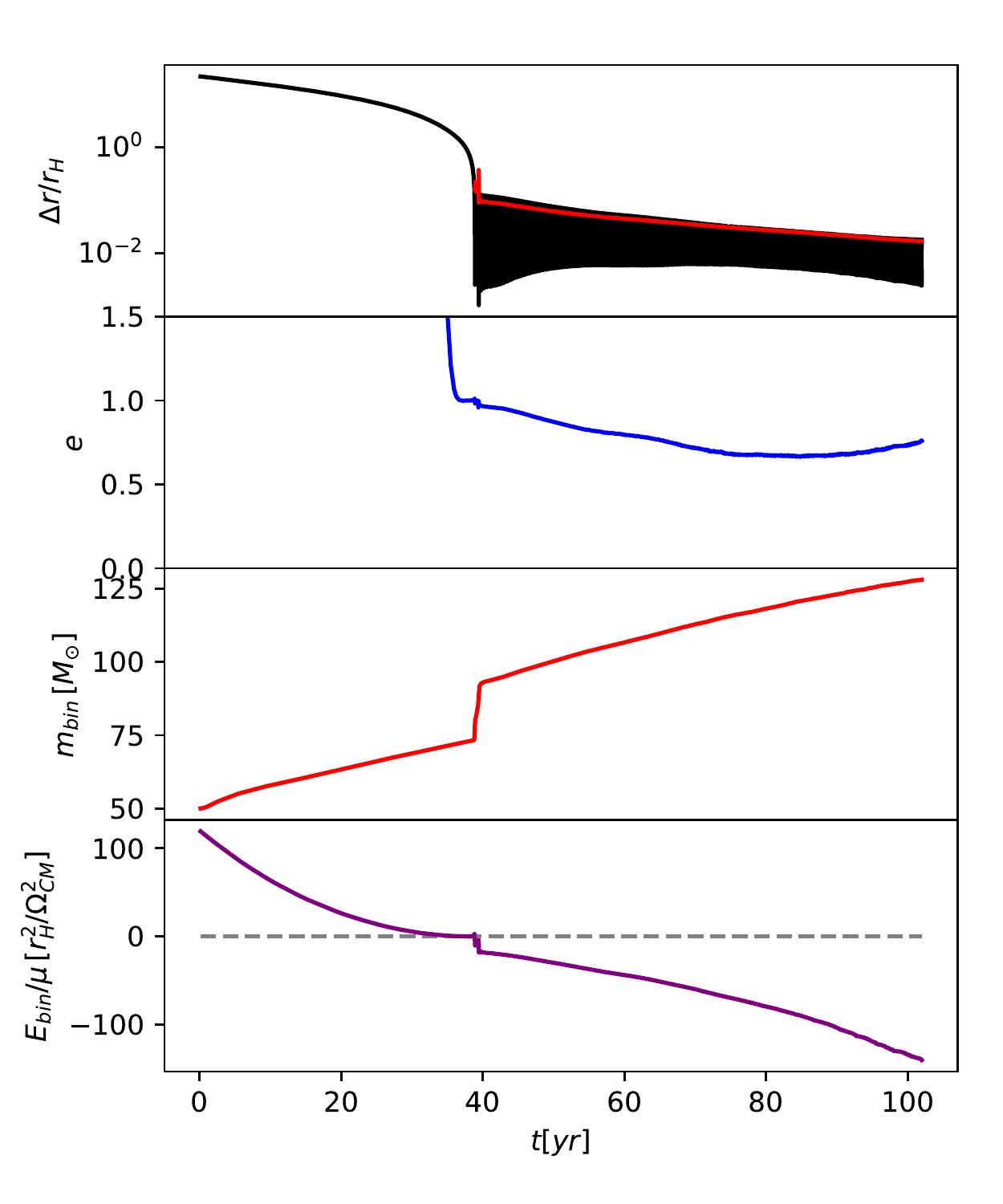}
    \caption{
    Binary orbital elements as a function of time. From top to bottom, panel (i) shows the separation as a fraction of the binary Hill sphere (black) with the semi major axis overlayed (red). (ii) The binary eccentricity showing the initial hyperbolic encounter with $e>1$. (iii) The binary mass as a function of time, showing a rapid accretion period during the encounter. Finally, (iv) the energy per unit mass of the binary which crosses the axis.}
    \label{fig:orb_elements_fiducial}
\end{figure}

Figure~\ref{fig:orb_elements_fiducial} shows the orbital elements of the simulated binary, including the eccentricity $e$, separation $\Delta r$ and semi-major axis $a$, in addition to the total binary mass $M_{\rm{bin}}$ and specific two-body energy as a function of time. We express our quantities in the natural units of the binary where $r_{\rm{H}}$ is the initial satellite Hill radius and $\Omega_{\rm{CM}}$ is the orbital period of the COM of the binary about the SMBH. The eccentricity is calculated via.
\begin{equation}
    \centering
    e = \sqrt{1+\frac{2E_{\rm bin}L^{2}}{G^{2}M^{2}_{\rm bin}\mu^{3}}}
    \label{eq:eccentricity},
\end{equation}
where $M_{\rm{bin}} = m_1+m_2$ is the binary mass, $\mu = m_1 m_2/M_{\rm{bin}}$ the reduced mass and G is the gravitational constant,  $L=\mu(\boldsymbol{r}_1-\boldsymbol{r}_2)\times(\boldsymbol{v}_1-\boldsymbol{v}_2)$ is the angular momentum with $\boldsymbol{v}_{i}$, $\boldsymbol{r}_{i}$ being the velocities and positions of the BH satellites $i=(1,2)$.
The two-body energy $E_{\rm bin}$ is calculated in the COM frame of the binary as

\begin{equation}
    \centering
    E_{\rm bin}=\frac{1}{2}\mu \|\boldsymbol{v}_1-\boldsymbol{v}_2\|^2 - \frac{GM_{\rm{bin}}\mu}{\|\boldsymbol{r}_1-\boldsymbol{r}_2\|},
    \label{eq:two_body_energy}
\end{equation}

\noindent Initially the BHs do not form a bound system, as implied by $\{e,E_{\rm bin}\}>0$. 
\newline\indent The binary forms at $t \sim 40\,\rm{yr}$, indicated by the sudden drop in separation and specific binding energy. At this point the energy of the binary becomes negative and the separation begins to oscillate as the BHs orbit their COM, indicating the BHs are now energetically bound and the capture is successful. After this point the binary is hardened, visible in the decreasing semi-major axis and orbital energy. As indicated by the drop in orbital energy, the energy dissipation during the first encounter is highly efficient. For capture to be successful this is necessarily so. Since the encounter occurs in such a short time, the positive energy of the binary must be dissipated within the time the binaries are in each other's Hill spheres, else they cannot remain bound. In practice, the energy must be reduced well below zero to form a permanently bound binary to ensure that the central SMBH does not disrupt loosely bound satellites. Furthermore, the panel on $M_{\rm{bin}}$ in Fig.~ \ref{fig:orb_elements_fiducial} shows that a significant amount of gas is also accreted during the chaotic outflows. This rate is considerably super Eddington at $\sim10 M_{\odot}/yr$. Such significant accretion driven mass-change effects can also enhance capture and stabilise satellite orbits \citep{Heppen77} by increasing the contribution of the negative two-body potential term in the internal energy equation in Eq.~\eqref{eq:two_body_energy} and the size of the Hill sphere of the satellites (see Eq.~\eqref{eq:hill}).
\newline\indent Our results agree well with analogous findings in the literature of proto-planetary disc collisions \citep[e.g.][]{Watkins1998a,Watkins1998b,Shen2010,Munoz2015} and even behaves similarly to the star-disc encounters of \citet{Boffin1998} provided the star actually crosses the accretion disc in their simulations like the BHs do in this case. In particular the sudden drop in orbital energy over one period we observe is consistent across the literature, provided there is intersection with the disc. Likewise, the sudden drop in eccentricity and semi major axis followed by a gradual decrease is also consistent with \citet{Munoz2015}, where the post encounter orbital evolution is also considered. The origin for the later increase in eccentricity that occurs after roughly 80 years in our simulation (equivalent to $\sim$ 1000 inner binary orbits) is unclear, however the range in the eccentricity is particularly interesting. In our case, we find the binary eccentricity does not decrease below 0.6, consistent with \citet{Munoz2015}. Provided AGN binary satellite models assume a gas capture formation origin, as \citet{Tagawa2020} predicts the majority will be, this draws the zero eccentricity assumption in most pre-existing gas-embedded binary simulations that the binary should be initialised with minimal eccentricity into question, in line with the analytic predictions of \citealt{Tagawa2021}.

\subsection{Orbital energy dissipation}
\label{sec:fiducial_E_diss}
To understand the change in orbital energy of the binary at capture the various mechanisms which can alter its energy must be understood. There are three such mechanisms in our simulations which we label and describe separately below as the energy per unit time removed/added to the binary:
\begin{enumerate}

    \item \textbf{SMBH interaction}, $\varepsilon_{\rm SMBH}$ - energy per unit time exchanged by the binary system and the SMBH.
    \item \textbf{Gas gravitational dissipation}, $\varepsilon_{\rm grav}$ - from the gravitational interaction with the surrounding gas.
    \item \textbf{Accretion}, $\varepsilon_{\rm acc}$ - due to conservation of linear momentum of accreted gas particles onto the BHs.
\end{enumerate}
\noindent These quantities represent the rate of work done on the binary which we calculate per unit mass of the binary
by taking the time derivative of eq.~\eqref{eq:two_body_energy}, see appendix \ref{app:appendix_derivations}. The result is
%
\begin{equation}
    \varepsilon = \frac{d}{dt}\bigg(\frac{E_{\rm bin}}{\mu}\bigg) = (\boldsymbol{v}_{1}-\boldsymbol{v}_{2})\cdot(\boldsymbol{a}_{1}-\boldsymbol{a}_{2}) - \frac{G\Dot{M}_{\rm bin}}{\|\boldsymbol{r}_{1}-\boldsymbol{r}_{2}\|}.
    \label{eq:dissipation_general_form}
\end{equation}
Here the first term represents the work done by all external forces (i.e. excluding their mutual attraction) and momentum transfer which drive a relative acceleration $\boldsymbol{a}_{1}-\boldsymbol{a}_{2}$ of the satellites ($\boldsymbol{v}_{1}$ and $\boldsymbol{v}_2$ is the relative velocity) and the second term is sourced by accretion, resulting from the increase in the second term in eq \eqref{eq:two_body_energy}.
The contribution corresponding to $\varepsilon_{\rm SMBH}$ is calculated from the force differential due to the gravitational interactions with the SMBH, 
%
\begin{equation} 
\varepsilon_{\rm SMBH} =(\boldsymbol{v}_{1}-\boldsymbol{v}_{2})\cdot
(\boldsymbol{a}_{1,\rm SMBH}-\boldsymbol{a}_{2,\rm SMBH})
\label{eq:work_SMBH}
\end{equation}
where the acceleration of BH $i=(1,2)$ due to the SMBH is
\begin{equation}
\boldsymbol{a}_{i,\rm SMBH} = GM_{\rm SMBH}\frac{(\boldsymbol{r}_{i}-\boldsymbol{r}_{\rm SMBH})}{||\boldsymbol{r}_{i}-\boldsymbol{r}_{\rm SMBH}||^{3}}\,.
\end{equation}
Here $M_{\rm SMBH}$ is the mass of the SMBH and $\boldsymbol{r}_{\rm SMBH}$ is its position, which is generally very close to centre of mass of the entire simulation. 

The gas gravitational dissipation term $\varepsilon_{\rm grav}$ is analogous to $\varepsilon_{\rm SMBH}$ by summing over the gravitational forces acting on the BHs from all $N_{\rm p}$ gas particles of mass $m_{\rm p}$, 

\begin{equation} 
\varepsilon_{\rm grav} =(\boldsymbol{v}_{1}-\boldsymbol{v}_{2})\cdot
(\boldsymbol{a}_{1,\rm gas}-\boldsymbol{a}_{2,\rm gas}),
\label{eq:work_gas}
\end{equation}
where the acceleration of BH $i=(1,2)$ due to the gas is
\begin{equation}\label{eq:work_grav}
\boldsymbol{a}_{i,\rm gas} = \sum_{p=1}^{N_p} Gm_{p}\frac{(\boldsymbol{r}_{i}-\boldsymbol{r}_{p})}{||\boldsymbol{r}_{i}-\boldsymbol{r}_{p}||^{3}}\, .
\end{equation}

The accretion term $\varepsilon_{\rm acc}$ represents the work done on the binary due to the change in the linear momentum of the BHs when they accrete gas particles. Upon accretion of a particle by a BH, the BH's position, velocity and acceleration are modified by a mass weighted average of all $N_{\rm acc}$ accreted particles during that timestep as
\begin{align}
    \Delta \boldsymbol{a}_{i} &= \frac{
    M_{i}\boldsymbol{a}_{i} + m_{\rm p}\sum_{j}^{N_{\rm acc}}\boldsymbol{a}_{\rm p,j}}{M_{i}+N_{\rm acc}m_{\rm p}}-\boldsymbol{a}_{i}\,.
    \label{eq:accacc}\\
    \Delta\boldsymbol{ v}_{i} &= \frac{M_{i}\boldsymbol{v}_{i}+m_{\rm p}\sum_{j}^{N_{\rm acc}}\boldsymbol{v}_{\rm p,j}}{M_{i}+N_{\rm acc}m_{\rm p}}-\boldsymbol{v}_{i}\,.
    \label{eq:accvel}
\end{align}
Here $\boldsymbol{a}_{i}$, $\boldsymbol{v}_{i}$ are the accelerations and velocities of the $i=(1,2)$ satellite BHs prior to accretion during that timestep (since accretion is evaluated last), $\Delta \boldsymbol{a}_{i}$ is the contribution from the SPH particles to the $i$th BH's acceleration just prior to accretion and $ \Delta{\mathbf  v}_{i}$ is the change in its velocity due to the impulsive momentum transfer upon accretion. Here $\boldsymbol{a}_{\rm p,j}$ and $\boldsymbol{v}_{\rm p,j}$ are the SPH particles' acceleration and velocity upon accretion. The change in velocity, whilst instantaneous, can be treated as an acceleration over the length of the timestep, $\Delta t$. Using eqns.~\eqref{eq:accacc} and \eqref{eq:accvel} dissipation due to accretion can be written as
\begin{equation}
    \varepsilon_{\rm acc} \approx(\boldsymbol{v}_{1}-\boldsymbol{v}_{2})\cdot(\boldsymbol{a}_{1,\rm acc} - \boldsymbol{a}_{2,\rm acc}) - \frac{G\dot{M}_{\rm bin}}{\|\boldsymbol{r}_1-\boldsymbol{r}_2\|}.
\end{equation}
where
\begin{equation}
    \boldsymbol{a}_{i,\rm acc}=\bigg(\Delta{\mathbf  a_{i}}+\frac{\Delta{\mathbf v}_{i}}{\Delta t}\bigg),
\label{eq:work_acc}
\end{equation}
assuming that the impulsive momentum transfer takes place at the end of the timestep. 
Note that this is a non exact value of $\varepsilon_{\rm acc}$ since the $\Delta{\mathbf  v}/\Delta t$ term is an average over the timestep. This approximation is more accurate the higher the particle accretion rate. Unlike in grid codes where a smooth sink \textit{rate} is used, SPH codes have the caveat that accretion relies on discrete particle accretion which can be highly volatile if the number of accreted particles per timestep is low. 

\begin{figure*}
    \centering
    \includegraphics[width=19cm]{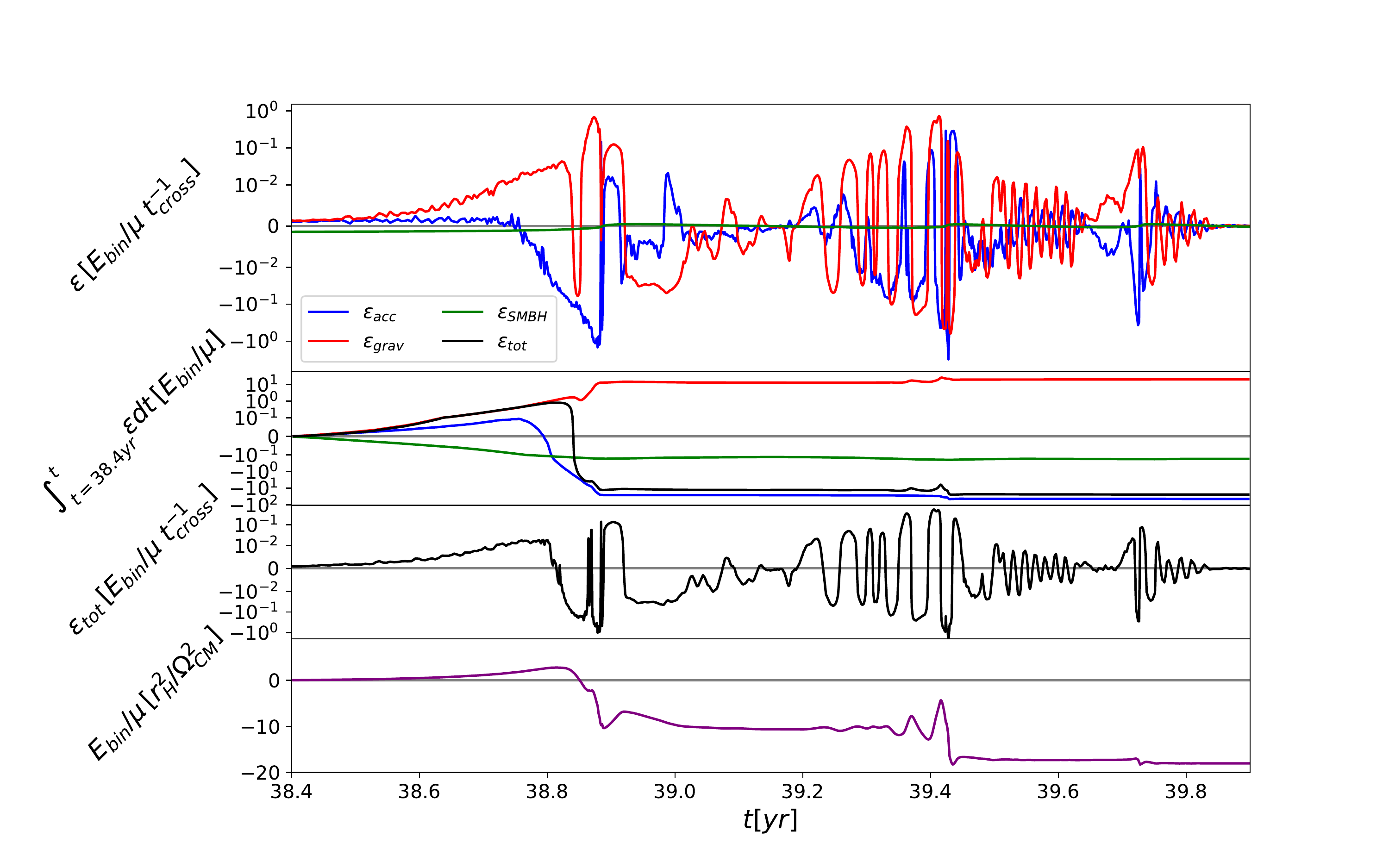}

    \caption{\textit{1st row:} The contribution to the rate of binary orbital energy dissipation, in units of initial binary orbital energy upon entering the Hill sphere per crossing time, for each individual processes: accretion ($\varepsilon_{\rm acc}$, Eq.~\ref{eq:work_acc}), gas gravity ($\varepsilon_{\rm grav}$, Eq.~\ref{eq:work_gas}), and SMBH tidal force ($\varepsilon_{\rm SMBH}$, Eq.~\ref{eq:work_SMBH}).\textit{2nd row:} The cumulative contribution of the work done by all three mechanisms including the cumulative total. \textit{3rd row:} The sum of the three components shown in the top panel, $\epsilon_{\rm tot}$. \textit{4th row:} The binding energy of the two BHs from eq \eqref{eq:two_body_energy}. The sharp dip in $\epsilon_{\rm tot}$ during the close approach leads to a significant permanent decrease of the orbital energy. 
    }
    \label{fig:diss_vs_t}
\end{figure*}
Figure \ref{fig:diss_vs_t} shows the rate of work due to each component during the binary capture process, including their cumulative value over the domain of the plot.
In the figure, dissipation due to the local gas gravity is initially positive, due to strong density pileups at the inner edge of their circumsingle discs (ahead of the individual BHs) generated as the BHs accelerate through the local gas. After the first encounter these inhomogeneities begin to orbit the BHs, leading to small oscillations in $\varepsilon_{\rm grav}$. Just prior to the first pericentre passage, which occurs at 38.8yr, strong accretion removes a significant portion of energy. This is the hallmark of the gas capture process. The energy lost at this moment allows for the energetic retention of the binary, as indicated in the bottom panel showing the total energy of the binary transitioning form positive to negative, dipping after each of the two first encounters. If the binary did not dissipate enough energy at the first encounter then it's energy would remain positive and the two objects would remain on an unbound orbits in the two body regime. We find the force differential due to the SMBH is insignificant compared to forces due to accretion and gas gravity. From the cumulative work done, the gas is shown to actually do net positive work on the binary while accretion leads to negative work, overcoming the positive contribution from $\varepsilon_{grav}$.

\begin{figure*}
    \includegraphics[width=0.32\textwidth]{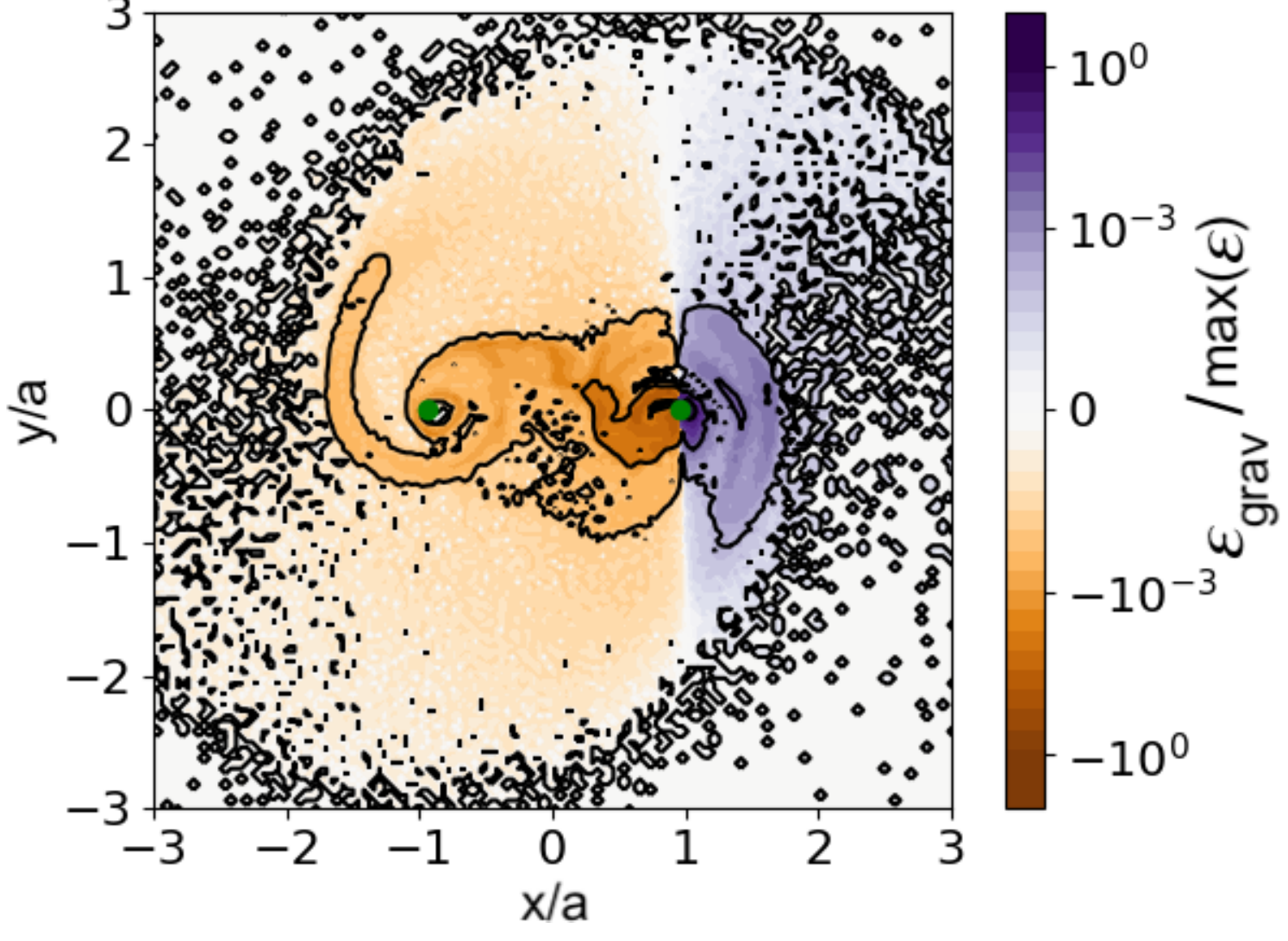} 
    \includegraphics[width=0.32\textwidth]{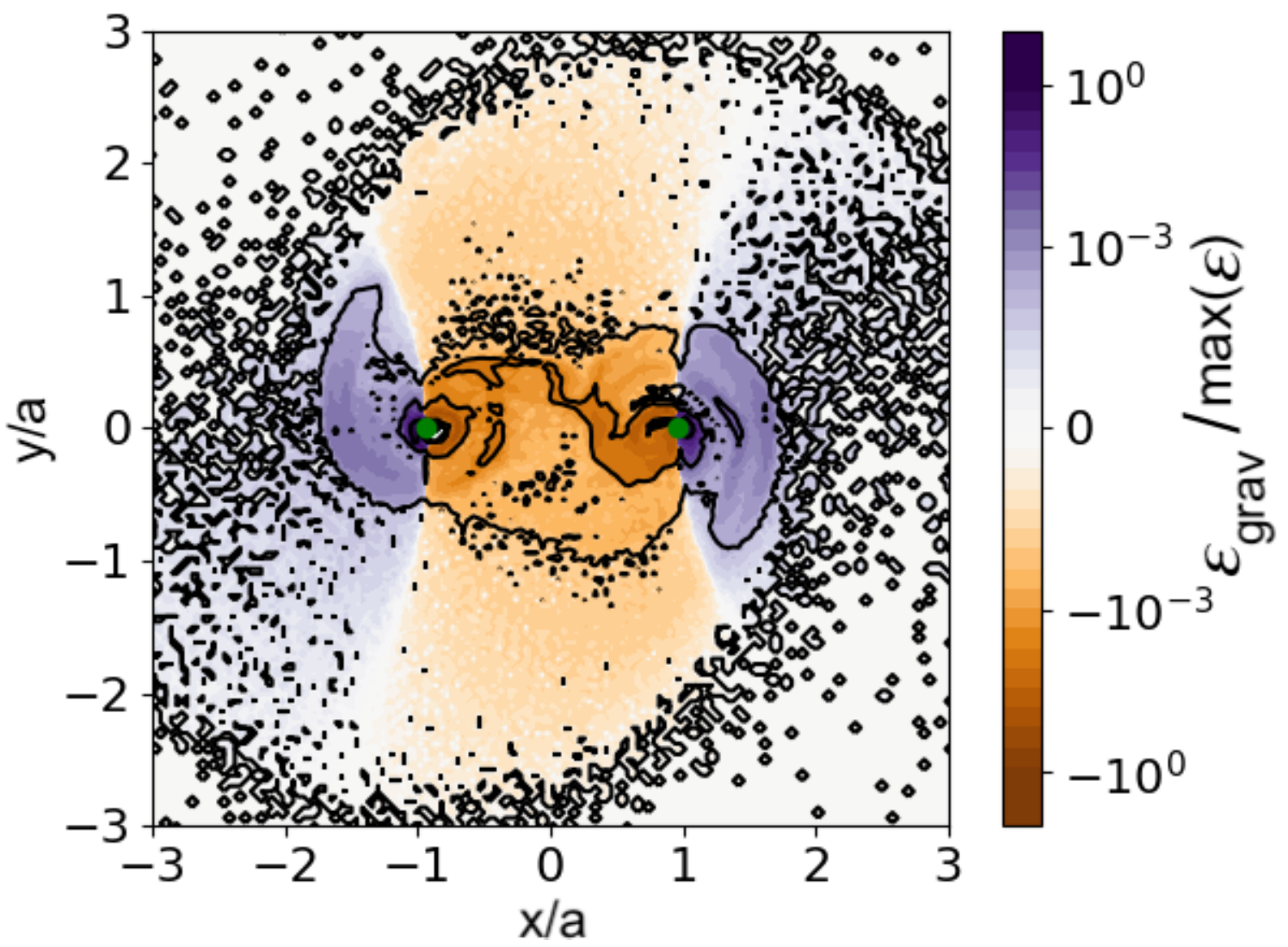}
    \includegraphics[width=0.32\textwidth]{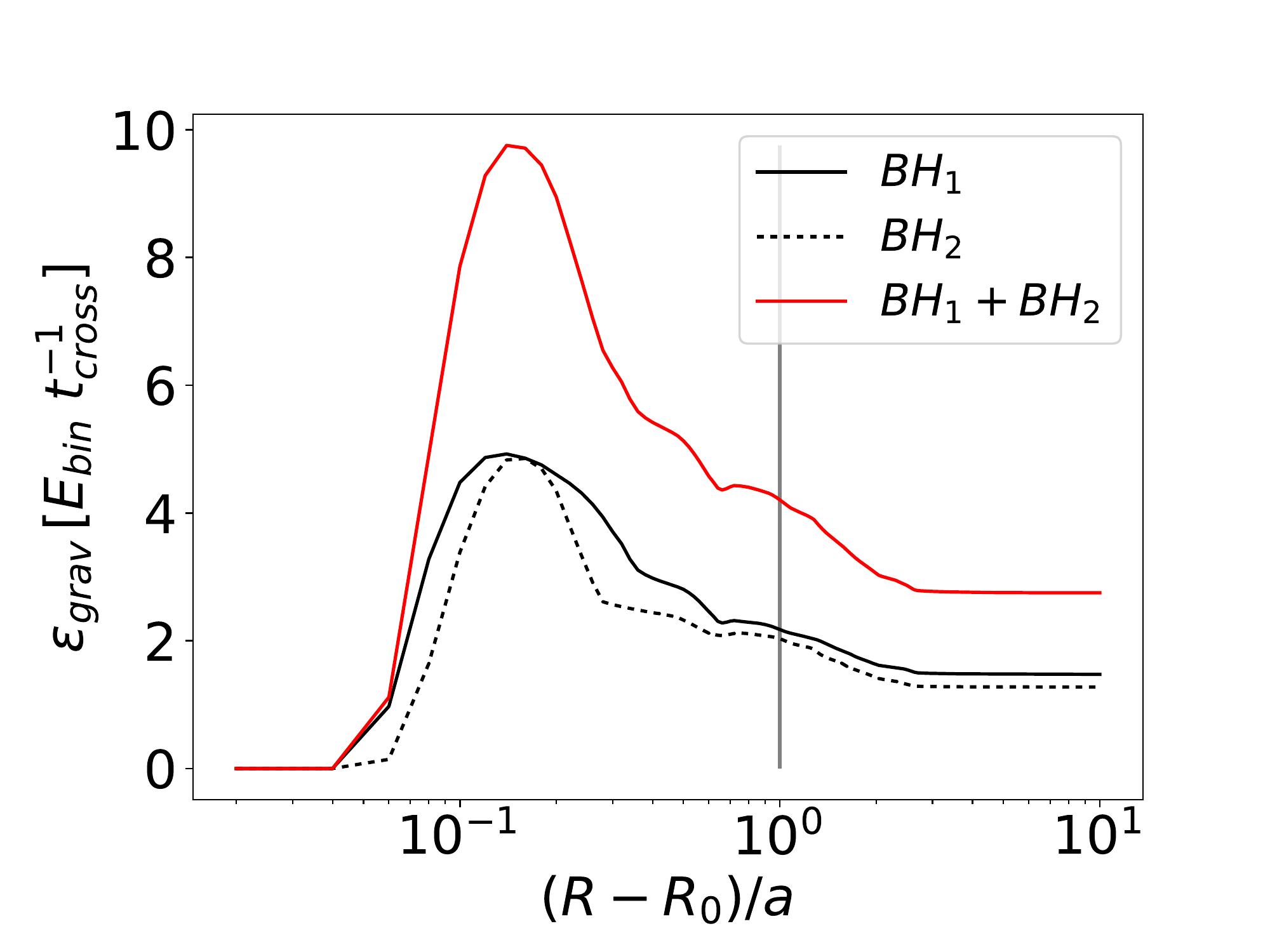} 
\caption{The rate of work done by the gas gravity, in units of the energy upon entering each other's Hill sphere per crossing time, on the binary ($\varepsilon_{\rm grav}$) at 39.01yr, shortly after the first close approach at 38.88yr. Left and middle panels show respectively the 2D maps of the gravitational work density by the gas on one binary component and the full binary at a representative timestep. The right panel shows the cumulative radial distribution of $\varepsilon_{\rm grav}$ on each BH, respectively as well as the total. Black holes are represented by green dots at (-1,0) and (1,0).}
\label{fig:diss_map}
\end{figure*}

The local gas gravity is shown to vary on scales shorter than the binary period. Figure~\ref{fig:diss_map} shows 2D maps of $\varepsilon_{\rm grav}$, the gas gravitational component to the rate of work or binary energy dissipation rate in the plane of the binary. To achieve this, gas particles are binned in a $200\times 200$ grid in $x$-$y$ and the sum in Eq.~\eqref{eq:work_grav} is restricted accordingly. In each of these bins, the dissipation can be positive or negative depending on whether the gas is lagging behind or ahead of the path of the BHs respectively. The 2D dissipation maps are shown separately with respect to only one of the BHs or both BHs to illustrate the morphology of the work done on a single object and the binary system during the first encounter. Also shown is the cumulative radial energy dissipation constructed by radially binning the particles in $R=\sqrt{x^{2}+y^{2}}$ from the centre of each BH.

The cumulative dissipation as a function of $R$ from the BHs is shown in Figure~\ref{fig:diss_map}, right panel in units of the binary energy when the BHs enter each other's Hill sphere per the crossing time $t_{cross}$ for the Hill sphere, where we have taken the crossing time to be $r_{H}$ divided by the relative speed when entering the Hill sphere.  The figure indicates a very strong energy transfer onto the binary from material very close to the BHs ($\sim 0.1a$), while the gravity of the gas removes energy outside this very small distance. We verify that during this period there is a resolution of $\sim$100,000 SPH particles within $r_{H}$, with particle densities being highest in the CMSDs. Therefore, we confidently rule out the variation emerging from small number statistics close to the sinks in the simulation.
Outside of $3a$ we find no significant effect of the gas on the binary energy. The two other panels showing the 2D distribution show very dense gas buildup just ahead of the BHs and (less) dense trails following them. Though the size of the trails are larger than the density pileups, they are not able to overcome the acceleration on the BHs due to material ahead of their path, leading to a net gain in energy of the binary due to the local gas as shown in the third panel. This picture reflects the findings and discussion of positive torque sources in the isolated binary simulations of \citet{Tiede2020}. 

\subsubsection{Angular momentum transfer}
The specific torques due to each of the three physical phenomena described earlier can be similarly labelled as $\tau_{\rm SMBH}$, $\tau_{\rm grav}$ and $\tau_{\rm acc}$. Their calculation is analogous to equations \eqref{eq:work_SMBH}, \eqref{eq:work_gas}, and \eqref{eq:work_acc} by changing the dot product to a cross product i.e 
\begin{align}
\tau_{\rm SMBH}&=(\boldsymbol{r}_{1}-\boldsymbol{r}_{2})\times 
(\boldsymbol{a}_{1,\rm SMBH}-\boldsymbol{a}_{2,\rm SMBH})\,, \label{eq:SMBHtorque}\\
\tau_{\rm grav}&=(\boldsymbol{r}_{1}-\boldsymbol{r}_{2})\times 
(\boldsymbol{a}_{1,\rm gas}-\boldsymbol{a}_{2,\rm gas}),\,
\label{eq:gravtorque}\\
\tau_{\rm acc} &\approx 
(\boldsymbol{r}_{1}-\boldsymbol{r}_{2})
\times
\left[
\left(
\Delta\boldsymbol{a}_{1}+\frac{\Delta\boldsymbol{V}_{1}}{\Delta t}\right)-
\left(
\Delta\boldsymbol{a}_{2}
+\frac{\Delta\boldsymbol{V}_{2}}{\Delta t}
\right)
\right]
\,.
\label{eq:acctorque}
\end{align}

\begin{figure*}
    \centering
    \includegraphics[width=19cm]{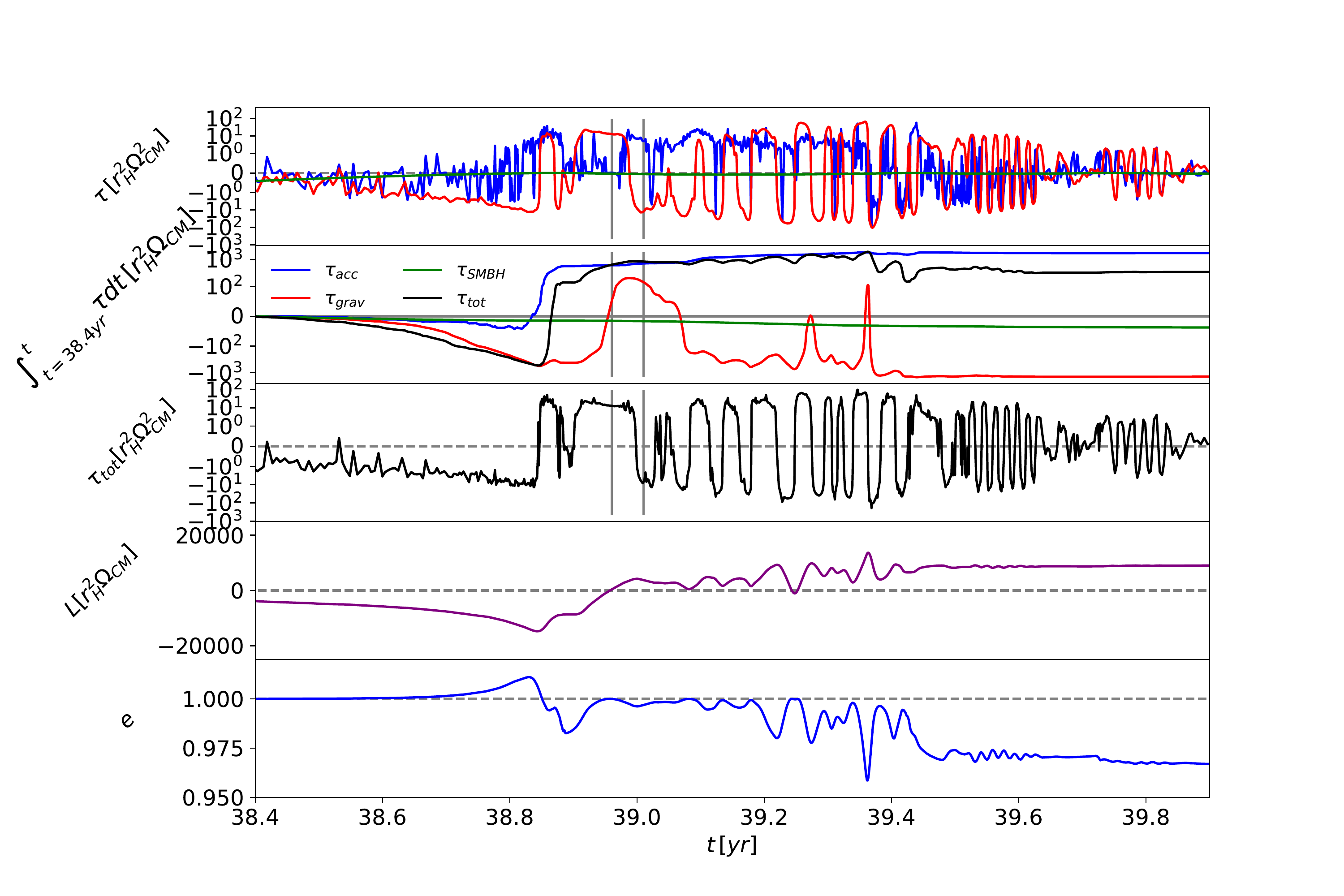}
    \caption{\textit{1st row:} The contribution of different processes to the specific torque:  accretion ($\tau_{\rm acc}$, Eq. \eqref{eq:acctorque}), gas gravity ($\tau_{\rm grav}$, Eq. \eqref{eq:gravtorque}), and presence of SMBH ($\tau_{\rm SMBH}$,  Eq. \eqref{eq:SMBHtorque}, practically negligible throughout). \textit{2nd row:} Cumulative contribution of each torque mechanism over the figure's time domain showing accretion dominates and circularises the binary. \textit{3rd row:} The net torque from the sum of the three components shown in the top. \textit{4th row:} The orbital angular momentum of the BH binary around its center of mass, its component perpendicular to the gas disk. \textit{5th row:} Binary eccentricity in the corotating frame (Eq. \eqref{eq:eccentricity}). Though the torques are considerably stochastic, there is a systematic increase in angular momentum and a decrease in eccentricity. The vertical grey lines indicate the positions where 2D snapshots of the $\tau_{\rm grav}$ are taken in Figure \ref{fig:torque_map_fiducial}.
    } 
    \label{fig:T_vs_t_enc}
\end{figure*}
Following the same procedure as in the previous section for the energy change. Figure \ref{fig:T_vs_t_enc} shows the instantaneous torque contribution of each component as a function of time. Initially, the gas gravitational torque ($\tau_{\rm grav}$) dominates the torque until the BHs cross each other's CSMDs at 38.8 yr and accretion ($\tau_{\rm acc}$) promptly dominates. Significant oscillations are seen in the gas gravitational torques acting on the binary, matching the periodicity of the energy dissipation rate in Figure~\ref{fig:diss_vs_t}. Shortly following the first encounter the net gas torque is dominated by the gravitational attraction of the gas in the immediate vicinity of the objects. On top of these oscillations, there is a net increase in angular momentum from accretion which helps circularise the binary. The eccentricity is then reduced during each encounter (see Figure \ref{fig:orb_elements_fiducial}). 

\begin{figure*}
\begin{tabular}{cc}
    \includegraphics[width=0.40\textwidth]{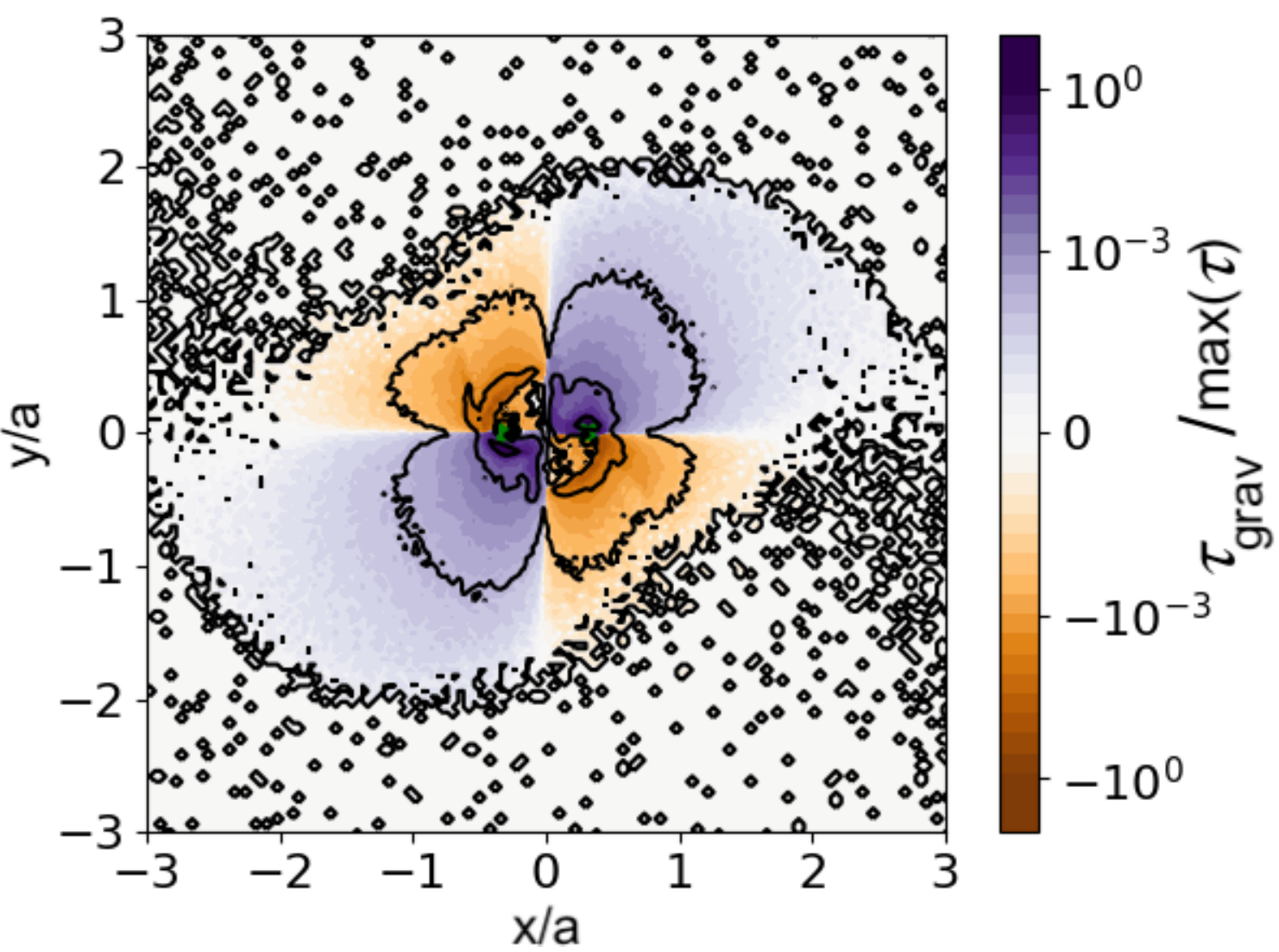} &

    \includegraphics[width=0.40\textwidth]{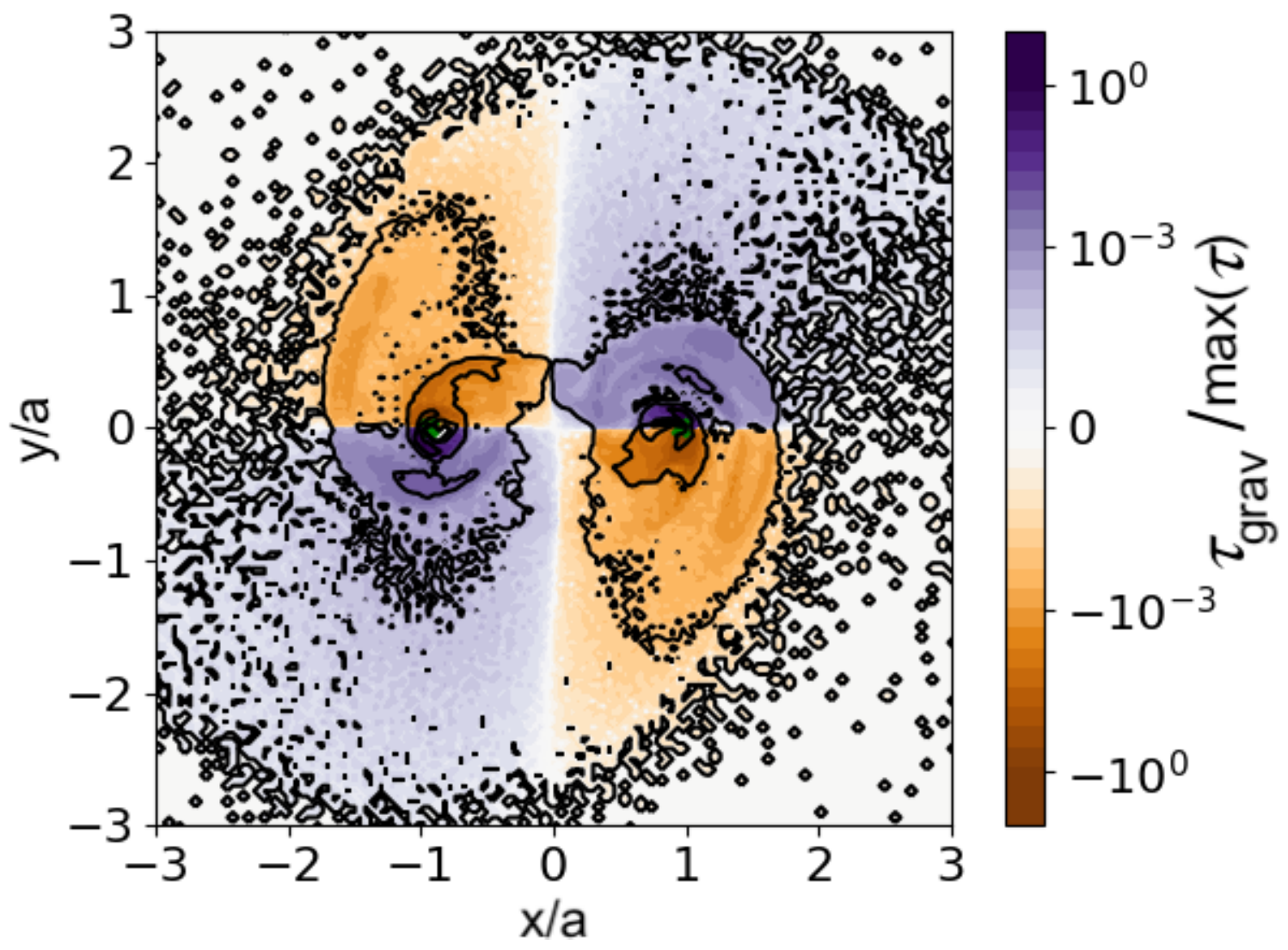} \\
    
    \includegraphics[width=0.40\textwidth]{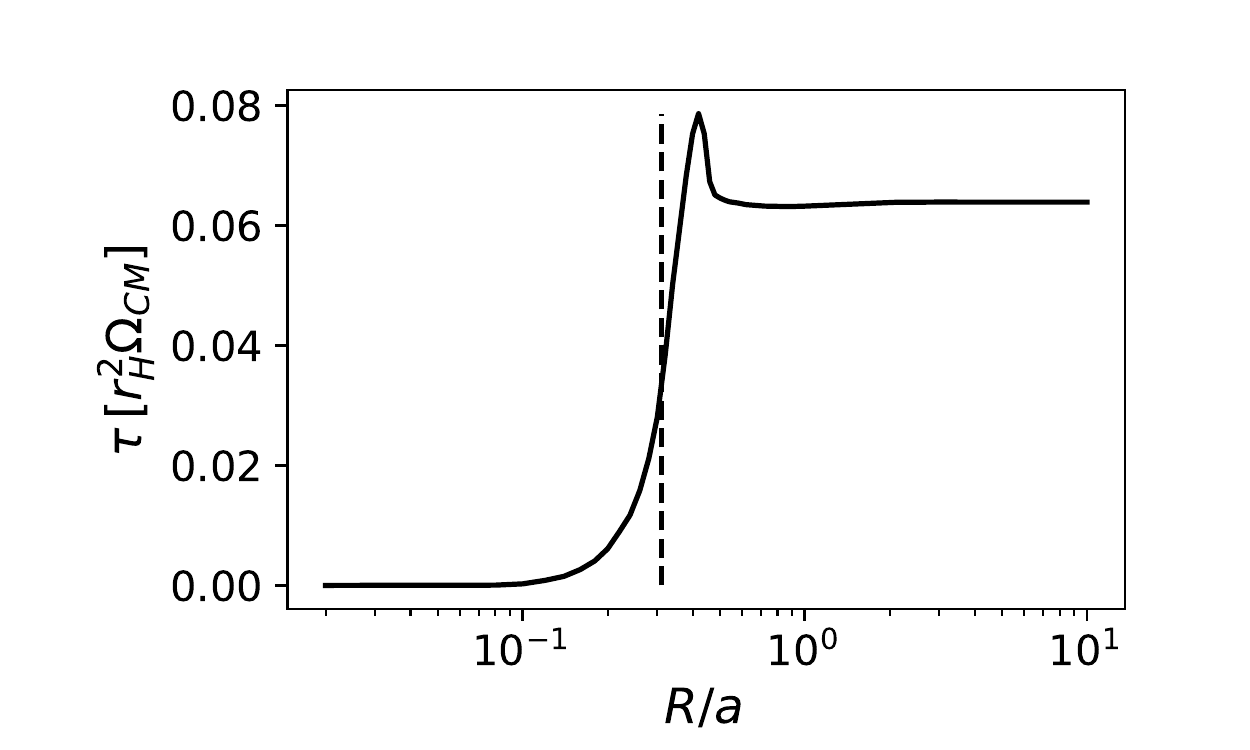} &
    \includegraphics[width=0.40\textwidth]{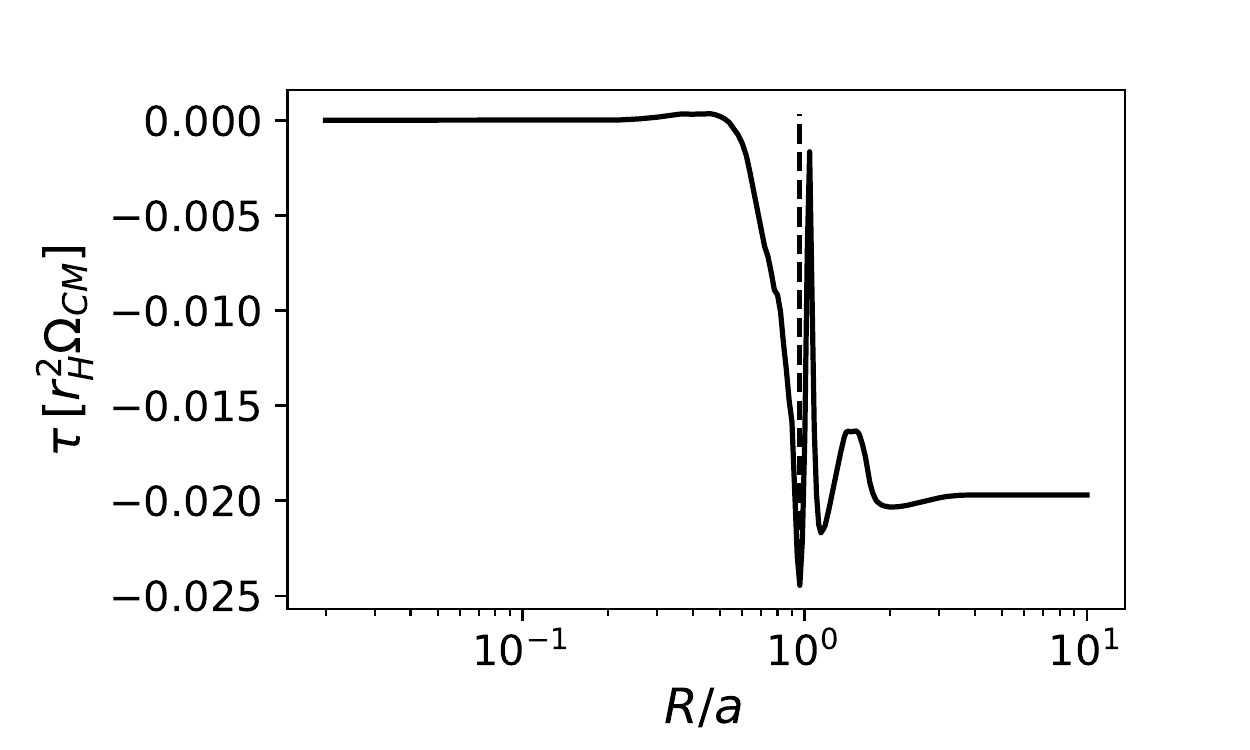}

\end{tabular}

\caption{(Top row, left to right) Gravitational torque per unit area, normalised to the maximum value within the 200x200 grid shown at 39.8 and 39.01 yrs respectively. Black contours are used to outline equally dense regions of gas. In the second panel, trailing gas structure begins to form behind the binary which is orbiting counter-clockwise (prograde with AGN disc). (Bottom row, left to right) cumulative specific torque on binary as a function of distance from the COM at the same respective times as the above 2D plots. Vertical dashed lines show the position of binary BHs in the snapshot. These show that the strongest contributions to the gravitational torque come from gas very close to the individual BHs and can change sign within a single orbit as the nonaxisymmetric perturbations to the CSMDs orbit their BHs.} 
\label{fig:torque_map_fiducial}
\end{figure*}
Figure \ref{fig:torque_map_fiducial} shows the 2D torque map at two different times within a single orbit (39.8 yrs and 39.01 yrs) to show the spatial source of the torques when their net value is positive and negative, along with the cumulative torque as a function of distance from the COM. The times of both snapshots are shown as vertical lines in Fig. \ref{fig:T_vs_t_enc}. In the left panel the net torque on the binary is positive, corresponding to where the torque is consistently positive, during the first encounter, before it begins to oscillate (as shown in Figure~\ref{fig:T_vs_t_fiducial}). The primary source of the positive contribution to the torque occurs at the separation of the BHs (dashed 1st line in the plot). This indicates that the main source of torque arises from structure at the orbital radius of the binary.
During this period, the aforementioned density pileups dominate the gravitational torques, shown in the top plot as the large purple wakes ahead of the BHs in the top left plot, outlined by the innermost black density contours. In the next 0.2 years the dense, truncated, CSMDs are significantly perturbed by the owners' sibling BH and in the frame of each BH a large cylindrically nonaxisymmetric lump of gas begins to orbit around each individual BH. At 39.01 yrs (right column of  Figure \ref{fig:torque_map_fiducial}) the cumulative torque is dominated by negative torques at similar distances to the BH as the positive torques in the left column. The 2D torque map at this later time shows that the gas density close to the BH has larger and more extended contours in the negative torque (yellow) quadrants. It is this asymmetry close to the BH which orbits around the BHs and leads to strong oscillations in the gravitational torque until the second close encounter where the discs are nearly entirely disrupted due to the extreme depth of the periapsis.  

The change in eccentrity during the encounter is consistent with findings for stellar accretion disc encounters (e.g \citet{Munoz2015} also finding an eccentricity decrease after each periapsis passage. Additionally, \citet{Bonetti+2020} also find that for massive black hole binaries subject to dynamical friction with mass ratios larger than $10^{-3}$, the eccentricity also decreases.

\subsection{Binary evolution}
In the previous section it was shown that the initial encounter(s) of the binary can reduce the energy (and also eccentricity) of the binary from being positive to negative and thus energetically bound via interaction with the gas. Here, we discuss the subsequent secular evolution of the torques, dissipation mechanisms and resultant effect on the binary orbital elements. 

\subsubsection{Secular Energy dissipation}
\begin{figure}
    \centering
    \includegraphics[width=8cm]{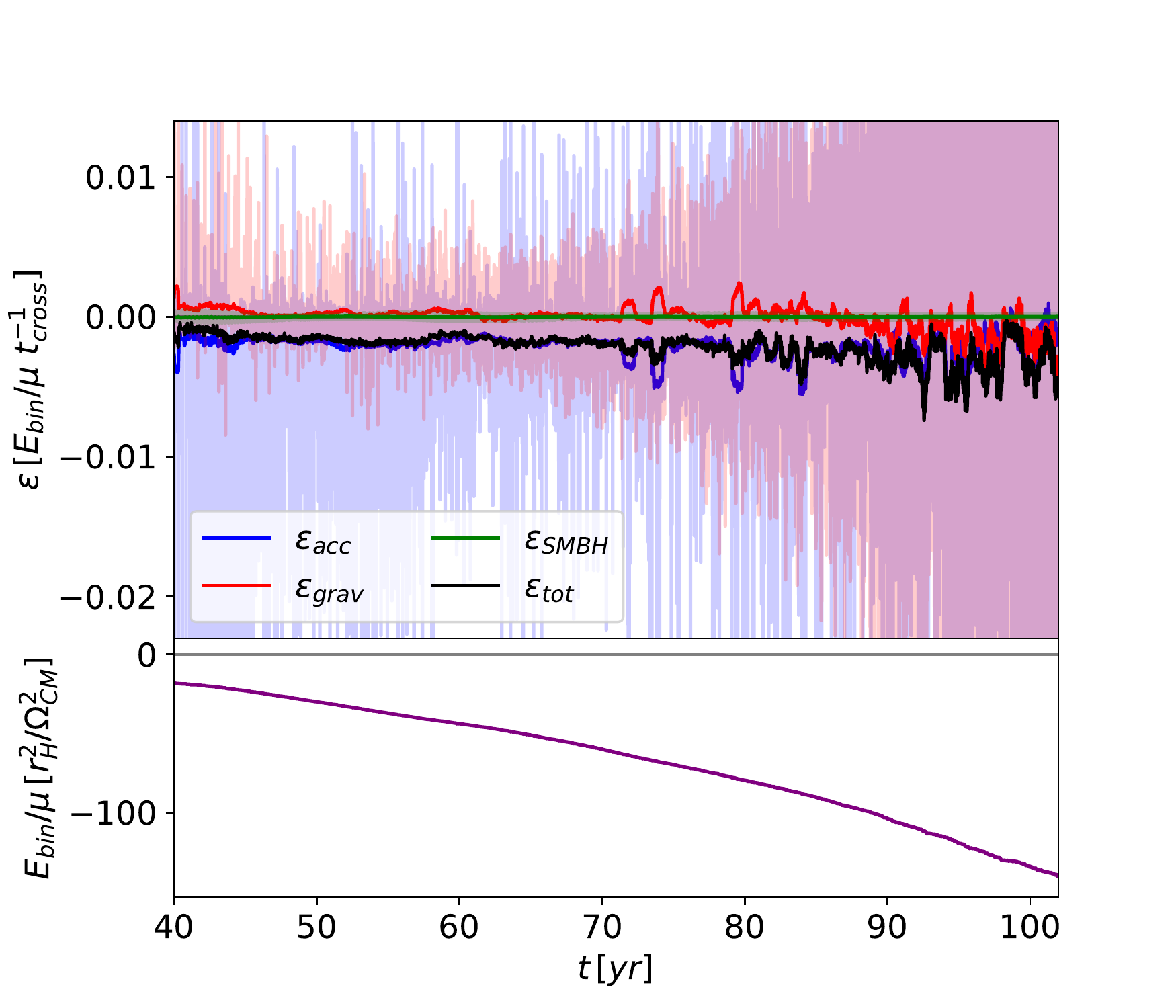}
    \caption{(Top panel) Energy dissipation as a function of time for each of the three mechanisms showing a net loss of energy from the binary. Bold lines show the running mean compared to the raw data in the background (opaque lines). (Bottom Panel) Binary energy as a function of time showing directly the energy loss over this time period.}
    \label{fig:Ediss_vs_t_fiducial}
\end{figure}

Figure \ref{fig:Ediss_vs_t_fiducial} shows the energy dissipation from each physical mechanism as a function of time for the period following the capture, alongside the binary energy. The results show a continual removal of energy from the binary predominantly due to accretion until $\varepsilon_{\rm grav}$ drops below zero. The reason for the flipping in $\tau_{\rm grav}$ is unclear. 
Comparing the raw data to the running mean, which has a window of one year, their values differ on average by about an order of magnitude, indicating significant variability in the forces the binary experiences. The running mean clearly indicates a systematic near linear rate of binary energy loss as a function of time, corroborated by the binary energy vs. time panel which displays relatively low variability during its gradual decrease over the simulation runtime. Since we find accretion to have a hardening effect on the binary, it suggests that its inclusion in works such as \citet{Li2021} that neglect accretion and have many outspiraling embedded binaries could have different results if accretion were included. 

\subsubsection{Secular Torques}
\begin{figure}
    \centering
    \includegraphics[width=8cm]{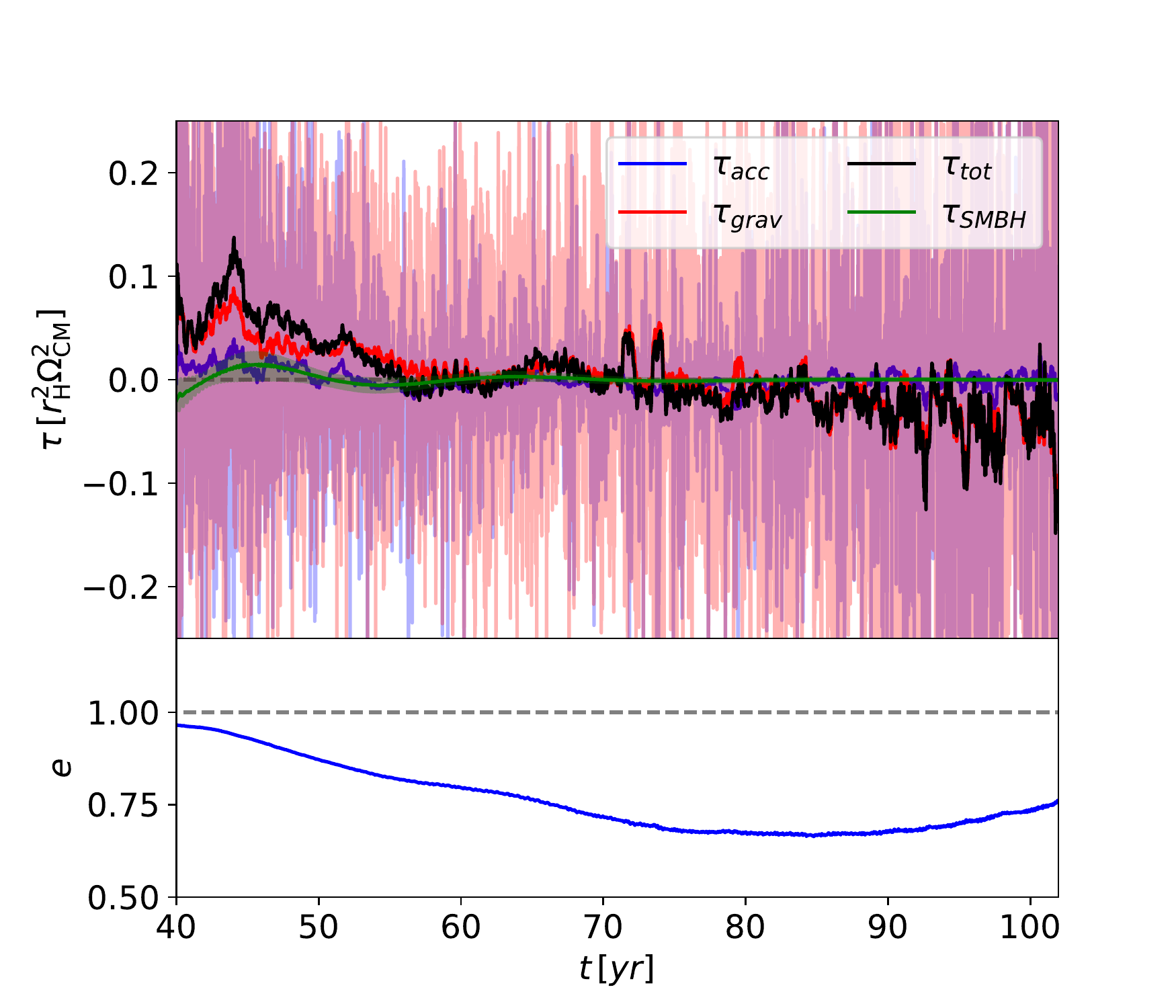}
    \caption{(Top panel) Torques as a function of time for each of the three mechanisms showing a net loss of angular momentum from the binary. Bold lines show the running mean compared to the raw data in the background (opaque lines). (Bottom Panel) Binary eccentricty as a function of time.}
    \label{fig:T_vs_t_fiducial}
\end{figure}
The torques as a function of time, alongside the eccentricity are displayed in Figure \ref{fig:T_vs_t_fiducial}. For our fiducial case, the torques are dominated by the gas gravity as shown by the behaviour of $\tau_{\rm grav}$. Curiously, for the first half of the evolution, up to around 80 years, the net torque is initially positive and later switches to negative values. The result of this shift is an initially decreasing eccentricity before plateauing and then increasing gradually. Comparing this result to works looking at pre-existing gas-embedded binaries, we find that this result disagrees with those the wind tunnel simulations of \citet{Li2022} where their initially eccentric binaries still experience damping, though their most eccentric binary ($e=0.5$) is still less eccentric than the minimum achieved here. The torque reversal is also inconsistent with the findings of \citet{Dempsey2022} since we find that for $a/r_{\rm H}<0.05$ the torques are positive and for $a/r_{\rm H}>0.05$ they are negative, just the opposite as in \citet{Dempsey2022}. However their study agrees on the separation where the torques reverse sign. In the fiducial case we find that the torques reverse when the binary has semi-major axis $a/r_{\rm H}=0.05$ which is reasonably close to the the turnover at $a/r_{\rm H}=0.1$ in \citet{Dempsey2022}. However, there are several important differences in these studies which could play a role in the binary evolution. Most notably, \citet{Dempsey2022} consider pre-existing low-eccentricity binaries, in a shearing box configuration with no viscosity implementation. With these differences they achieve higher resolution around their binaries than shown here in this paper.

\begin{figure*}
\begin{tabular}{cc}
    \includegraphics[width=0.40\textwidth]{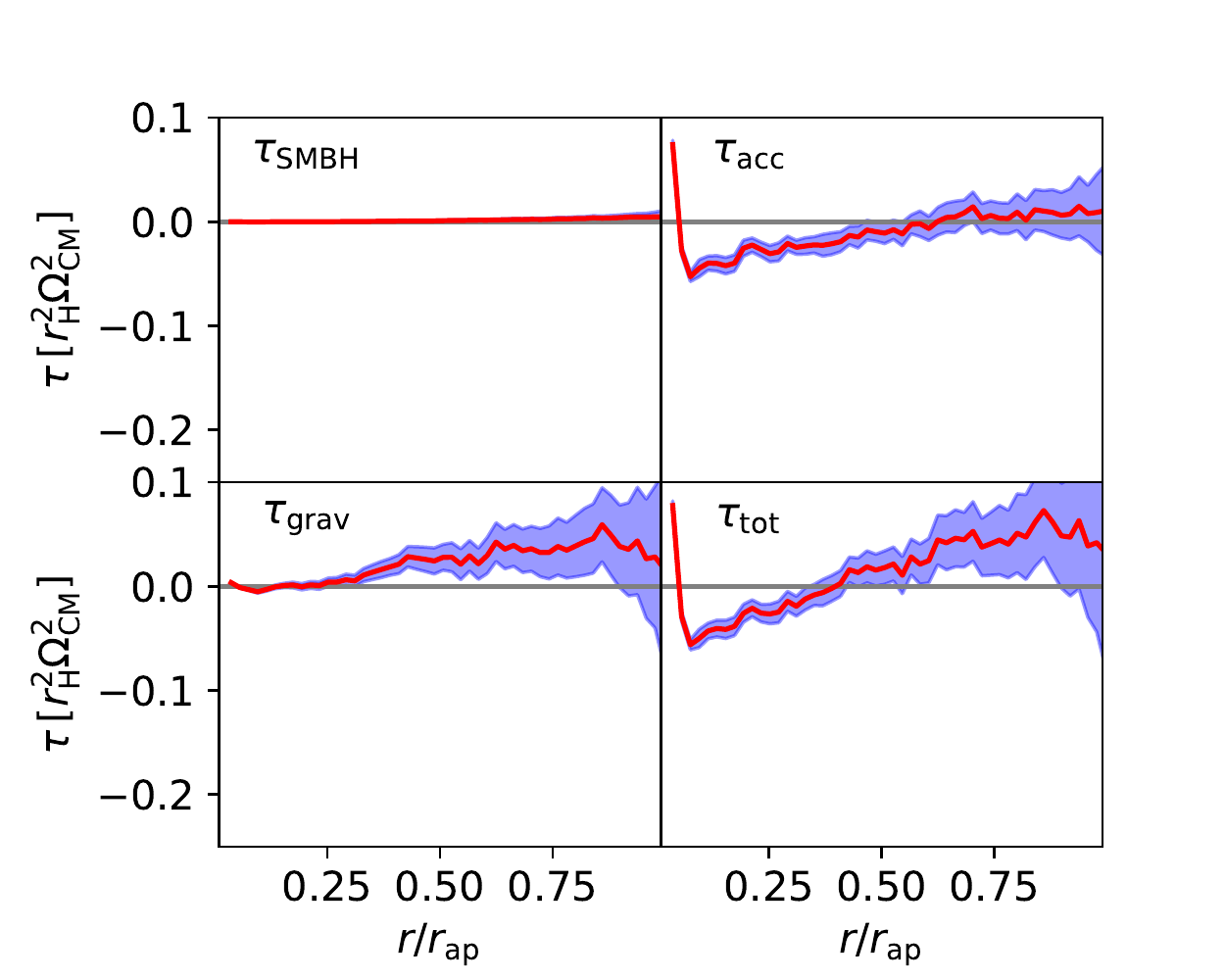} &

    \includegraphics[width=0.40\textwidth]{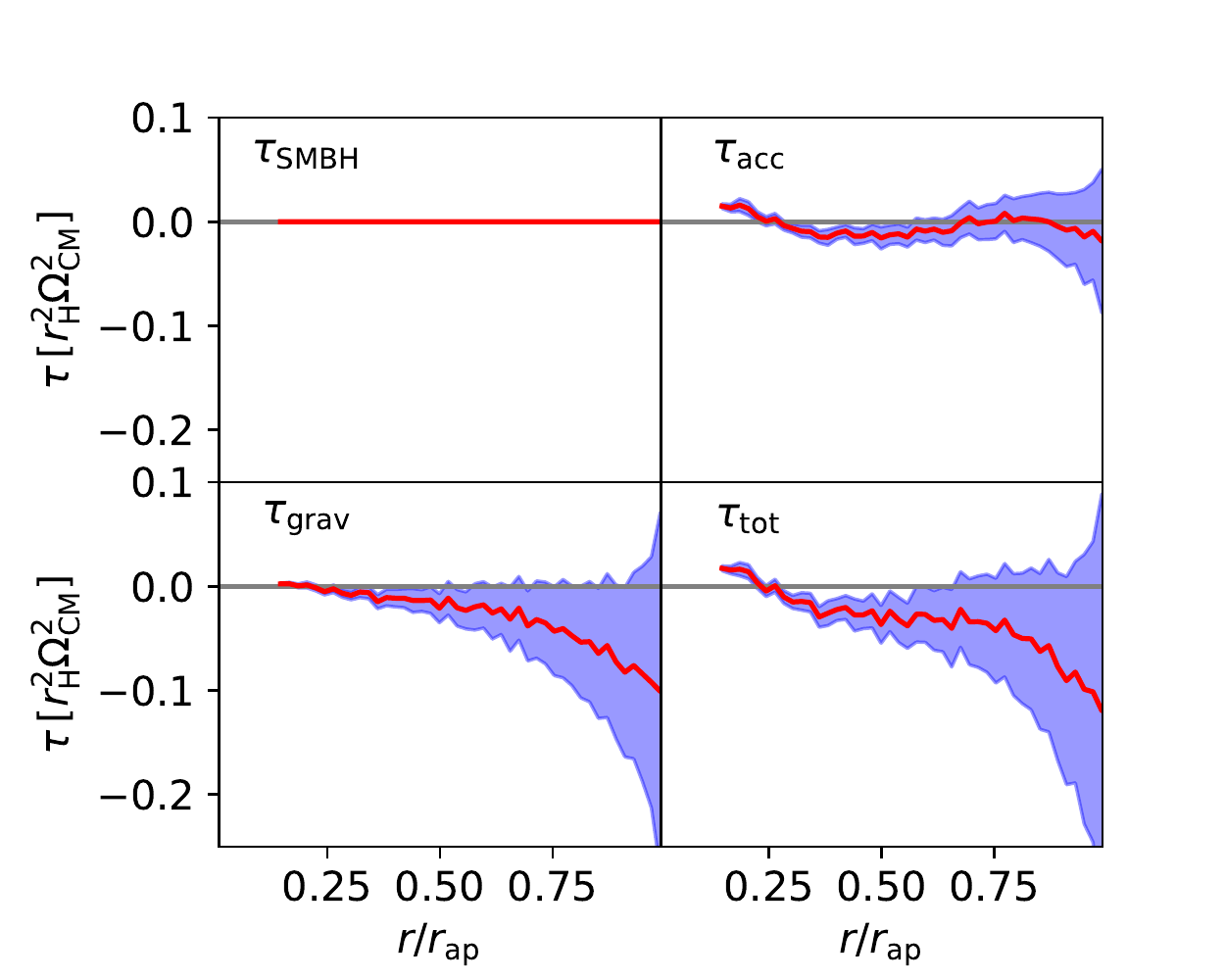}

\end{tabular}
\caption{Torque as a function of binary separation due to the SMBH, accretion, gas gravity, as well as the total. Shaded blue regions show the $1-\sigma$ variance of the torques about the mean (red). Values at each radius, in units of the instantaneous orbital periapsis, are averaged over two time frames. On the left, the period of time where the torque is positive (42-80yrs) and on the right, when the torque is negative (80-end of simulation). Torque is dominated by gas gravity in both cases when the binary is near apoapsis where the gravitational torque skews positive at earlier times and switches to negative after 80 yrs. } 
\label{fig:T_vs_r}
\end{figure*}

As our binary is highly eccentric, the separation of the BHs varies drastically between $\Delta r=a(1\pm e)$ where the gas density is highly nonlinear. Thus one might expect violent variations in the gaseous torques and the resulting evolution to be stochastic. To investigate the distribution, we calculate the average strength of each torque source in bins of orbital separation, shown in Figure \ref{fig:T_vs_r}. The torques vs binary separation are shown in units of the apoapsis so the bounds of the radial binning has a fixed range between zero and unity. In this representation one can observe how the strength of the torque varies with orbital separation by binning these quantities over many orbits. We average these torques over two periods, the first when $\tau_{\rm grav}$ is positive in Figure \ref{fig:T_vs_t_fiducial} (42-80yr) and when $\tau_{\rm grav}$ is negative (80+yr). As separation is directly coupled to the phase $\phi$ of the orbit via $\Delta r = a(1+e)/(1+e\cos(\phi))$, this is akin to probing the torque as a function of the binary orbital phase. 

It is reaffirmed from Figure \ref{fig:T_vs_r}  that the torques are dominated by $\tau_{\rm grav}$ at all times and that it is positive between 42-80yr and negative thereafter. The strength of the gravitational torque is indeed dependent on radius, being maximal when the binary is at apoapsis and minimal at periapsis. The physical mechanism driving this is the interaction between the binary and the gas as the objects reach their maximum separation. At apoapsis the objects approach higher density gas flows near the cavity wall and therefore perturb the cavity more strongly. In conjunction, the velocity of the binary is also minimal at this point and with such a high eccentricity, is far less than the orbital motion of the gas disc, which is orbiting in the opposite direction to the binary. Hence the accreted material is accreted from behind (i.e. trailing the binary) and induces a positive torque on the binary when the momentum of the accreted gas is added to the binary via Equations \ref{eq:accacc} and \ref{eq:accvel}. Gas that is perturbed by the BHs' close proximity to the cavity wall, that is not accreted, arcs in front of the BHs and also tugs them forward due to the mass of the gas, serving to add even more angular momentum to the system. Though the average total torque is negative at smaller at separations of $\Delta r/r_{\rm a}\leq0.3$, the binary spends only a small amount of time at close separation due to its high eccentricity and so the resulting angular momentum change is minimal. While the gravitational torque is positive for the first half of the evolution between 42-80yrs the sign flips for the second half and becomes negative when the binary is near apoapsis which is in agreement with Figure \ref{fig:T_vs_t_fiducial}. In their paper, \citet{Zrake2021} find that their isolated eccentric binaries flip also torques, at a slightly lower eccentrity of $e\sim0.45$ compared to $e\sim0.6$ in this case.

\subsection{Summary of Fiducial Model}
To summarise, the fiducial model demonstrated that BHs can form a bound binary system following a dynamical encounter when the BHs are embedded in an accretion disc, even though their two-body energy when first crossing each other's Hill spheres is greater than zero and their eccentricity is greater than unity. For this model, this is primarily achieved through very strong accretion against the motion of the binary (Figure \ref{fig:Ediss_vs_t_fiducial}) where momentum conservation from head on accretion leads to a drag on the BHs and reduces their relative energy so they remain energetically bound.
We observe strong oscillations in the gravitational torque and energy exchange between the gas and BHs following the first encounter due to strong asymmetries in the CSMDs, which are destroyed through tidal forces by the deep second encounter. After their destruction the torques become more stochastic and evolve the binary secularly. This result challenges the assumption that dynamical friction will efficiently dissipate the two-body energy of the binary, as analytically proposed in \citet{Tagawa2020} since in the first encounter we find inverse dynamical friction when the BHs cross each others CSMDs due to a pileup of gas in front of the BHs. However, after the first crossing of the CSMDs and after a significant amount of gas is expelled, the gas gravity acts in tandem with accretion to dissipate the binary energy on the second encounter. We therefore conclude that whether dynamical friction helps or hinders binary formation in this channel depends \textit{strongly} on the local gas morphology and rather than simply a scalar background (AGN disc) gas density and relative motion. We note that the significantly super Eddington accretion during the encounter will likely invalidate the assumption of local isothermal equilibrium and negligible radiative effects, however including such effects are prohibitively expensive. We also note that enhanced circumsingle and circumbinary disc temperatures are shown to actually more rapidly harden binaries \citep[i.e][]{Li_2022_hot_discs}. During the first encounter, torques from the local gas circularise the binary, reducing the eccentricity whilst giving energy to it. This energy increase is mitigated by accretion drag which dominates the energy change of the binary and reduces the total orbital energy in the COM frame of the two BHs to below zero so that a bound binary is formed. 
\newline\indent Following the encounter, a prograde binary is formed and all binary properties evolve on the timescale of hundreds of binary orbits. We observe a gradual removal of orbital energy through accretion and the gas gravity that further hardens the binary. Interaction from the gas continues to circularise the binary until reversing when the binary reaches an eccentricity of 0.65. The reason for this is unclear, though we identify that this torque (regardless of sign) is maximal when the binary is at its apoapsis due to its stronger interaction with the cavity wall.

\section{Results (Parameter study)}
\label{sec:results_all}
In this section, we now consider all 15 of our models consisting of the 5 different initial orbital radial separations between the two interacting black holes and 3 different AGN disc densities.
\subsection{Capture}
Snapshots from the 15 simulations are shown just prior to merger, at first encounter and at the termination of each simulation in Figures \ref{fig:prior}, \ref{fig:frstenc} and \ref{fig:simend} respectively in the appendix. Figure \ref{fig:prior} illustrates individual BH gas discs and their tidal streams in all our models. All BH satellite discs are prograde with the SMBH disc as expected.
\begin{figure*}
    \centering
    \includegraphics[width=19cm]{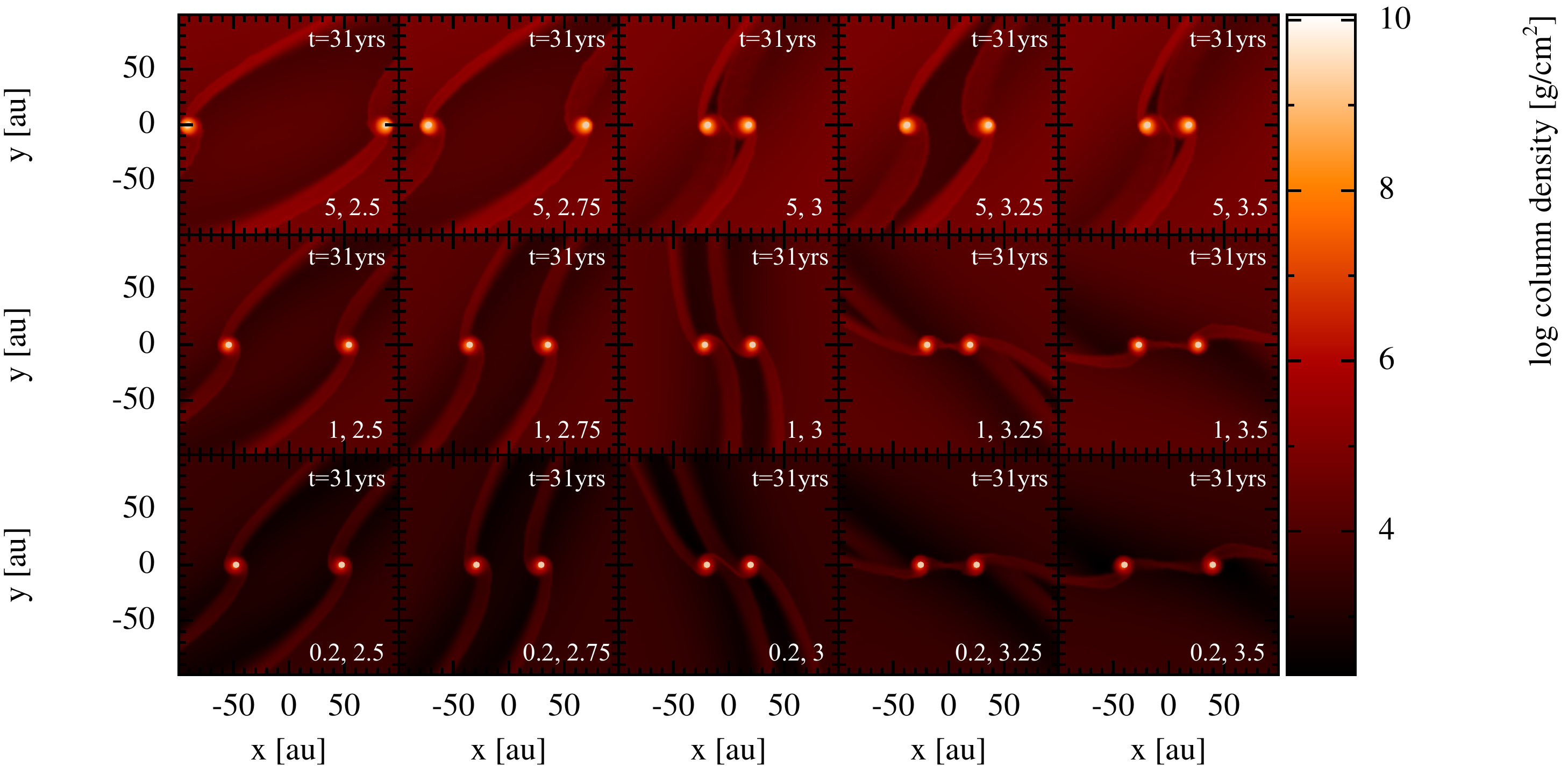}
        \caption{Surface density plots of all 15 models just prior to the first encounter of BH satellites. The arrangement follows increasing initial radial separation from left to right and decreasing AGN disc mass from top to bottom. Each model is labelled in the top left of their panel.}      
        \label{fig:prior}
\end{figure*}
\begin{figure*}
    \centering
    \includegraphics[width=19cm]{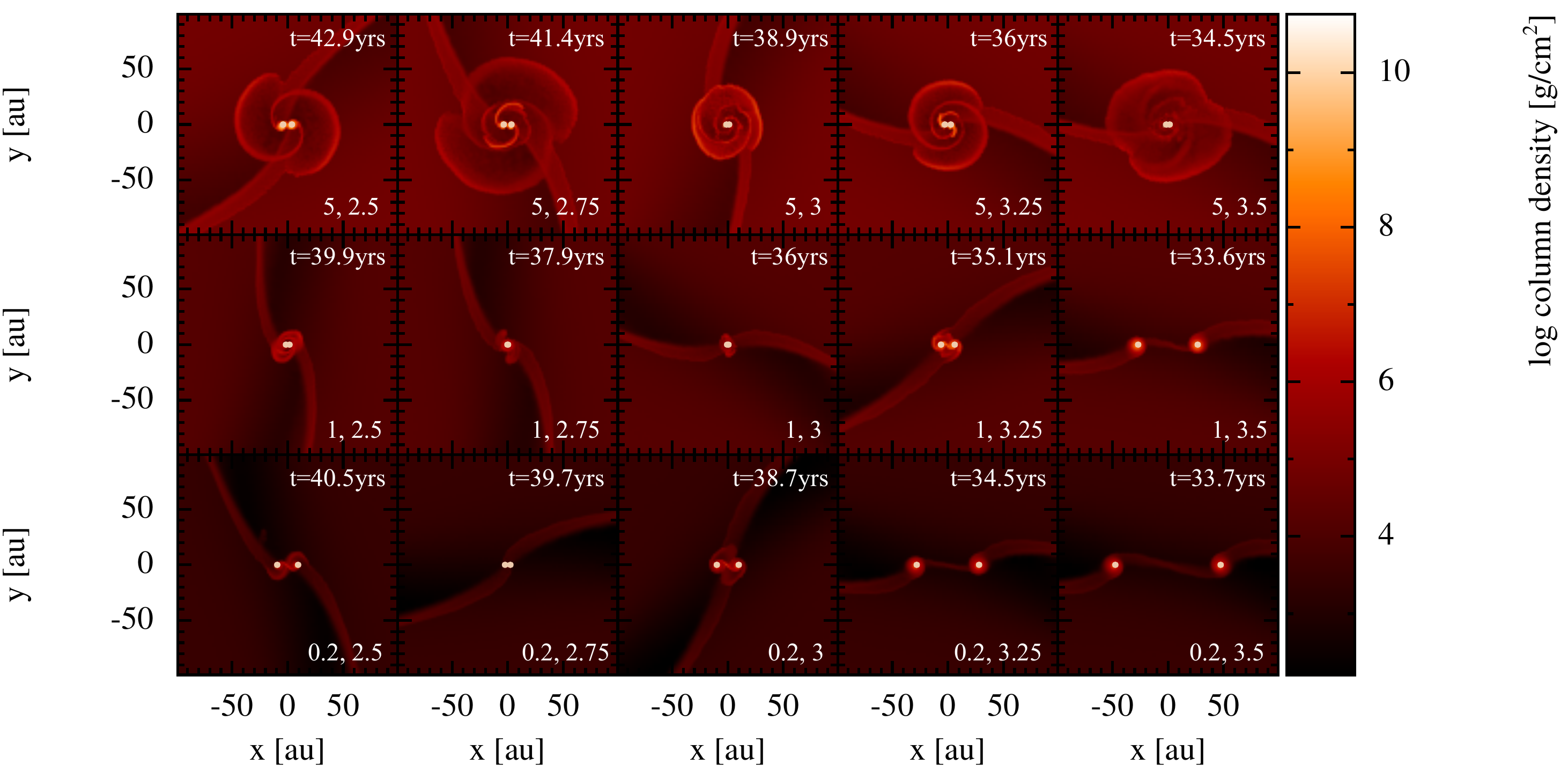}
        \caption{Surface density plots of all 15 models at apoapsis immediately following their first encounter, demonstrating significant mass ejection in the high disc mass cases (top row).}
        \label{fig:frstenc}
\end{figure*}
\begin{figure*}
    \centering
    \includegraphics[width=\textwidth]{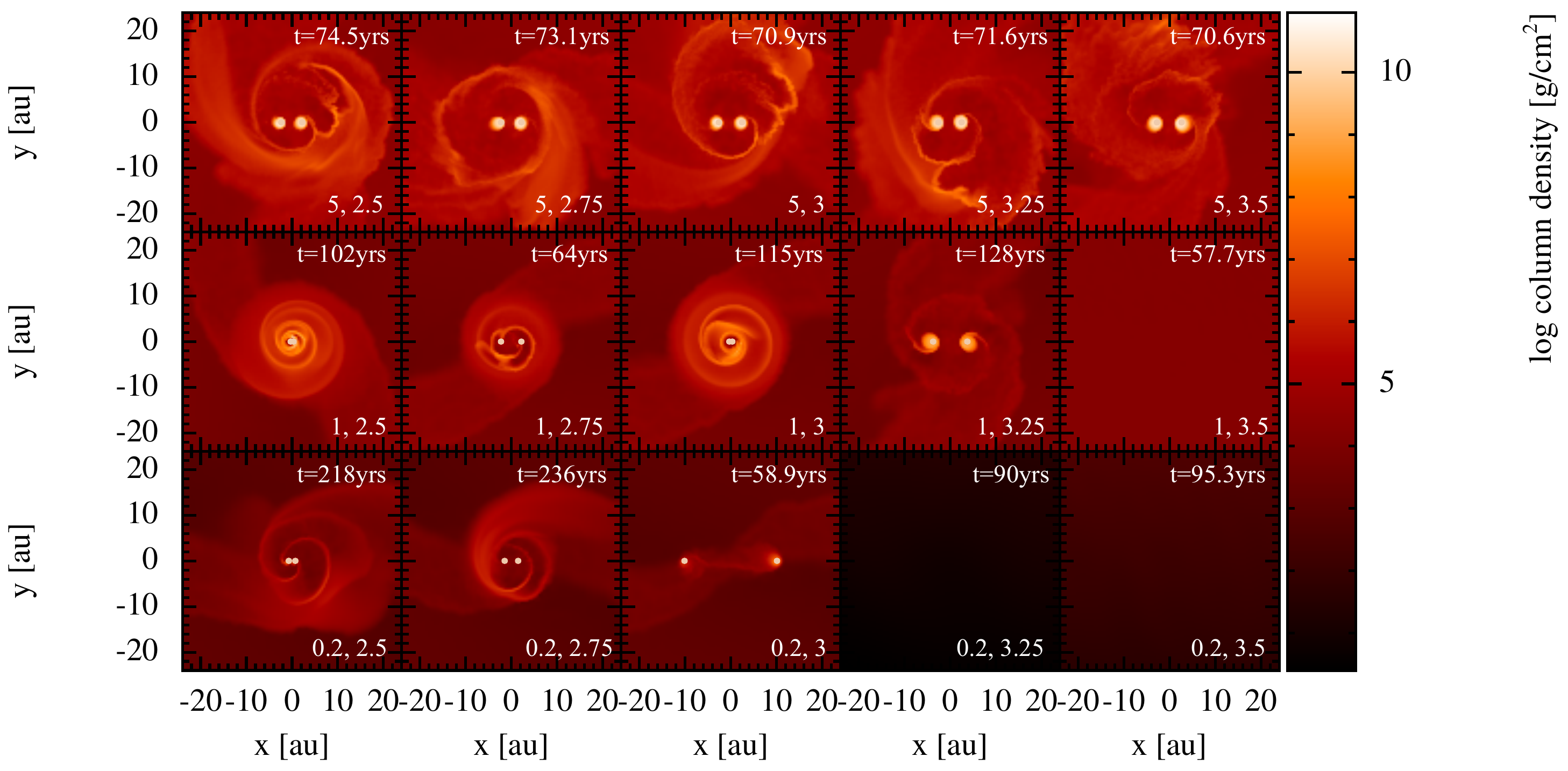}

        \caption{Surface density plots of all 15 models at their end point, i.e the decoupling of the binary or upon termination of the simulation) The final results show a variety of systems, highly asymmetric flows in the top row, three smooth circumbinary discs, and three failed captures in the bottom right corner.}    
        \label{fig:simend}
\end{figure*}
\begin{figure*}
    \centering
    \includegraphics[width=17cm]{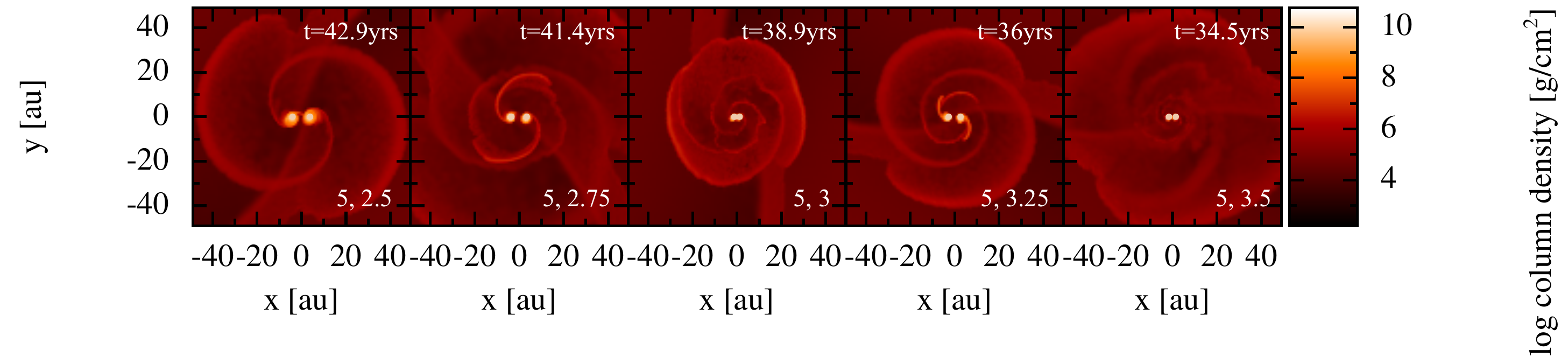}
    \caption{Left to right, simulations with AGN disc masses of $5M_{\rm d}$ with increasing initial radial separation. The gas distribution after the first close encounter exhibits strong spiral outflows and circumsingle discs. This phase corresponds to the fourth panel of the cartoon in Fig.~\ref{fig:cartoon}.}
    \label{fig:enc_M5}
\end{figure*} 
In the high disc mass cases very large gas outflows are observed which remove a portion of mass on the same order of magnitude as the mass of the binary system itself, visible through the sharp drop of the enclosed gas mass as a function of time in Figure \ref{fig:massloss}.
\begin{figure*}
    \centering
    \includegraphics[width=17cm]{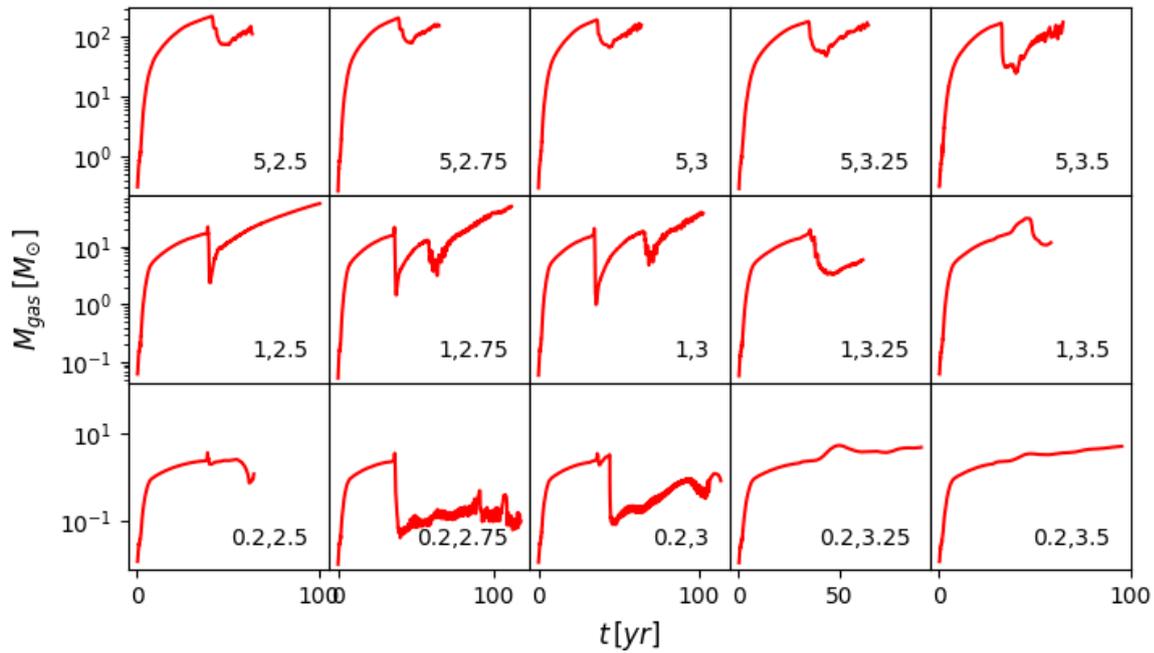}
    \caption{Enclosed gas mass around the binaries within the defined volume of eq. \eqref{eq:surface} as a function of time for all 15 models. The initial spike in enclosed mass is due to the volume enclosing a greater extent of each BHs streamers as the volume evolves from approximately two spheres around each BH to a single volume of larger radius about their COM.}
    \label{fig:massloss}
\end{figure*}
\noindent In both the snapshots and Figure \ref{fig:massloss} it is clear that the mass loss scales with the gas density of the AGN disc, due to the larger amount of mass bound to the objects before capture with increasing $M_{\rm d}$. While not surprising, the trend has implications for the efficiency of the positive energy transfer from the two BHs to the surrounding gas.   
\newline\indent The binary separation, $\Delta r$, eccentricity, $e$, and $\hat{z}$ component of the specific angular momentum, $L_{z}/M_{b}$ as a function of time is shown for all models in Figures \ref{fig:sep}, \ref{fig:ecc} and \ref{fig:angmom} respectively. Together they describe the orbital evolution of the binary.
\begin{figure*}
    \centering
    \includegraphics[width=17cm]{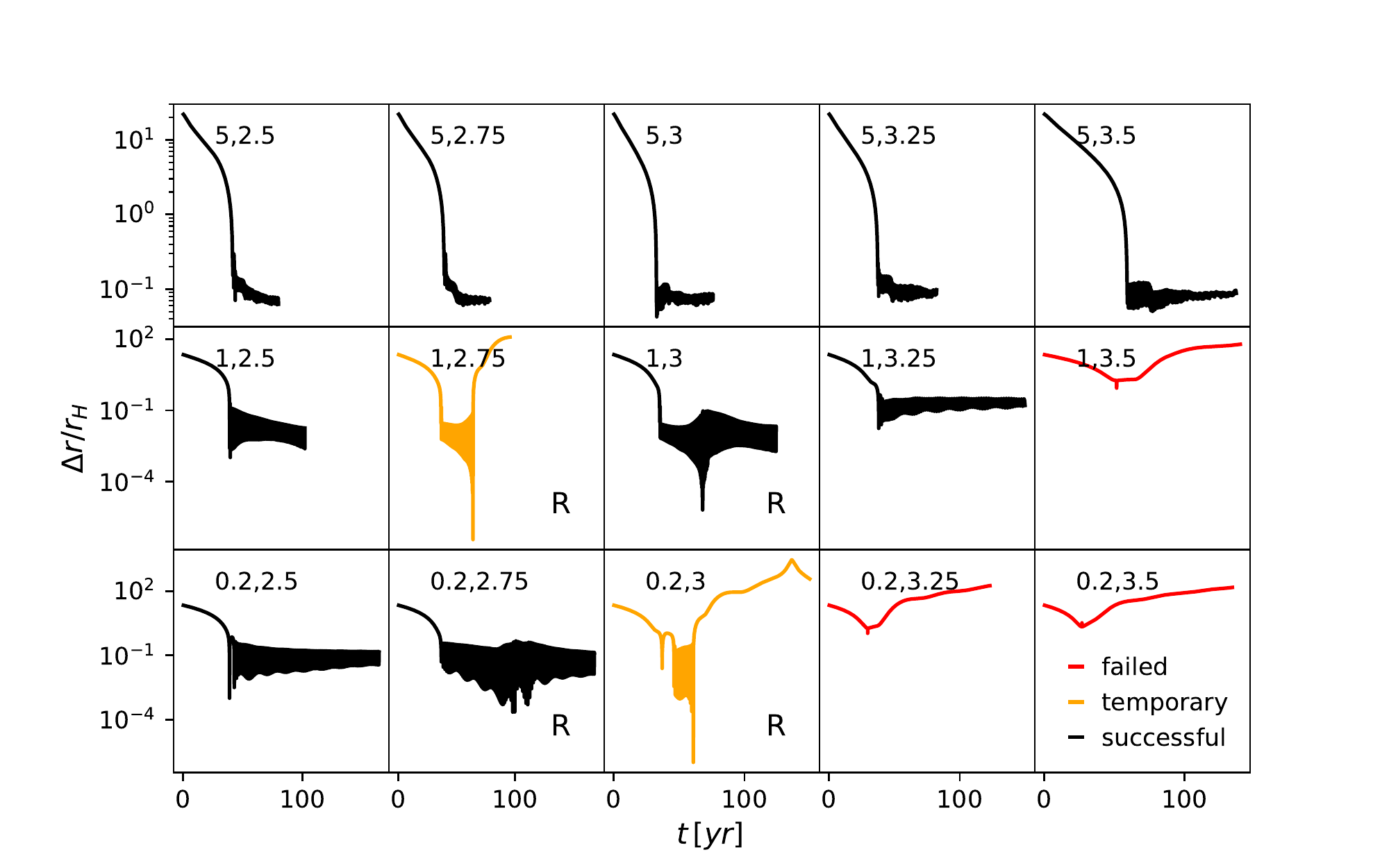}
    \caption{Binary separation in units of single-object Hill radius as a function of time. The panels correspond to the 15 models with varying disc mass (by row) and initial orbital separation (by column), represented by the two numbers in each panel). We label failed, temporary and successful captures in red, orange and black respectively. The R labels indicate which models are retrograde binaries.}
    
    \label{fig:sep}
\end{figure*} 
\begin{figure*}
    \centering
    \includegraphics[width=17 cm]{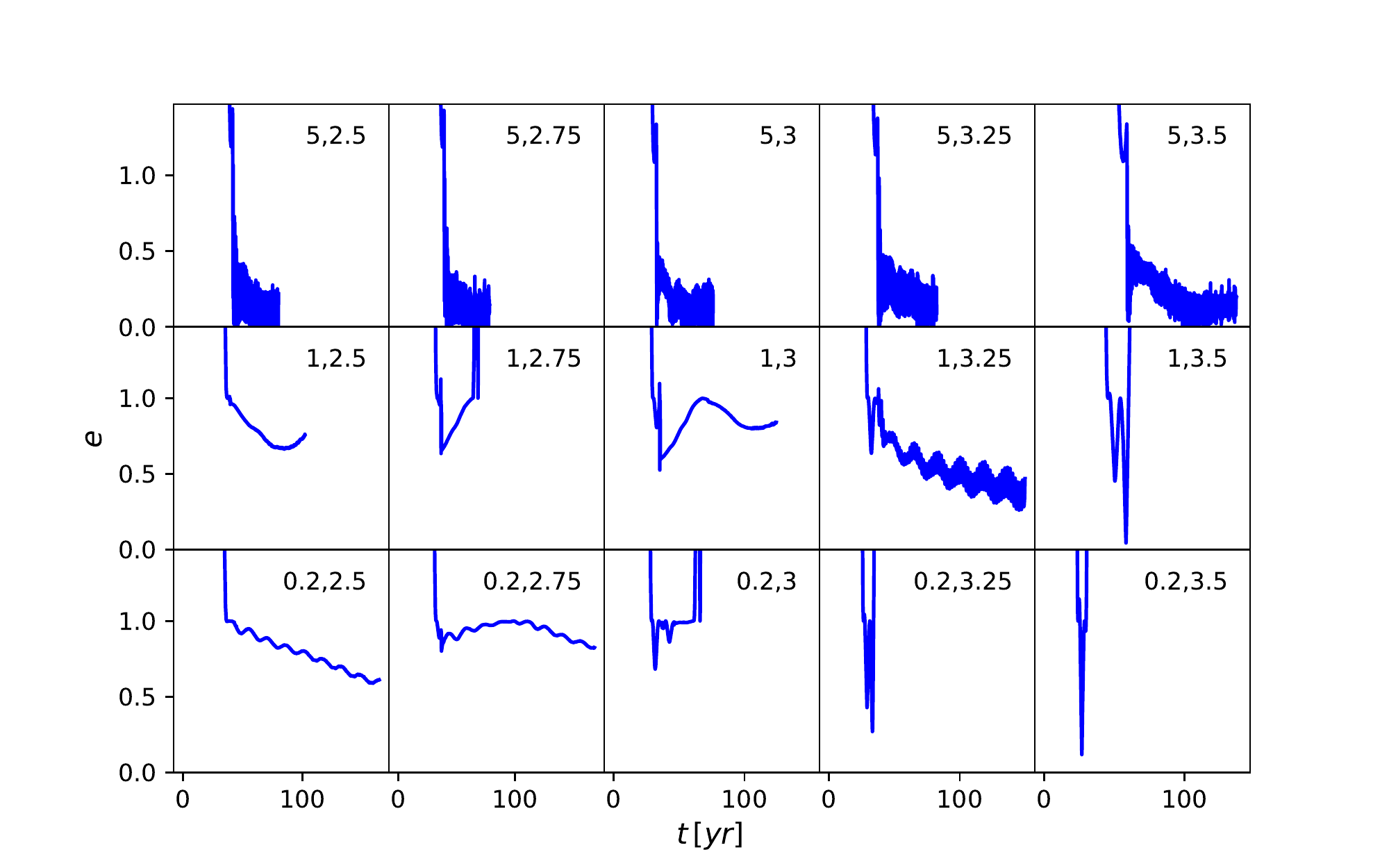}
    \caption{Binary eccentricity as a function of time for all 15 models.}
    \label{fig:ecc}
\end{figure*} 
\begin{figure*}
    \centering
    \includegraphics[width=17 cm]{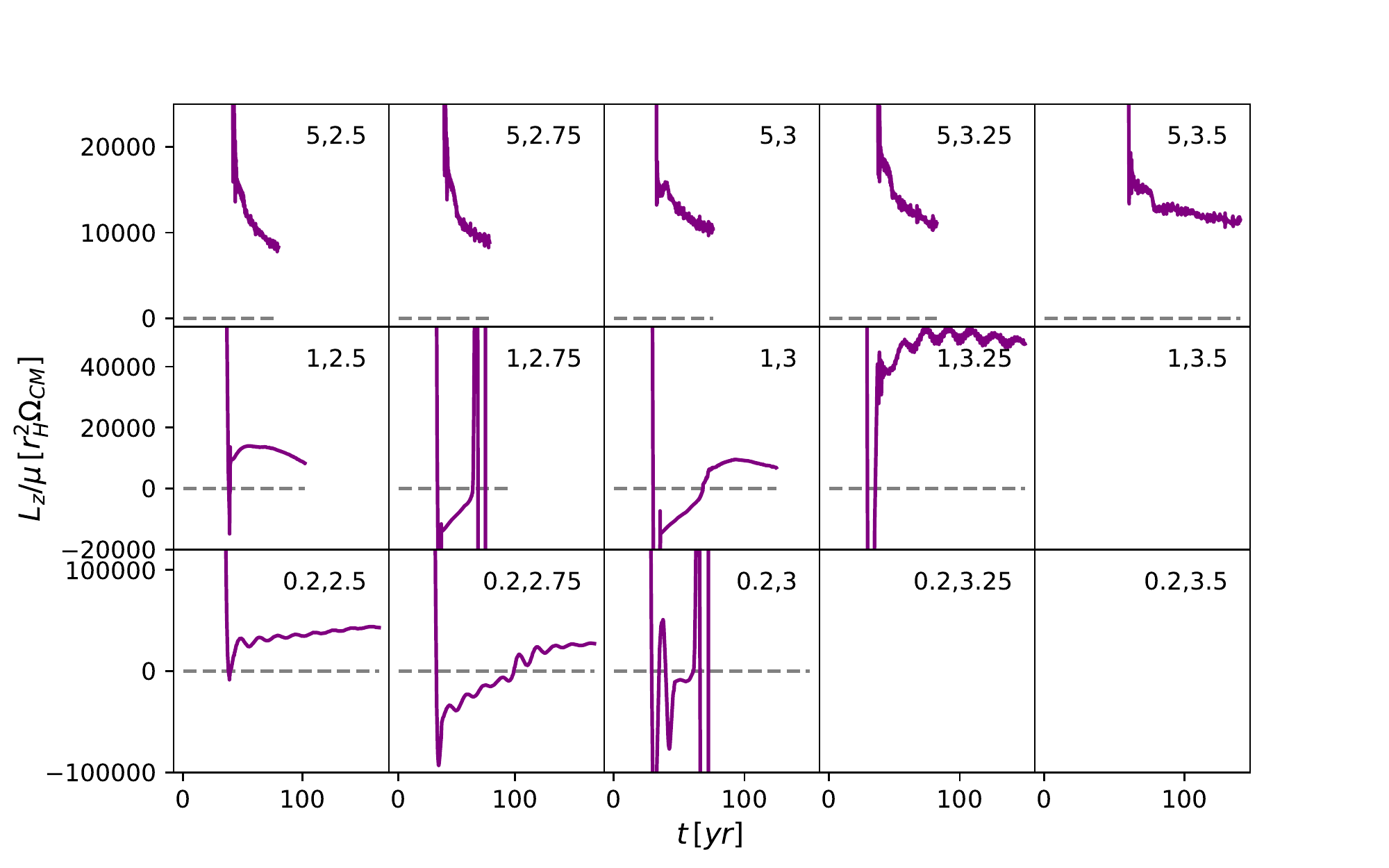}
    \caption{Binary $z$ component of specific angular momentum, as a function of time for all 15 models. Here, positive and negative values of $L_z$ correspond to prograde and retrograde orbits, respectively.}        
    \label{fig:angmom}
\end{figure*} 
Of first note is the decoupling of $\mathit{Cap_{1,2.75}}$ and $\mathit{Cap_{0.2,3}}$ and three failed captures, $\mathit{Cap_{1,3.5}}$, $\mathit{Cap_{0.2,3.25}}$ and $\mathit{Cap_{0.2,3.5}}$. The failed captures are a consequence of their impact parameters at first encounter being comparable or larger than the binary Hill radius $r_{\rm H}$, so these represent weak encounters between the objects and are immediately perturbed by the SMBH. In the parameter space of $\Delta R_{i}$ and $M_{\rm d}$, represented by the grid itself, these three models populate the bottom right corner. If we consider that an increasing $\Delta R_{i}$ will inevitably lead to a higher chance for a high impact parameter (since the minimum value is already outside twice the starting Hill radii of the objects), and models with a higher $M_{\rm d}$ will accrete more mass by the point the objects encounter each other, this explains this "island of failed captures" in our grid. In addition to accretion, we would also expect the mass of gas within the hill sphere of each BH for the $0.2M_{d,0}$, $1M_{d,0}$ and $5M_{d,0}$ models to reflect the mass difference in the AGN disc, i.e 0.2, 1 and 5 times as much mass. This leads to an increased gravitational attraction and focusing between BHs during their approach, where BHs in higher AGN disc masses can be deflected towards each other earlier along their keplerian orbits. This, in tandem with the increased accretion explains the reason all five $5M_{d,0}$ models lead to a capture.

Black holes in models with higher $\Delta R_{i}$ and low $M_{\rm d}$ are far more likely to have a weak encounter as they do not accrete as much mass and therefore don’t increase their Hill radius as much when they encounter each other. So the impact parameter in units of Hill radii is larger than in the $5M_{\rm d}$ models and hence more likely to be a weak encounter. Additionally, the reduced gas mass interacts less strongly with the binary, affecting its efficiency of energy removal. This explains the failed captures in the bottom right corner. From the three plots, the $5M_{\rm d}$ models (top row) have far more similar behaviour across the five simulations compared to the other two rows which have greatly different morphologies. This could suggest that higher AGN disc masses lead to more consistent subsequent evolution, however we instead expect the similar results to be due to the fact their initial periapsies are very large. Recall that \citet{Munoz2015} find the depth of the first encounter to be extremely important for the later evolution of the binary.
\newline\indent Looking at the eccentricities of Figure \ref{fig:ecc} there is a strong contrast between the eccentricity evolution of the $5M_{\rm d}$ models and the others. Firstly we observe that all systems in this row have damped eccentricities, unlike the other two rows. This coincides with the fact that all binaries in the top row are \textit{prograde} binaries (indicated by having positive angular momentum, see Figure \ref{fig:angmom}) unlike the other two which have some retrograde binaries: $\mathit{Cap_{1,2.75}}$, $\mathit{Cap_{1,3}}$, $\mathit{Cap_{0.2,2.5}}$ and $\mathit{Cap_{0.2,3}}$. For all our captures, those in prograde orbits experience eccentricity damping while retrograde binaries are excited to higher eccentricities after an initial damping from the first encounter. This reiterates one of the major distinctions between our simulations here and those of pre-existing binaries in accretion discs, where we cannot reproduce the highly circular initial conditions used for disc embedded binaries. The $5M_{\rm d}$ models also have a larger variability in eccentricity over each orbit. We attribute this to the far more volatile, higher density flows around the binary perturbing the binary over the course of each orbit, especially considering their larger separations since torque scales with separation. This is also observed later in $\mathit{Cap_{1,3.25}}$ as its separation approaches $0.5r_{\rm H}$. In all of our retrograde binaries, specific angular momentum is lost with time as the eccentricity increases, we find this continues until either decoupling ($\mathit{Cap_{1,2.75}}$, $\mathit{Cap_{0.2,3}}$) or the binary \textit{flips} orbital direction and begins orbiting prograde ($\mathit{Cap_{1,3}}$, $\mathit{Cap_{0.2,2.75}}$). For the binaries that flip, the eccentricity behaviour also transitions from increasing to decreasing at the same point the binary rotation flips, such that they then behave identically to our initially prograde binaries. Longer term oscillations are also present in the eccentricity due to resonant behaviour with the SMBH. When the binary eccentricity vector is aligned with the SMBH this coincides with the period of higher eccentricity due to increased tidal forces from the SMBH and lower eccentricity when perpendicular as the gas re-circularises the binary. Leading to a beat period of two times the orbit about the AGN.
\subsection{Summary of Models}
\begin{table*}
    \centering
    \begin{tabular}{|c|c|c|c|c|c|c|c|} 
    Sim Label & $\frac{M_{\rm d}}{M_{\rm d,0}}$ & $\frac{\Delta R}{r_{\rm H}} (t=0)$ & capture & pro/retrograde & endpoint & endpoint (inc. GWs) & $e$ evolution \\
    \hline 
    $\mathit{Cap_{5,2.5}}$ & 5 & 2.5 & yes & pro & inspiralling & unchanged & damped \\
    \hline
    $\mathit{Cap_{5,2.75}}$ & 5 & 2.75 & yes & pro & inspiralling & unchanged & damped  \\
    \hline
    $\mathit{Cap_{5,3}}$ & 5 & 3 & yes & pro & stalled & unchanged & damped \\
    \hline
    $\mathit{Cap_{5,3.25}}$ & 5 & 3.25 & yes & pro & stalled & unchanged & damped \\
    \hline
    $\mathit{Cap_{5,3.5}}$ & 5 & 3.5 & yes & pro & stalled & unchanged & damped \\    \hline
    $\mathit{Cap_{1,2.5}}$ & 1 & 2.5 & yes & pro & inspiralling & unchanged & damped \\    \hline
    $\mathit{Cap_{1,2.75}}$ & 1 & 2.75 & yes & retro & decoupled & merger & excited\\    \hline
    $\mathit{Cap_{1,3}}$ & 1 & 3 & yes & retro$\rightarrow$pro & inspiralling & merger & excited$\rightarrow$damped \\    \hline
    $\mathit{Cap_{1,3.25}}$ & 1 & 3.25 & yes & pro & stalled & unchanged & damped\\   \hline
    $\mathit{Cap_{1,3.5}}$ & 1 & 3.5 & no & N/A & flyby & unchanged & N/A\\
    \hline
    $\mathit{Cap_{0.2,2.5}}$ & 0.2 & 2.5 & yes & pro & stalled & unchanged & damped\\    \hline
    $\mathit{Cap_{0.2,2.75}}$ & 0.2 & 2.75 & yes & retro$\rightarrow$pro & inspiralling & unchanged & excited$\rightarrow$damped \\    \hline
    $\mathit{Cap_{0.2,3}}$ & 0.2 & 3 & yes & retro & decoupled & merger & excited\\    \hline
    $\mathit{Cap_{0.2,3.25}}$ & 0.2 & 3.25 & no & N/A & flyby & unchanged & N/A\\   \hline
    $\mathit{Cap_{0.2,3.5}}$ & 0.2 & 3.5 & no & N/A & flyby & unchanged & N/A\\

    \end{tabular}
    \hspace{0.1cm}
    \caption{Summary of all models including initial conditions, capture success, rotation relative to SMBH, eccentricity behaviour and the end result of their evolution with and without including GW disipation.}
    \label{tab:sum}
\end{table*}
\indent To summarise our models, shown in tabular form in Table \ref{tab:sum}, we have 12 successful captures, two of which later decouple. For the remaining 10, the majority of our binaries are either inspiralling slowly or stalled, with the exception of $\mathit{Cap_{1,3.25}}$ and $\mathit{Cap_{0.2,2.5}}$ that appear to be outspiralling. Note that when we refer to inspiralling or outspiralling we do so in reference to the evolution of $\Delta r$ in units of $r_{\rm H}$, so while $\Delta r$ could be increasing, $\Delta r/r_{\rm H}$ can still decrease. We refer to such a case still as an inspiral as although the physical separation may be increasing, the binary is still hardening as the Hill radius is continuing increase at a faster rate, due to accretion. Simultaneously the Schwarzschild radii $r_{s}$ and innermost stable orbits (what the BHs must cross in order to merge) of one or both the objects, which scales with $M$ rather than $M^{1/3}$ for the Hill radius is therefore increasing at an even faster rate than the increase in $\Delta r$. So in addition to hardening the binary is also coming closer to merging. If we define gravitational wave dissipation, the merger timescale $t_{merger}\equiv a/(da/dt)$ then for fixed $a$ this increases with $M^{3}$ (\citealt{Hansen1972}). So non negligible accretion can aid merger despite even an increasing separation. The units of $\Delta r/r_{\rm H}$ are then more appropriate than $\Delta r/r_{s}$ as it is possible to have $\Delta r/r_{s}$ decreasing but $\Delta r/r_{\rm H}$ increasing for a binary with increasing $\Delta r/r_{s}$ if their derivatives have $\Dot{r_{s}} > \Dot{\Delta r} > \Dot{r_{\rm H}}$. In such a case the binary may still decouple when the separation approaches the Hill radius even though the separation relative to the merger radius is decreasing, so we maintain $\Delta r/r_{\rm H}$ as our metric to label our binaries as inspiralling/outspiralling.

\subsection{Gas dissipation of energy}
\label{sec:gasdissipation}
We have demonstrated that captures are possible for all disc masses, yet more successfully in the higher-mass cases. This indicates that the local gas plays a direct role in aiding capture and preventing the objects decoupling via increasing eccentricity. As indicated by Figure \ref{fig:massloss} there is significant mass loss from the region at capture. In this section this energy exchange is quantified directly by considering the gains and losses of the energy stored in the BH orbits and the surrounding gas.
\newline\indent First, a boundary must be defined, from which we can measure the inflow/outflow of gas and its energy relative to the centre of mass of the binary. Due to the inherently chaotic nature of disc-disc collisions, care must be taken with regards to assumptions of symmetry and evolution of the region when defining a boundary. To account for both, the region considered is defined as an evolving volume defined where Eq.~\eqref{eq:deriv_surface} is satisfied, where $M_{1}$, $M_{2}$ \& $M_{\rm SMBH}$ are the masses of the two stellar mass BHs and SMBH, $r_{1}$, $r_{2}$ \&$r_{\rm SMBH}$ similarly are the positions and $G$ is the gravitational constant:

\begin{equation}
    \centering
   \frac{Gm_{\rm p}M_{1}}{||\boldsymbol{r}-\boldsymbol{r}_{1}||^{2}} + \frac{Gm_{\rm p}M_{2}}{||\boldsymbol{r}-\boldsymbol{r}_{2}||^{2}} -\frac{Gm_{\rm p}M_{\rm SMBH}}{||\boldsymbol{r}-\boldsymbol{r}_{\rm SMBH}||^{2}} \geq 0.
    \label{eq:deriv_surface}
\end{equation}

\noindent That is, the volume inside which a particle experiences a greater gravitational force, in magnitude, by the binary system than the SMBH. The alternative of summing the vector forces of the binaries prior to calculating its magnitude is avoided as this would lead to quantities of gas between the binary objects being excluded due to cancellation of the forces between the two objects. Given that PHANTOM uses fixed particle masses and $M_{\rm SMBH}>>M_{1},M_{2}$, this then reduces to 

\begin{equation}
    \centering
    \frac{M_{1}}{||\boldsymbol{r}-\boldsymbol{r}_{1}||^{2}} + \frac{M_{2}}{||\boldsymbol{r}-\boldsymbol{r}_{2}||^{2}}- \frac{M_{\rm SMBH}}{||\boldsymbol{r}||^{2}} \gtrsim 0.
    \label{eq:surface}
\end{equation}

\noindent With this description, the evolution of the system in time is accounted for, including its increasing size due to the increasing Hill radius of the binary due to accretion onto the BHs.
\newline\indent Using this boundary, particles are checked to see if they are part of the binary system at each timestep. In this region the kinetic energy of both the BHs and gas particles, $K_{\rm BH}$ and $K_{g}$ respectively as well as the potential energy associated with the sinks $U_{BH-BH}$ and the combined sink-gas component $U_{BH-g}$ are calculated from a standard N-body summation in the centre-of-mass frame of the binary:

\begin{equation}
    \centering
    K_{\rm BH} = \frac{1}{2}(M_{1}V_{1}^{2}+M_{2}V_{2}^{2}),
    \label{eq:K_BH}
\end{equation}
\begin{equation}
    \centering
    K_{g} = m_{\rm p}\sum_{i=1}^{N_{enc}} \frac{V_{p,i}^{2}}{2} + \frac{3}{2\mu m_{H}}k_{B}T_{i},
    \label{eq:K_gas}
\end{equation}

\begin{equation}
    \centering
    U_{BH-BH} = -\frac{GM_{1}M_{2}}{||\boldsymbol{r_{1}}-\boldsymbol{r}_{2}||},
    \label{eq:W_gas}
\end{equation}

\begin{equation}
    \centering
    U_{BH-g} = -m_{\rm p}\sum_{i=1}^{N_{enc}} \frac{GM_
    {1}}{||\boldsymbol{r_{i}}-\boldsymbol{r}_{1}||} + \frac{GM_
    {2}}{||\boldsymbol{r_{i}}-\boldsymbol{r}_{2}||},
    \label{eq:W_BH_gas}
\end{equation}

\noindent Where $V_{1}$, $V_{2}$ are the velocities of the two binary objects, $V_{\rm p}$ and $r_{\rm p}$ are velocity and position of a gas particle respectively, and $m_{H}$ is the mass of monatomic hydrogen. The summation in Eq.~\eqref{eq:W_BH_gas} is performed over all $N_{enc}$ enclosed gas particles in the volume. The thermal contribution of a gas particle with temperature $T_{i}$ is added to the kinetic energy using the usual ideal gas energy equation, where $k_{B}$ is the Boltzmann constant. These values then construct the total energy associated with the sinks, gas and entire system, labelled $E_{BH-BH}$, $E_{BH-g}$ and $E_{tot}$ respectively. These are easily calculated in Equations \ref{eq:B_BH_BH}-\ref{eq:B_tot}:

\begin{equation}
    \centering
    E_{BH-BH} = K_{\rm BH} + U_{BH-BH},
    \label{eq:B_BH_BH}
\end{equation}

\begin{equation}
    \centering
    E_{BH-g} = K_{g} + U_{BH-g},
    \label{eq:B_BH_gas}
\end{equation}

\begin{equation}
    \centering
    E_{tot} = E_{BH-BH} + E_{BH-g}.
    \label{eq:B_tot}
\end{equation}
\begin{figure*}
    \centering
    \includegraphics[width=18cm]{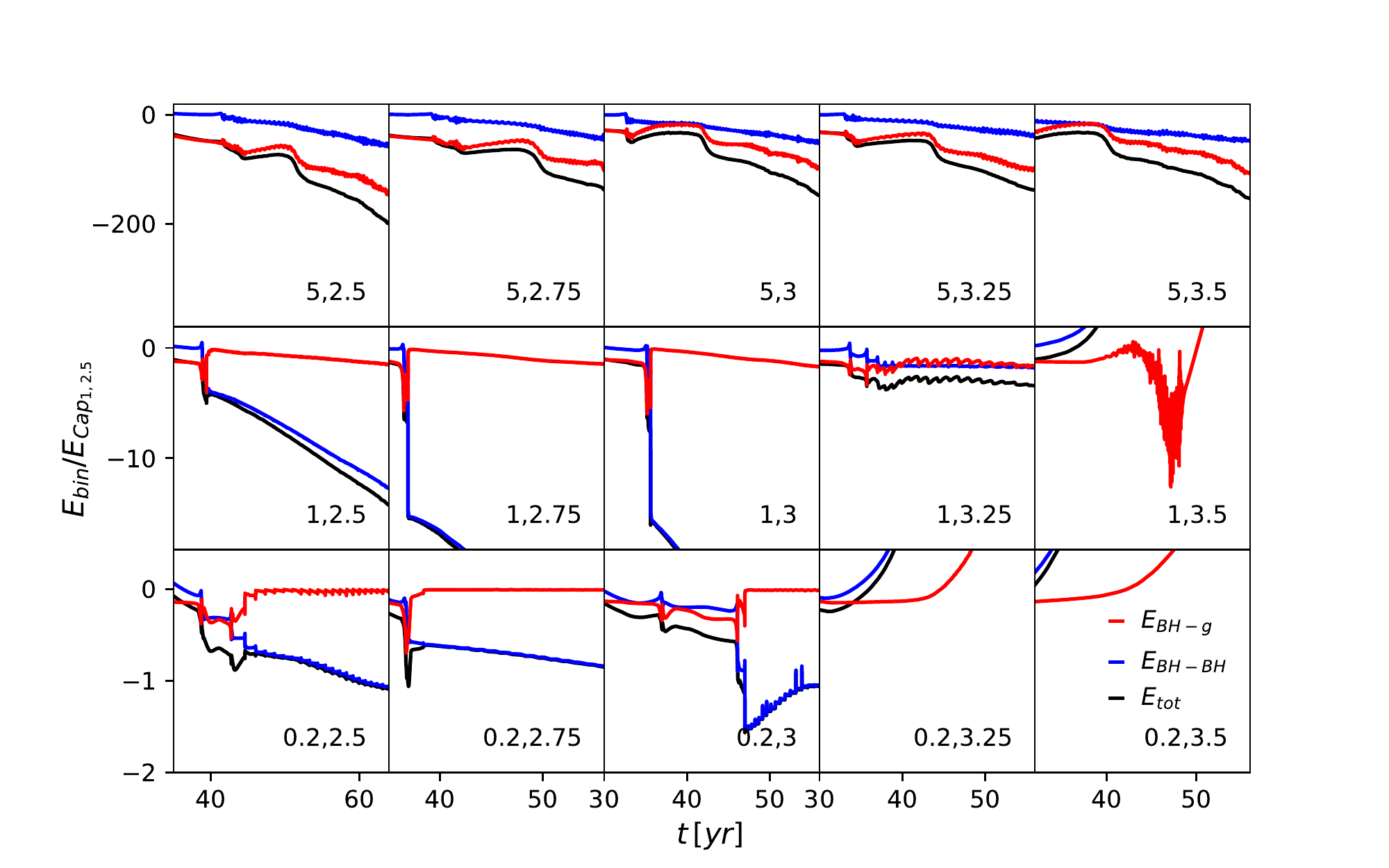}
    \caption{Two body energy of the BHs,  $E_{BH-BH}$ (blue) (eq~\ref{eq:two_body_energy}), between BHs and gas, $E_{BH-g}$ (red) (eq~\ref{eq:B_BH_gas}) and total energy $E_{tot}$ (black) as a function of time, showing transfer of energy from sinks to gas and energy loss due to gas being expelled during capture for the $5M_{\rm d}$ cases. All energies are normalised to the initial two body energy at Hill the sphere crossing in our fiducial model $Cap_{1,2.5}$}.
    \label{fig:bindings}

\end{figure*}
\noindent Note that since we neglect self-gravity there is no potential term \textit{between} gas particles in this calculation.
\newline\indent In Figure \ref{fig:bindings} we show each of these energies as a function of time for each model. The efficiency of gas dissipation and accretion drag during the capture can be characterised by two factors. Firstly, the initial drop in total energy is observable as the initial sudden drop in $E_{BH-BH}$ and $E_{tot}$, followed by a more gradual exchange should the binary remain bound after the first encounter. For the $1M_{\rm d}$ and $0.2M_{\rm d}$ models that are successful, the first effect can be clearly seen as $E_{BH-g}$ has a sudden increase while $E_{BH-BH}$ shows a decrease of the same magnitude. Comparing the fiducial model ($Cap_{1,2.5}$) panel to the energy dissipation of the binary in Figure \ref{fig:Ediss_vs_t_fiducial} the initial drop in $E_{BH-gas}$ corresponds to the energy deposited back into the binary during the first encounter followed by the opposite during the second encounter where a large amount of energy is removed by the gas, represented by the jump up to nearly zero. This occurs due to strong gravitational drag when they pass deeply into each other's CSMDs for the second time, visible in Figure \ref{fig:accretion} as an enormous jump in the total mass of the binary $M_{\rm{bin}}$. During this brief episode the accretion rate is significantly super-Eddington. Considering the magnitude of the energy change and mass accretion in such a short timescale, these instances can be treated more as soft collisions, as opposed to a dynamical exchange of energy.  
\begin{figure}
    \centering
    \includegraphics[width=9cm]{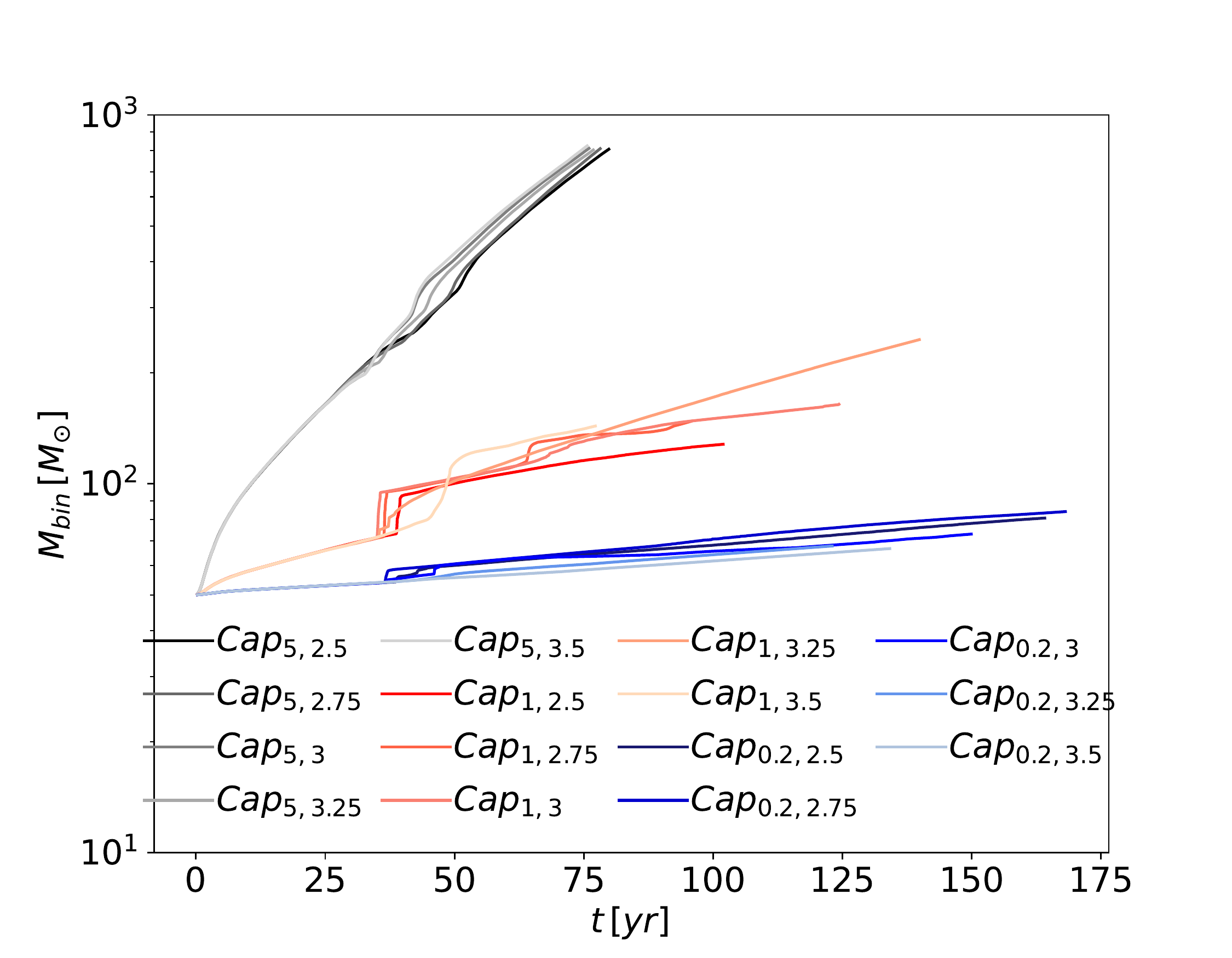}
    \caption{Total binary mass as a function of time for all 15 models, showing a large accretion of mass at the first encounter for models with $1M_{\rm d}$ and $0.2M_{\rm d}$.} 
    \label{fig:accretion}
\end{figure}
\newline\indent After this episode, $E_{BH-g}$ remains steady, near zero, while $E_{BH-BH}$ continues to decrease. This demonstrates the second phase of a continuous transfer of positive energy from the binary objects to the gas or accreted (recall accretion is not energy conserving since the collision is inelastic with no emission of the lost kinetic energy), which despite re-accumulating in the binary region (See Figure \ref{fig:massloss}) retains a near constant, loose coupling to the binary while the binary binding energy continues to decrease. For the $5M_{\rm d}$ models the local gas mass exceeds the BH masses and is the larger contributor to the binding energy. In these cases the drop in $E_{BH-BH}$ is still observable but $E_{BH-g}$ remains the dominant contributor to $E_{tot}$. After this $E_{BH-g}$ continues to decrease at roughly the same rate as $E_{BH-BH}$ after an initial period where it rises immediately following the capture. In contrast to the lower disc mass cases there is no discernible jump in binary mass at the first encounter. This can be explained by observing from Figure~\ref{fig:sep} that in all high-mass cases the first encounter is not as close and therefore they do not cross each others CSMDs as deeply, if at all. As a result, the initial drop in $E_{BH-BH}$ is far smaller, the retention of the discs leads to a more gradual stripping of the CSMDs via the trailing streams of gas, which then form spirals out from the edge of each BH's accretion disc opposite from the COM in a spiral manner as in the top panel of Figure \ref{fig:frstenc}. This spiralling continues in an ordered manner for a few tens of orbits. The imprint of this is seen in the decrease in energy and the period where $M_{enc}$ continues to decrease before plateauing and reversing. After this period the gas flows around the binary become more chaotic, marking the shift where $E_{BH-g}$ begins to decreases alongside $E_{BH-BH}$. 
\newline\indent Going from bottom to top, paying attention to the vertical scale, the energy of the gas gained and the energy of the BHs lost appears to scale with the AGN disc mass. To visualise the scaling of the dissipation with disc mass, the rise in $E_{BH-g}$ and loss in $E_{BH-BH}$ over the encounter timescale are taken from Figure \ref{fig:bindings} and plotted for each value of $M_{\rm d}$. Here the encounter timescale is taken to be the time needed for the spiral structure to dissipate;  specifically when the local gas environment settles and $E_{BH-g}$ either begins to decrease again or remain at a steady level. In the $5M_{\rm d}$ cases this requires a longer amount of time due to the development of the strong and prolonged spiral structure. The results are shown in Figure \ref{fig:diss_vs_Md}. 
\begin{figure}
    \centering
    \includegraphics[width=9cm]{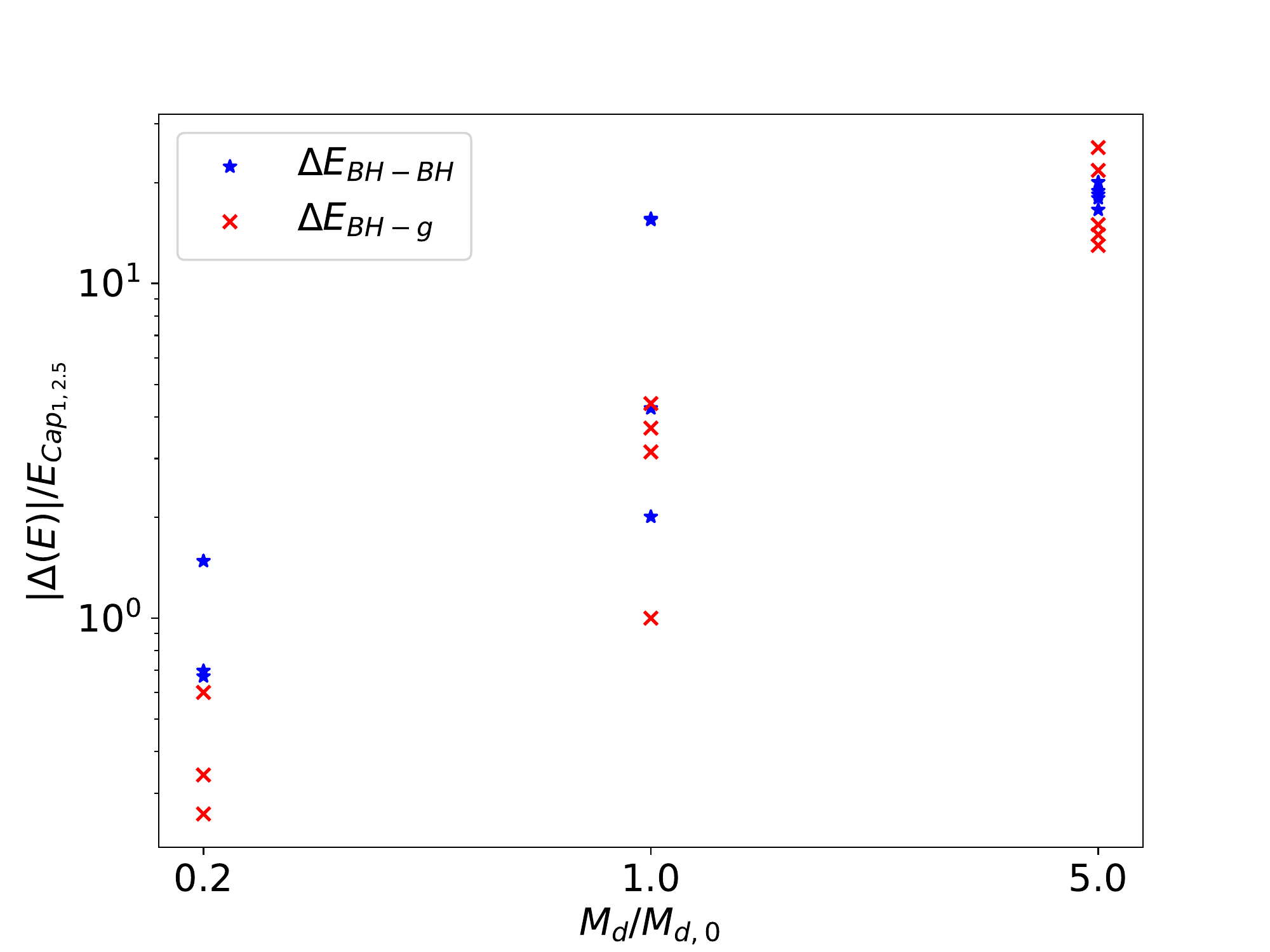}
    \caption{Loss in energy of BHs (blue) and gain in energy stored in the local gas (red), showing that for higher AGN disc masses, more binary energy is dissipated and transferred to the gas, a lot of which is then lost over time for the high-mass cases. All energies are normalised to the initial two body energy at Hill the sphere crossing in our fiducial model $Cap_{1,2.5}$.}
    \label{fig:diss_vs_Md}
\end{figure}
\newline\indent In the figure, it becomes clear that an increased AGN disc mass (resulting in higher ambient and BH disc gas density) is able to gain a larger amount of energy from the binary. Normalisation to the binary mass softens but does not remove the trend from $E_{BH-BH}$. For all $5M_{\rm d}$ models, we find that they dissipate more $E_{BH-BH}$ than the $M_{\rm d}/5$ cases. The two very high blue points in the middle column correspond to $\mathit{Cap_{1,2.75}}$ and $\mathit{Cap_{1,3}}$. This is a result of the aforementioned extreme accretion in the first periapsis of the capture. Given the discrepancy between the nature of the first encounters between these two and the rest of the captures, it may be that if their approach did not result in this very deep crossing of their discs, then this would result in a more clear trend in the dissipation across $M_{\rm d}$, but we do not have enough data points to confirm this. The changes in $E_{BH-BH}$ and $E_{BH-g}$ are not one to one since gas is still flowing in to the region and becoming bound from streams parallel to the binary motion around the SMBH. Therefore there is a tendency to have $\Delta E_{BH-BH}>\Delta E_{BH-g}$. As a fraction of the energy in the system, this gap is far smaller for the $5M_{\rm d}$ and for one case the opposite is true. This is a result of the more significant mass loss in these models where gas particles that exit the volume defined in Eq.~\eqref{eq:surface} cease to be counted in the summation of Eq.~\eqref{eq:W_BH_gas}. 
\newline\indent To further highlight the efficiency of the gas capture, the effect of removing the gas just prior to the encounter is explored. From the four successful captures in the fiducial models, the instantaneous positions, velocities and masses are recorded at the moment the satellite BHs' Hill radii intersect. Using these as initial conditions, the models are repeated as purely 3 body problems. As another test, the instantaneous enclosed gas mass at the Hill sphere intersection is also recorded and added to the mass of the BHs in proportion to their individual masses. By doing this we mimic the 3-body encounter more accurately as the mass of the gas in the circumsingle discs, which alters the approach, is encorporated into the BH masses. This scenario more accurately reflects the energy of the system to the gas case, but removes the ability for the objects to transfer this energy to a background medium via dynamical friction, or slow via direct accretion onto the BHs. The separations as a function of time for these two tests are shown in Figure \ref{fig:nogas}.

\begin{figure*}
    \centering
    \includegraphics[width=19cm]{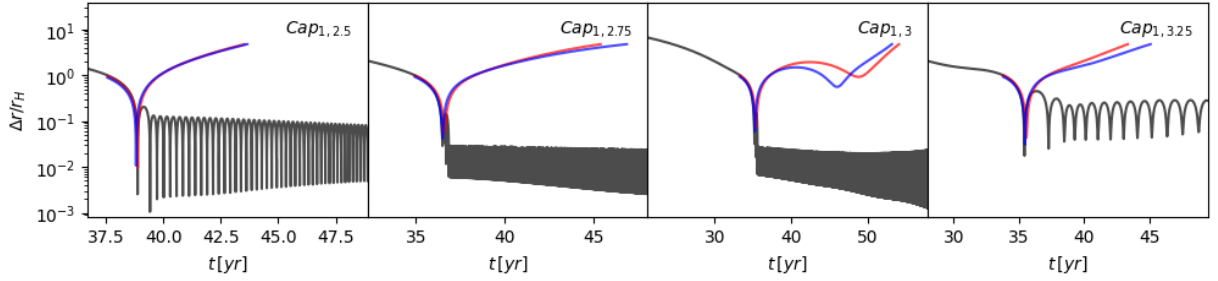}
    \caption{Separation as a function of time for the four successful captures from the fiducial AGN disc mass models (black) compared with captures with gas removed just prior to capture where the local gas mass is not added to the BHs (red) and when it is (blue). In both gasless scenarios the binary makes only one encounter and is immediately disrupted.}
    \label{fig:nogas}
\end{figure*}

Looking at the results in the Figure, it is clear that removing the gas dissipation leads to qualitatively different results. In all cases, for both gasless tests, the two BHs have one close encounter before being disrupted as they go on to exit each other's Hill sphere. The closest approaches of the gasless cases are also less deep due to the absence of dynamical friction which helps remove orbital energy on the approach to the first encounter. These results show qualitatively that it is the gas which is responsible for retaining the two objects as a binary. More specifically, concerning the second gasless setup, we show that it is the mechanism where the binary objects do work on the gas that is crucial for this binary formation pathway, rather than simply having an increased amount of mass in the binary Hill sphere. Otherwise one would expect the 3-body re-runs where the enclosed mass is added to the binary in Figure \ref{fig:nogas} to more closely match the original model. 

As a short exercise, we derive an upper limit for the minimum required background gas density for a binary to remain bound due to accretion alone, see appendix \ref{sec:acc_cap}.

\subsection{The Retrograde Case}
\begin{figure}
    \centering
    \includegraphics[width=9cm]{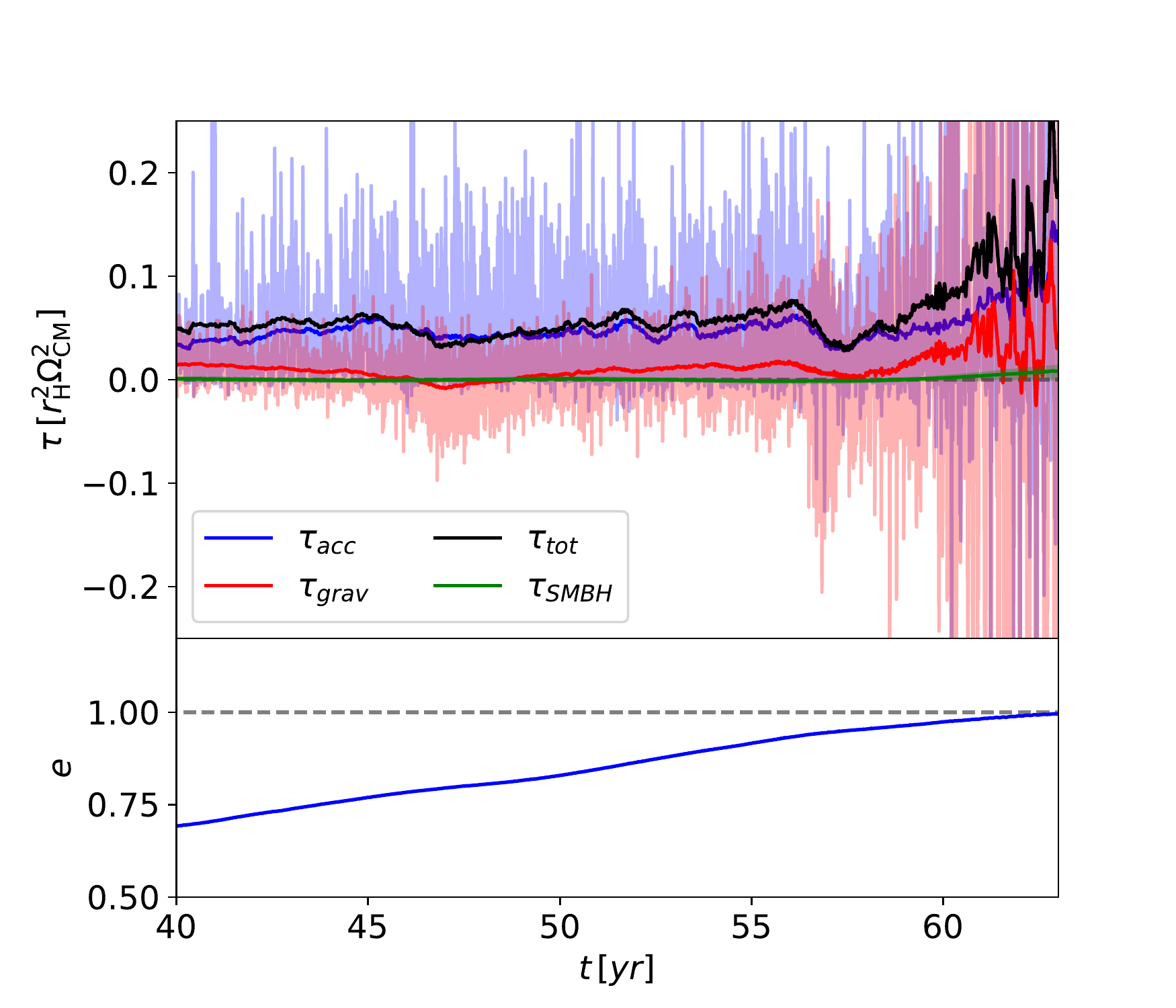}
    \caption{For $Cap_{1,2.75}$, the specific torque components and net torque (top) and specific angular momentum (bottom) as a function of time for the time frame where binary separation is less than the binary Hill radius. The value of $\tau_{\rm SMBH}$ remains negligible until the separation approaches the Hill radius. The strongest source of torque in this model comes from the positive accretion torque which for this retrograde case, is driving inspiral.}
    \label{fig:torque_v_time}
\end{figure}

\begin{figure}
    \centering
    \includegraphics[width=9cm]{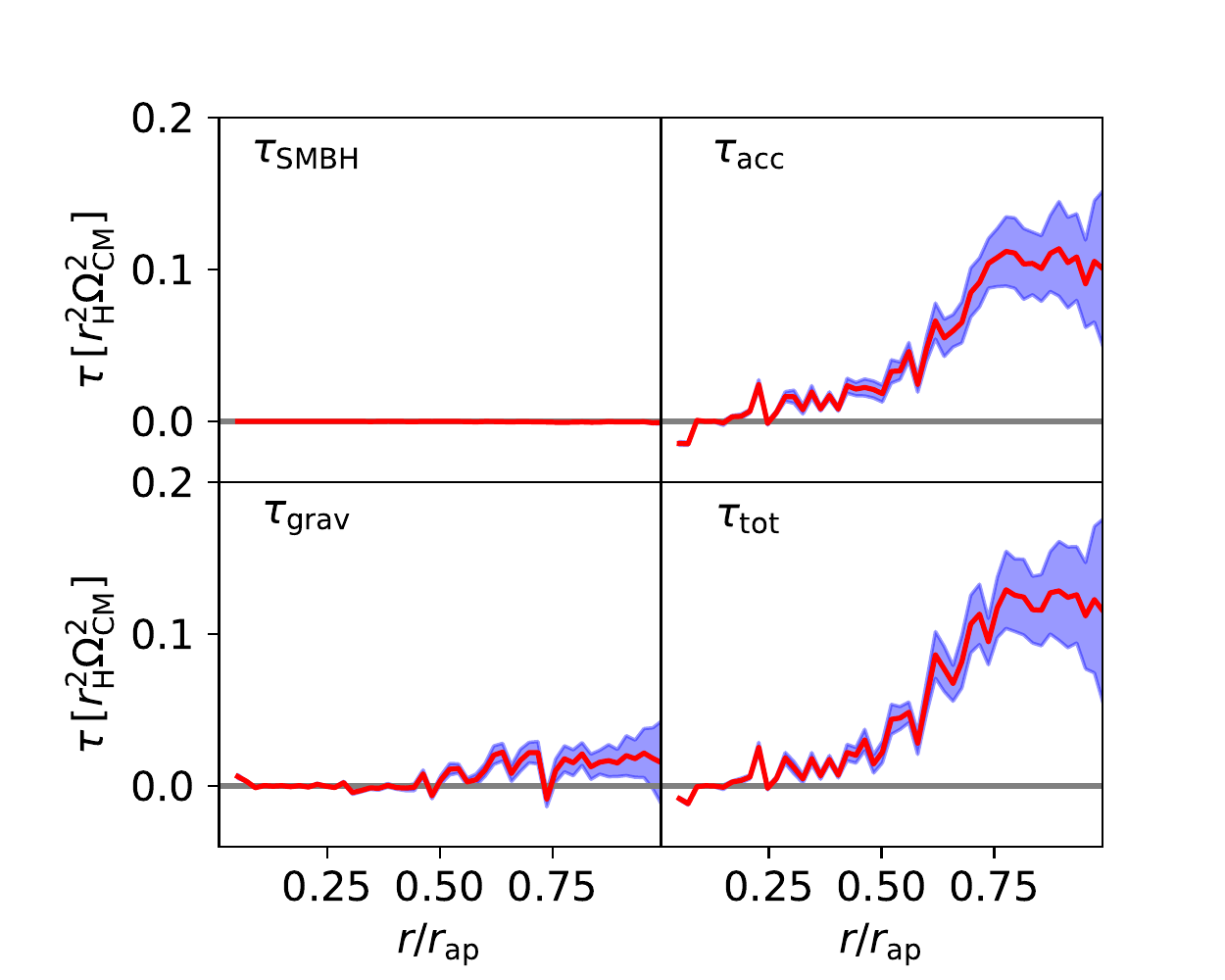}
    \caption{Radial dependence of torque sources for retrograde binary $Cap_{1,2.75}$. Accretion torque as binary interacts with cavity wall (top right panel) dominates the net torque on the binary. Shaded blue region represents the $1-\sigma$ variation the torques.}
    \label{fig:torque_v_r_retrograde}
\end{figure}

To compare the difference between prograde and retrograde binaries, we discuss the outcome of $Cap_{1,2.75}$, which forms a retrograde binary. The formation of retrograde binaries in our simulations occurs when the inner object passes the outer on its orbit around the AGN, before doubling back and around the outer BH in a clockwise manner. The form of these orbits are explored in great detail in \cite{Boekholt_2022}, see Figure 6 within.  Other than their origin, capture process for retrograde binaries shows no differences in behaviour to the prograde models, however we find the following evolution to differ significantly. In Figure \ref{fig:torque_v_time} we plot the time evolution of the post capture torques, deconstructed into each component, $\tau_{\rm grav},\tau_{\rm SMBH},\tau_{\rm acc}$ and the net value, $\tau_{tot}$. 
As the binary is rotating in retrograde (clockwise for our problem), positive torque increases the negative angular momentum of the binary towards zero, i.e inducing inspiral. We identify the dominant torque contribution comes from the accretion torque which drives a rapid inspiral of the binary. The nature of this inspiral can be described by the reduction of the angular momentum alongside an increasing eccentricity towards unity, until ultimately decoupling the binary as the apoapsis, $r_{a}=a(1+e)$, of the orbit exceeds the binary Hill radius where it is significantly perturbed by the SMBH and decouples. However, while the apoapsis increases, the periapsis,  $r_{\rm p}=a(1-e)$, continues to decrease to significantly small values as $e\rightarrow 1$. Should this separation be small enough, orbital energy dissipation via GW waves can be significant enough to begin to circularise the orbit such that periapsis does not continue to increase which would otherwise lead to the binary decoupling. 
\newline\indent We show the average torque as a function of  orbital separation for our retrograde binary generated over the period of time the binary is bound in Figure \ref{fig:torque_v_r_retrograde}
\noindent In all panels there is an increase in the spread of the torques in each bin of $dr$ and with the exception of $\tau_{\rm acc}$ are highly symmetric. This can be explained by the highly stochastic torques (see the raw torque values in Fig \ref{fig:torque_v_time}) being enhanced by the $(\boldsymbol{r}_{1}-\boldsymbol{r}_{2})$ terms in Equations \ref{eq:SMBHtorque}, \ref{eq:gravtorque} and \ref{eq:acctorque} when the binary is near apoapsis. The highly symmetric spread of the torques leads to the mean being far closer to zero than the overall spread, particularly for $\tau_{\rm grav}$. Indicating the binary experiences very strong but competing torque sources from the surrounding gas over its evolution. As pointed out, the accretion torque has a steep inclination from symmetric to positive torques when the binary separation is greater than $~0.2r_{ap}$, implying a bias for the accretion of particles \textit{against} the direction of the BH's velocity vector. The physical mechanism driving this is the interaction between the binary and the gas as the objects reach their maximum separation. At apoapsis the objects approach higher density gas flows near the cavity wall and therefore accrete a larger amount of mass. In conjunction, the velocity of the binary is also minimal at this point and with such a high eccentricity, is far less than the orbital motion of the gas disc, which is orbiting in the opposing direction to the binary. Hence the accreted material is accreted in opposition to the binaries direction of movement and induces a drag force on the binary when the momentum of the accreted gas is added to the binary via Equations \ref{eq:accacc} and \ref{eq:accvel}. Some amount of gas that is perturbed by the BHs close proximity to the cavity wall, that is not accreted, arcs behind the BHs and leads to a gravitational drag on the object that also removes angular momentum from the binary. Furthermore the radial bias of the torque excites the binary to higher eccentricities as the torque becomes resonant with the orbital phase of the binary. Explaining the eccentricity increase of the binary in Figure \ref{fig:ecc} as well as the increasing strength of both gravitational and accretion torques as the binary apoapsis increases in size. 
\newline\indent The torques shown here for a retrograde binary are an inversion of the prograde case. Accretion drag dominates over gas gravity which is the inverse of the prograde case. The net torque also behaves inversely, acting in opposition of the orbital motion of the binary. This corroborates the eccentricity excitation of the retrograde binaries in Figure \ref{fig:ecc}. The physical description of the scenario is also the inverse of the prograde case. Here the binary is orbiting in the opposite direction of the CB. Since the binary is initially highly eccentric and remains so for the duration of its evolution despite the damping then its orbital apoapsis extends close to the cavity wall and the orbital velocity of the binary is significantly slower than the local gas. The same processes, analogous to the prograde case discussed in \ref{sec:results_fiducial} then occur. The difference in the retrograde case is that the gas orbits in the opposing direction of the binary, so accretion onto the front of the BHs decelerates the binary rotation by reducing the angular momentum. Similarly gas is perturbed behind the BHs as they approach the cavity wall, tugging them backwards. Why the gravitational drag dominates over accretion in the prograde binaries, in contrast with the retrograde ones, is unclear and requires further investigation. One possible interpretation can be made through analogy with the standard Ostriker prescription for dynamical friction, $F_{DF}$, in a uniform medium (\citet{Ostriker1999}) expressed in Eq.~\eqref{eq:dynfric} for our parameters.

\begin{equation}
    \centering
    F_{DF} = \frac{4\pi G^{2}M_{\rm BH}^{2}\rho_{0}}{v^{2}_{rel}} I(\mathcal{M}).
    \label{eq:dynfric}
\end{equation}

\noindent In this description the dynamical friction is directly proportional to the background density of the medium $\rho_{0}$ and inversely proportional to the square of $v_{rel}$, relative velocity of the perturber and the medium. The $I(\mathcal{M})$ term, is a function of the Mach number $\mathcal{M}$. We can approximate the relative velocity as the difference between the apsidal velocity and the orbital velocity of gas at the cavity wall. For prograde binaries, this quantity will be less as the binary is orbiting in the same direction of the gas, so one would expect the dynamical friction due to gas curving round a perturber to be greater, as observed here.

\subsection{Dependence on disc mass and initial separation}
\begin{figure*}
\begin{tabular}{ccc}
   Prograde & Retrograde  & High AGN disc mass\\

    \includegraphics[width=0.33\textwidth]{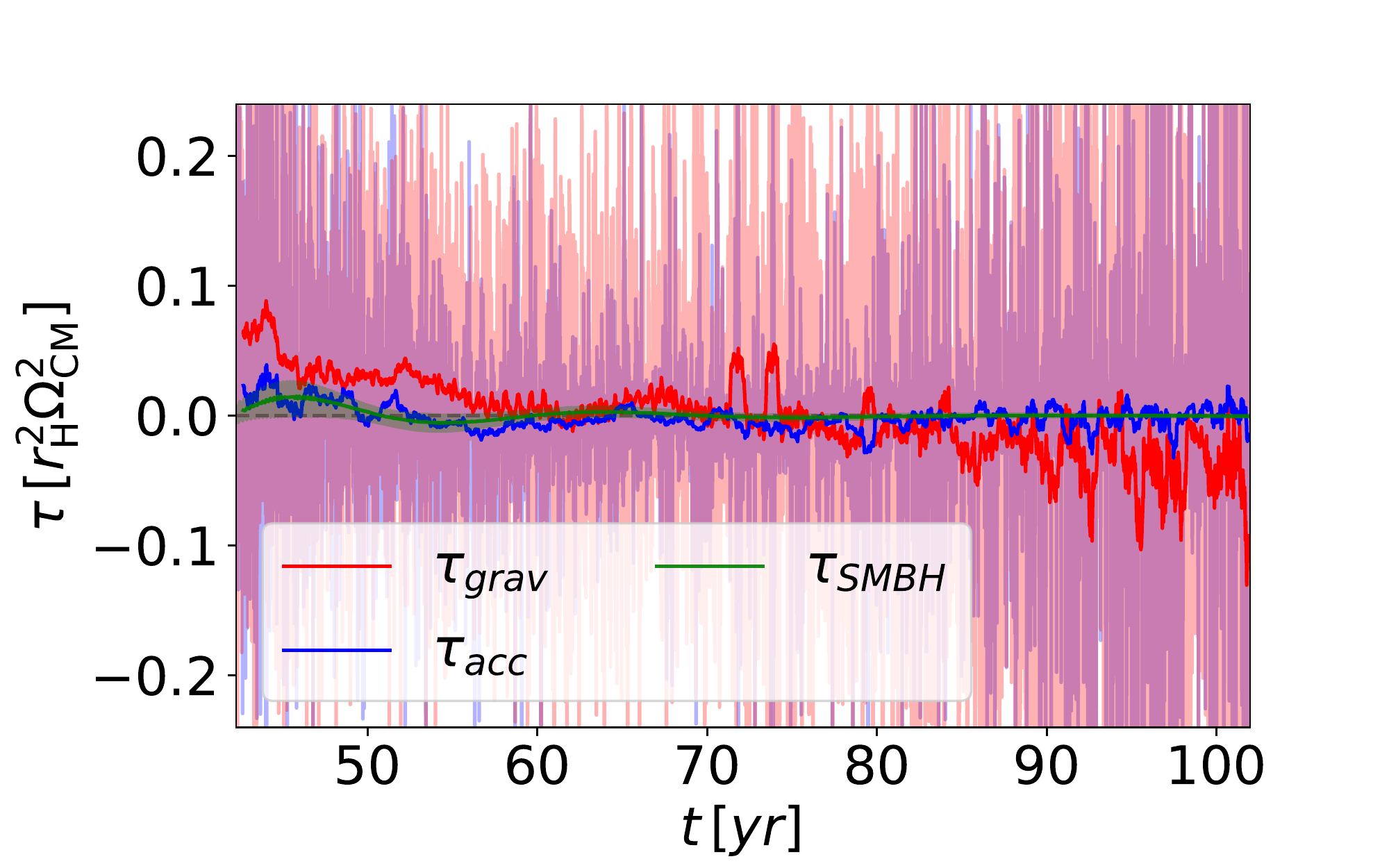} &
    \includegraphics[width=0.33\textwidth]{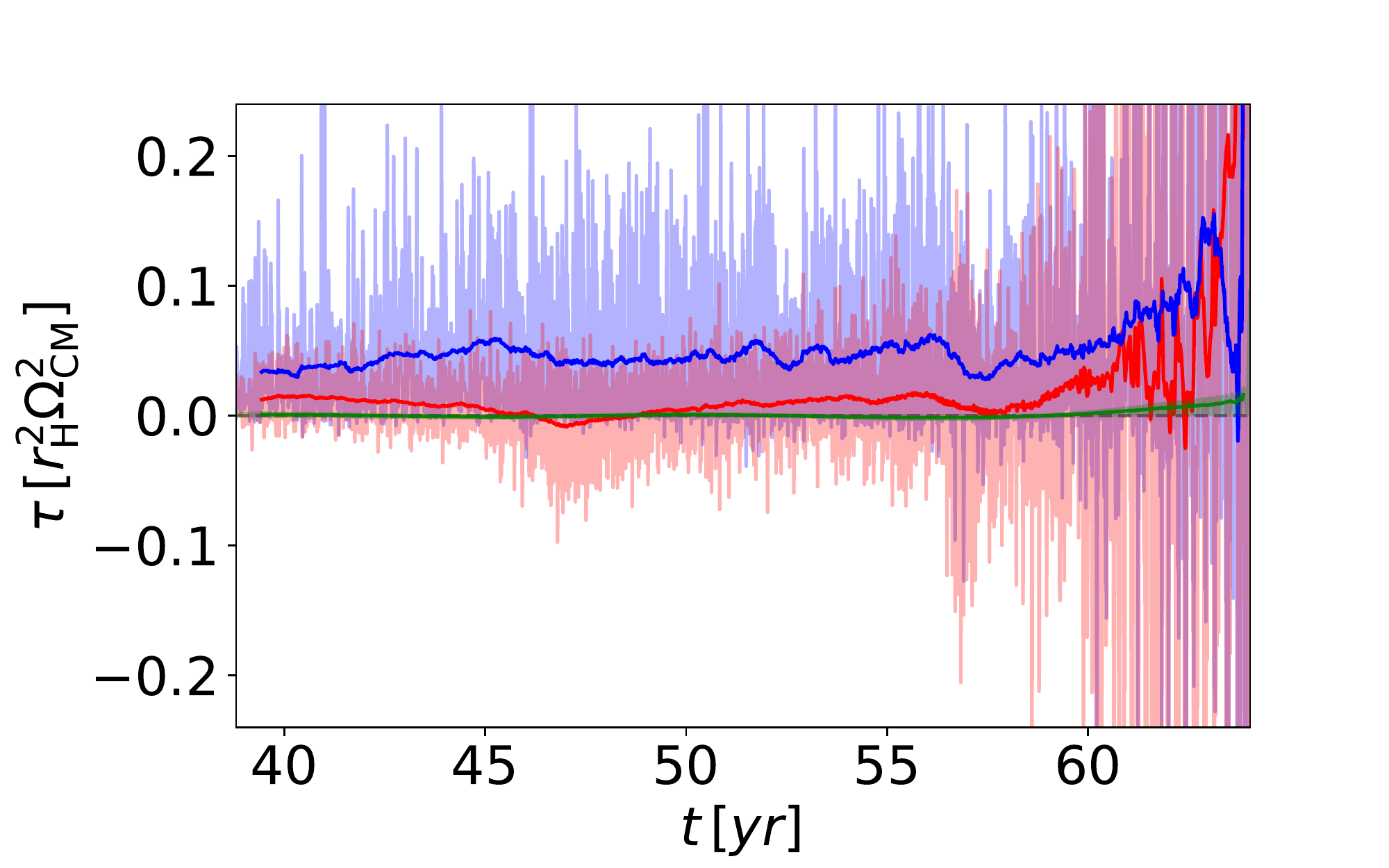} &

    \includegraphics[width=0.33\textwidth]{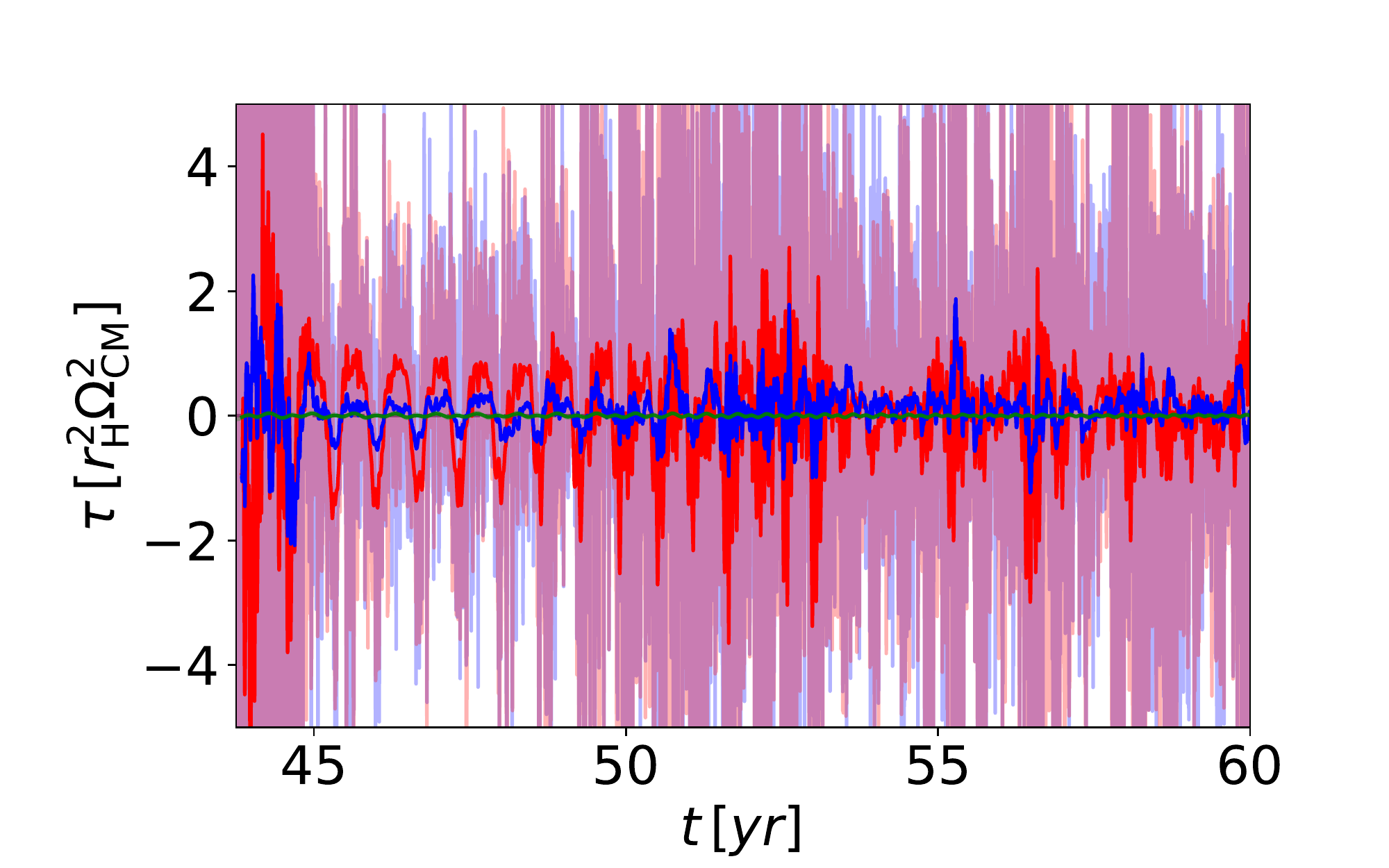}

\end{tabular}
\caption{Torque as a function of time for the three physical sources, SMBH (green), gas gravity (red) and accretion (blue) shown for three models. These models (left to right) are a prograde binary with fiducial $M_{\rm d}$ ($\mathit{Cap_{1,2.5}}$), a retrograde binary with the same $M_{\rm d}$ ($\mathit{Cap_{1,2.75}}$) and a prograde binary from the high $M_{\rm d}$ simulations ($\mathit{Cap_{5,2.5}}$). The results are qualitatively the same for other models with the same disc mass and orbital rotations.}
\label{fig:T_vs_t_comparison}
\end{figure*}

Now considering our other models, starting again with the torque evolution, we compare the torque components vs time in Figure \ref{fig:T_vs_t_comparison} for three different scenarios. These are i) a prograde binary of fiducial $M_{disc}$, ii) a retrograde binary of fiducial $M_{disc}$ and iii) a prograde binary of the 5$M_{disc}$ models. 

Moving on to the third panel of Figure \ref{fig:T_vs_t_comparison}, the high mass binaries show significant oscillation of the gas gravitational torques for the initial encounter, this continues until the strong gas spiral structure transitions into more chaotic behaviour. While the torques oscillate about zero during this period, the torque remains positive for longer at apoapsis and an order of magnitude stronger than the $1M_{\rm d}$ examples. This causes the very rapid damping in eccentricities of binaries in the high AGN disc mass simulations. The increase in strength is attributed directly to the higher local gas density, which enhances the gas dynamical friction acting on the binary. After the gas morphology becomes more chaotic the accretion and gravitational torques also become highly disordered and the binary eccentricities vary between $\sim 0-4$. Though the binary separation evolution varies considerably per model, the level of eccentricity damping/excitation scales consistently with $M_{\rm d}$ in all our models. The damping of $e$ for prograde binaries and excitation of $e$ in retrograde binaries is increasingly significant for higher $M_{\rm d}$. This is in line with the expected enhancement of $\tau_{\rm grav}$ due to increased densities, as expected from the Ostriker formula in Eq.~\eqref{eq:dynfric}, in conjunction with enhanced $\tau_{\rm acc}$ due to more linear momentum transfer through accreting in a denser medium.

\subsection{Work done}
\begin{figure*}
\begin{tabular}{ccc}

   Prograde & Retrograde  & High AGN disc mass\\
    \includegraphics[width=0.33\textwidth]{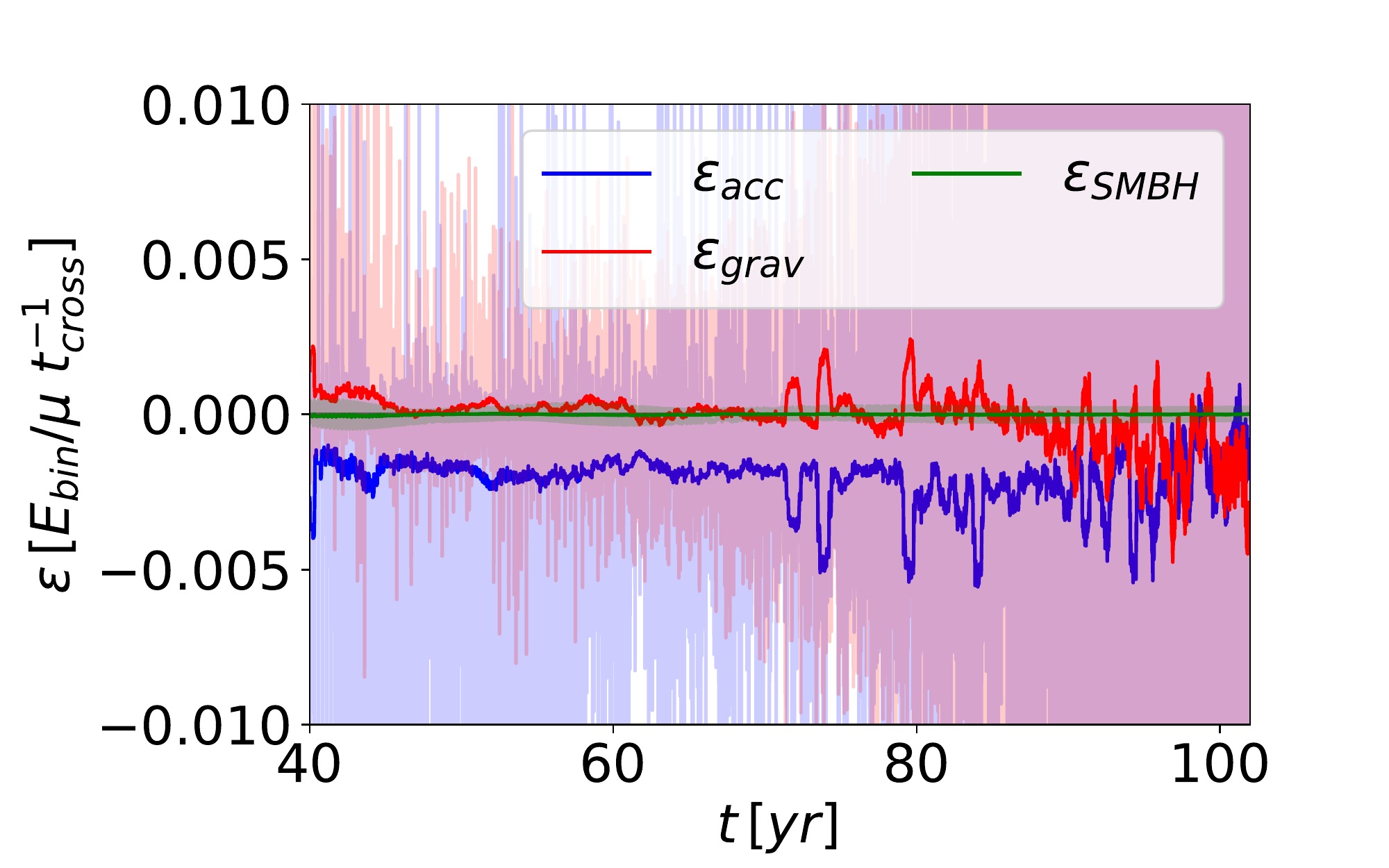} &
    \includegraphics[width=0.33\textwidth]{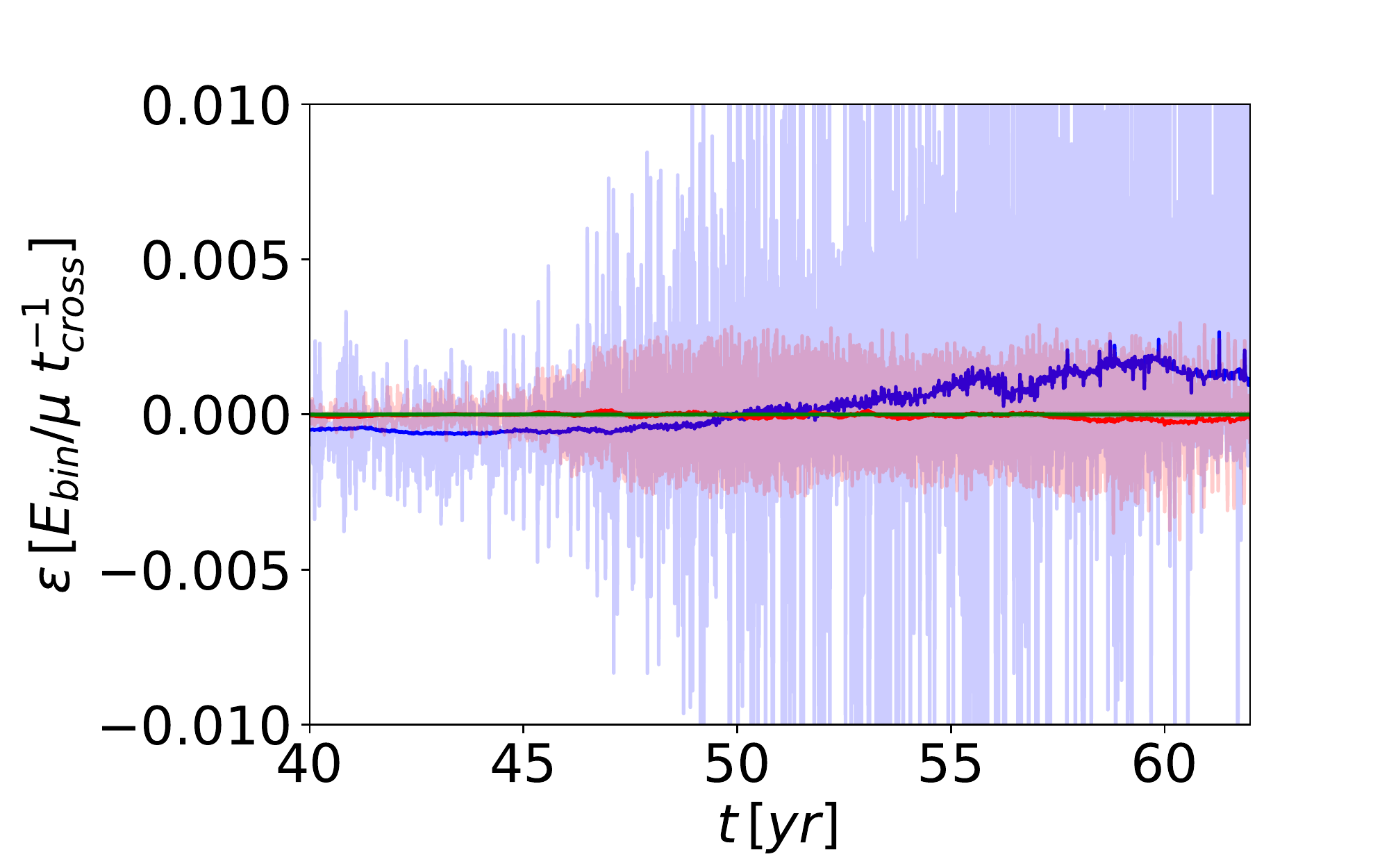} &

    \includegraphics[width=0.33\textwidth]{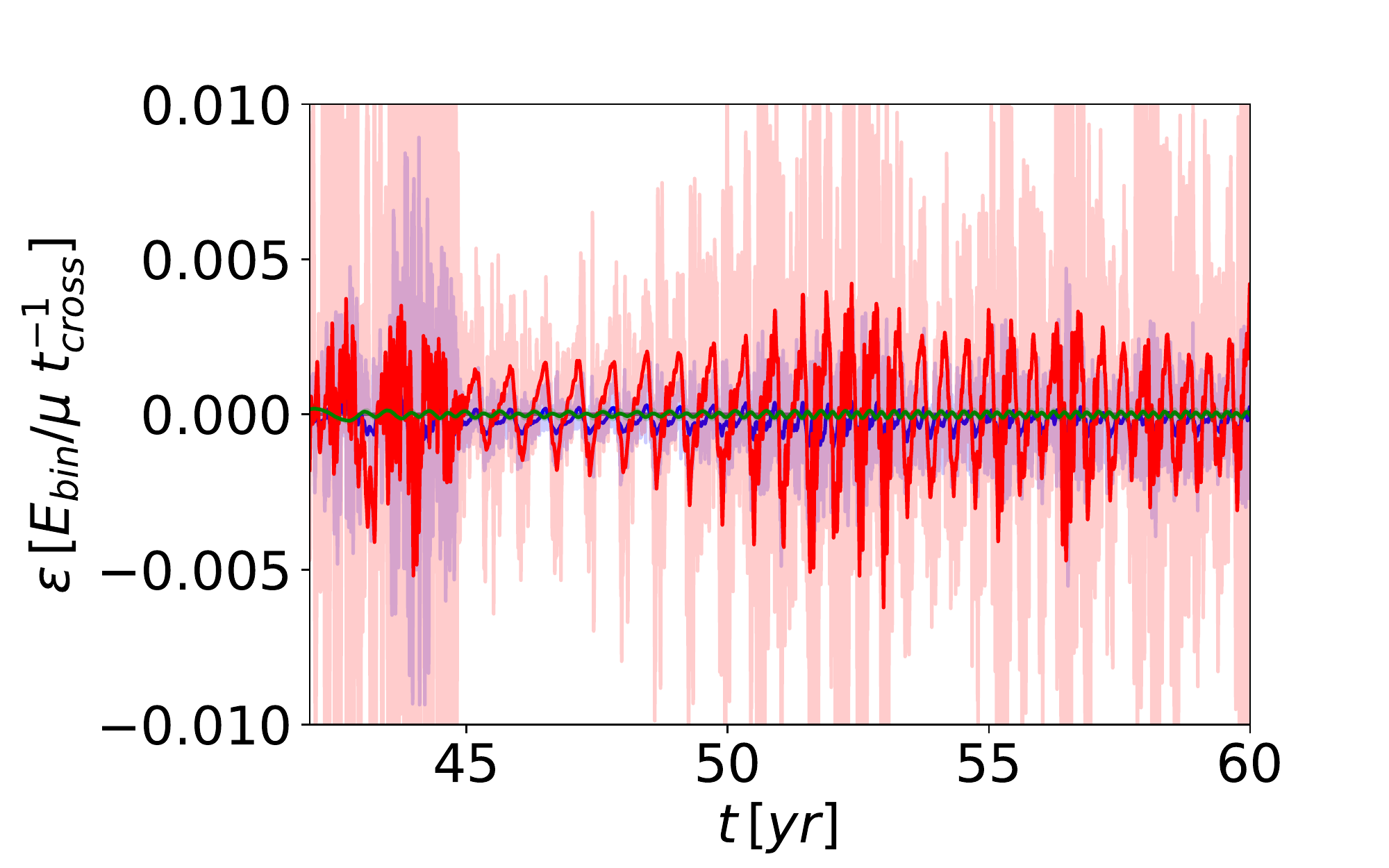}

\end{tabular}
\caption{Orbital energy dissipation as a function of time shown for three models for the three physical sources, SMBH (green), gas gravity (red) and accretion (blue). These models (left to right) are a prograde binary with fiducial $M_{\rm d}$ ($\mathit{Cap_{1,2.5}}$), a retrograde binary with the same $M_{\rm d}$ ($\mathit{Cap_{1,2.75}}$) and a prograde binary from the high $M_{\rm d}$ simulations ($\mathit{Cap_{5,2.5}}$). The results are qualitatively the same for other models with the same disc mass and orbital rotations.} 
\label{fig:workdone}
\end{figure*}

Since our binary evolution problem begins with an initial highly eccentric (sometimes hyperbolic) encounter, the initial eccentricities are all very high (>0.5). Therefore the assumption for circular binaries that forces acting radially with respect to the binary COM on the BHs are negligible is invalid.The work done per unit mass per unit time is shown for three examples, prograde and retrograde models with equal $M_{\rm d}$ and for a high mass prograde case in Figure \ref{fig:workdone}.

\indent For both prograde and retrograde binaries the net dissipation through gravitational interaction with the gas is minimal as $\varepsilon_{\rm grav}$ oscillates about zero. In a similar manner to the torques, the orbital energy evolves differently for prograde binaries depending on how close the first encounter is and whether a well defined CBMD can form. For the cases this is true, accretion drives a slow but steady reduction in the orbital energy of the binary (see left panel of Figure \ref{fig:workdone}). For prograde binaries that do not, all contributions to $\varepsilon$ oscillate about zero after the initial encounter, the orbital energy reaches a steady value and the binary stalls. Take $\mathit{Cap_{0.2,2.5}}$ for example in Figure \ref{fig:sep}. For our high mass models, recall all of which are prograde, do not form defined CBMDs, similarly have variability in $\varepsilon_{\rm grav}$ and $\varepsilon_{\rm acc}$ about zero. The increased AGN disc mass leads to stronger, well defined oscillations in both $\varepsilon_{\rm grav}$ and $\varepsilon_{\rm acc}$ in resonance with the orbital period. The variation in the former is a result of the far more massive circumsingle discs of the BHs dominating the gravitational forces in Eq.~\eqref{eq:work_grav}. When the BHs are enroute to apoapsis, the force differential dotted with the relative velocity vector is strongly negative, zero at apoapsis, before flipping as the binary approaches periapsis. Retrograde binaries have an interesting property where there is a period of time post capture, when the eccentricity is lower, when the components of $\varepsilon$ are identical to the prograde case, until the eccentricity reaches considerably high ranges of $\sim0.8$ and above and $\varepsilon_{\rm acc}$ rapidly transitions from negative to strongly positive and the binary semi-major axis increases. The reason for this is unclear though likely related to behaviour changing at the cavity wall, in the same way the torques increase rapidly at the same point.

\subsection{GW Dissipation}
\label{sec:GWdiss}
For merger, periapsis must be smaller than the sum of the innermost stable circular orbits of the BHs, for equal masses this is a merger radius of $r_{merge}=12GM_{\rm BH}/c^2$. For our parameters this is an extremely small distance of $~5\times10^{-6}r_{\rm H}$. However we note that in $\mathit{Cap_{1,2.75}}$ the final apoapsis of the binary prior to decoupling passes within this value, thus if we were to include GW dissipation we can conclude this system would undergo merger. We can also consider the increasing GW dissipation of orbital energy in the leadup to decoupling. To do this we compute the orbital energy lost at periapsis using the expression in \citet{Peters1964,Hansen1972}, see also \citealt{Samsing2018}:
\begin{equation}
    \Delta E_{GW} \approx  \frac{85\pi}{12\sqrt{2}}\frac{G^{7/2}}{c^5}\frac{M_{1}^{2}M_{2}^{2}\sqrt{M_{1}+M_{2}}}{r_{per}^{7/2}}.
    \label{eq:GWdiss}
\end{equation} 
\noindent We then express this quantity as the fraction of orbital energy, $E_{BH-BH}$, that is dissipated during a pericentre passage and label this quantity $\eta$:
\begin{equation}
    \centering
    \eta \equiv \frac{\Delta E_{GW}}{E_{BH-BH}}
    \label{eq:GWdissrate}
\end{equation} 

\begin{figure*}
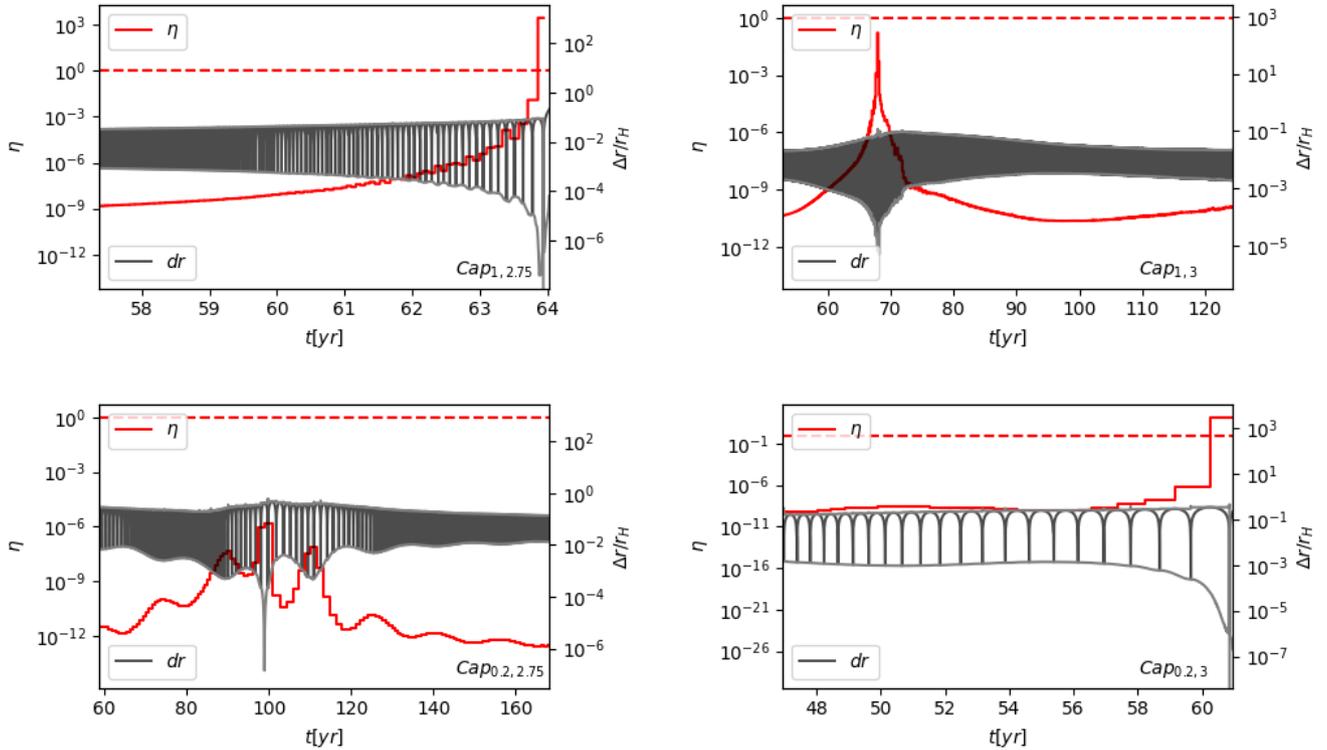

\begin{tabular}{cc}
    \includegraphics[width=0.48\textwidth]{6_7_GWdiss_cap1_2_75.pdf} &
    \includegraphics[width=0.48\textwidth]{6_7_GWdiss_cap1_3.pdf} \\

    \includegraphics[width=0.48\textwidth]{6_7_GWdiss_cap0_2_2_75.pdf} &
    \includegraphics[width=0.48\textwidth]{6_7_GWdiss_cap0_2_3.pdf}

\end{tabular}
\caption{Orbital energy dissipation as a fraction of binary orbital energy per periapsis pass $\eta$ (red) along side separation (black) for our four retrograde binaries. Grey lines represent the apoapsis and periapsis of the binary orbit calculated from the instantaneous orbital elements. The Binaries in models $\mathit{Cap_{1,2.75}}$ and $\mathit{Cap_{0.2,3}}$ undergo definite merger as $\eta$ exceeds unity.} 
\label{fig:GWdiss}
\end{figure*}

Where an $\eta$ value of one or greater implies the binary would be able to dissipate all of its orbital energy in one periapsis pass and undergo merger. Akin to the binary objects crossing the merger distance $r_{merge}$. We plot the value of $\eta$ as a step function evaluated at each periapsis pass along side the separation of the binaries for all 4 retrograde systems in Figure \ref{fig:GWdiss}. We find no prograde binaries at close enough separations to warrant this analysis at the time of termination for our simulations.

In Figure \ref{fig:GWdiss} we confirm the result that $\mathit{Cap_{1,2.75}}$ corresponds to a successful merger since $\eta$ exceeds unity during the last orbit. Though the orbital energy of the binary is actually increasing at this point via the accretion driven radial forces described in the previous section, at this point it is outrun by GW dissipation. The assumption made here is that we don't expect including GW dissipation live in our simulations to somehow significantly decrease $\eta$ at this point. If one considers the $\eta$ values in the lead up to the final event, they are orders of magnitudes below unity, so we argue this is a safe assumption. From the $\eta$ step function we also reaffirm our earlier short calculation of a direct merger as it exceeds one for the last periapsis pass (expectedly where $r_{per}$ drops below $5\times10^{-6}$). $\mathit{Cap_{0.2,3}}$ can also be identified as a merger, while $\mathit{Cap_{0.2,2.75}}$ fails to merge before the binary flips to prograde and the eccentricity begins to be damped by accretion and dynamical friction. The fate of the final retrograde binary, $\mathit{Cap_{0.2,3}}$, remains ambiguous. At the closest approach during the its evolution at around 70 years, it loses 30\% of its energy before becoming prograde. This is quite considerable indeed, though the critical question is whether this dissipation would lead to another significantly close encounter and finally merge the binary, or whether it will still flip to prograde and re-circularise. Since GW effects are calculate in post, this remains unknown. Since the rate of eccentricity excitation in retrograde binaries speeds up when the its value nears unity, they have a very small number of orbits where they have high enough eccentricity such that $r_{per}$ is small enough for GW emission to be significant before the binary either decouples ($\mathit{Cap_{1,2.75}}$,$\mathit{Cap_{0.2,3}}$) or flips to prograde ($\mathit{Cap_{1,3}}$,$\mathit{Cap_{0.2,2.75}}$). Therefore the specific moment where the binaries make their closest approach, in relation to the instantaneous orbital elements is \textit{extremely} influential to whether merger takes place or not. For example, in the $\mathit{Cap_{0.2,3}}$ panel of Figure \ref{fig:GWdiss}, the instantaneously calculated $r_{per}$ passes within $r_{merge}$ but the binary begins to re-circularise before the next periapsis passage and so GW dissipation reduces sharply due to its steep $1/r_{per}^{3.5}$ dependence and the binary fails to merge.

\section{Discussion}
\label{sec:discussion}

\subsection{Comparison to similar studies}

Within the final stages of our paper's production, three preprints looking into the specific problem of binary formation through gas drag in AGN discs appeared on the arXiv. \cite{Rozner+2022} and \cite{DeLaurentiis2022} both simulate the orbits of two BHs interacting around a SMBH assuming a gas drag term in their orbital integrations given by the Ostriker prescription (\citealt{Ostriker1999}), rather than performing hydrodynamical simulations.  These studies adopt an undisturbed background gas, and therefore do not capture the gas dynamics during the close interaction of the binary components, as we do here. They nevertheless both conclude that capture is possible and should be common, and provide insight into the role of dynamical friction in the capture process.
\cite{DeLaurentiis2022} perform an academic exercise in which the drag from dynamical friction is artificially switched off once the trajectories of the orbits (in the binary comoving frame) become significantly perturbed, i.e once the binary strongly interacts. This is motivated by the fact that after this stage, the velocity of the objects relative to the gas becomes dominated by motion about the binary COM, and Ostriker's assumption of unaccelerated linear motion relative to the background medium manifestly breaks down. They find that energetically well bound binaries are still able to form, typically with eccentric orbits that precess, but no longer shrink their semi-major axes beyond this point. Their estimates of the timescales required for capture in AGN via dynamical friction include the formation time of our binaries, albeit we find accretion to dominate the negative drag. There is also excellent agreement with our predictions on the initial eccentricities typically being $>\sim0.9$, as well as on the formation of both prograde and retrogade binaries. When dynamical friction is switched off after the first encounter, the binaries in \cite{DeLaurentiis2022} remain at these high eccentricities as there is no source of post-capture torque to further alter the eccentricity, as there is in our simulations here. 

\citet{Rozner+2022} take a similar approach to \cite{DeLaurentiis2022} and derive analytic functions for the energy dissipation during the a close fly-by between two compact objects.  Their treatment looks at two-body encounters, which may be applicable more broadly, and do not include the tidal forces from massive third body (i.e. they solve the two-body problem with the gas drag force added in).  They also find binary capture through dynamical friction when the relative velocity prior to the encounter is below a critical threshold.  They allow the dynamical friction to affect the binary orbits beyond the first encounter which leads to continued efficient energy removal post capture, as well as rapid eccentricity damping (also found in the majority of cases examined by \citealt{DeLaurentiis2022}).
In our hydro simulations we find our binaries shrink far more slowly (with respect to the outer binary orbital period) via a combination of gas gravity and accretion which is also highly stochastic and in some cases stall completely. However we consider binaries orbiting roughly 100 times closer to the SMBH compared to their work, i.e. at $\sim 0.01\,\rm pc$ instaed of $\sim 1\,\rm pc$, which could affect the result.

Most closely related to our study, \cite{Li_Dempsey_Lai+2022} explore the binary capture problem in AGN using 2D hydrodynamical simulations. 
They also find successful captures for binaries with mass ratios very similar to ours, formed via gravitational interaction with the gas alone with no accretion. Interestingly, they form predominantly retrograde binaries which harden on a similar time frame to our prograde binaries. The reason that they find far more retrograde binaries, whereas we (as well as the studies above, based on the Ostriker gas drag) find both prograde and retrogade binaries is unclear, but could possibly be because \cite{Li_Dempsey_Lai+2022} examine only a single impact parameter. The eccentricity evolution of their prograde binaries evolves inversely to ours, where eccentricity is damped similarly to our prograde binaries and at a similar rate to our $M_{d}=M_{d,0}$ cases. A possible reason for the hardening and damped eccentricities could be the softening lengths around the sinks. Comparing our sink/softening scales with \citet{Li_Dempsey_Lai+2022}, we resolve and simulate for sinks with a softening/accretion radius of approximately one fifth the size, where smaller softening lengths have been shown to lead to larger torques originating from the regions very close to the BHs (\citealt{Li2021,Li_2022_hot_discs}). During capture, we find that oscillatory gravitational work and torque at capture (Figures \ref{fig:diss_vs_t} \& \ref{fig:T_vs_t_enc}) generally does positive work and torque on the binary initially and originates from the inner regions around the BHs in our simulations, which may not be simulated for higher softening lengths. This could suggest softening plays a strict role during capture as well as the subsequent evolution of the binary. Additionally, \citet{Li_Dempsey_Lai+2022} allow the BHs to build up their circumsingle discs prior to their first encounter for far longer than shown here, which could potentially change the strength and behaviour of gas gravitational drag at the encounter. 

Due to the lengthy runtimes involved, we do not consider repeated scattering events like in \citet{Li2022_inc}, i.e where a decoupled binary may interact again as the inner BH catches back up to the outer BH. This possibility, along with the fact that in reality there is a population of $\sim10,000$ BHs in the AGN disc, allows us to  hypothesise that failed captures could be deflected into the trajectory of other orbiting black holes and lead to higher binary formation rates.

\subsection{Caveats}
In our study we have made several assumptions, which are to be relieved in follow up studies:
\begin{itemize}
    \item We assume gas heating/thermal shocks are minimal over the entire simulation domain. While this is a good assumption for the main body of the annulus, the high speed intersections of the accretion discs of the BHs will in practice lead to \textit{significant} heating of their material due to their high density and relative velocity. The effect of this heating in such an encounter on the forces (dynamical friction etc) the binary experiences is unknown. Though we posit that it will enhance the gas loss in the region due to the increased pressure associated with higher temperatures achieved in the local. This would be represented by an increase in the thermal term of  term in Eq.~\eqref{eq:K_gas}. 
    \item Our simulations predict accretions rates orders of magnitude higher than the Eddington limit. This outcome also stems from our omission of feedback mechanisms (radiative and heating) in the simulations. Very recently, \citet{Tagawa_feedback_2022} show that including these effects results in less rapid accretion during the encounter due to enhanced supportive pressure from radiative feedback from the inner CSMDs.
    \item Mostly for computational reasons we neglected the self-gravity of the gas in our simulations. For our models where the binary mass remains larger than the total gas mass within their mutual Hill sphere, then this assumption is reasonable. However, for our most massive disc model, we are in the opposite regime, and we expect that the gas dynamics during capture will differ and perhaps lead to easier binary capture if there is an additional binding term between gas particles that can prevent as much mass loss.
    \item The initial distribution of BHs being limited to circular orbits with zero velocity dispersion or inclination is a simplification of their true distribution. In reality the BH population of an AGN will have a range of eccentricities and inclinations. Encounters with variations of these parameters may alter the ease of the gas capture process shown here due to varying relative velocities upon intersecting each other's Hill sphere. As suggested by \citet{Li2022_inc}, allowing for non zero inclination reduces the probability of close encounters.
\end{itemize}

\section{Summary and Conclusions}
\label{sec:conclusions}
In this work we performed 15 simulations of two isolated BHs, encountering each other in an AGN disc. Of these 15, 12 successfully form binaries, 4 retrograde and 8 prograde. Two of our retrograde binaries go on to merge within the timescale of our models when GWs are considered. We summarise the key findings below:
\begin{itemize}
    \item The gas dissipation binary formation channel is efficient enough to form binaries from BHs on hyperbolic orbits. The amount of energy dissipated and torques scale with the density of the local medium, due to increased dynamical friction and accretion drag.
    \item The depth of the initial encounter greatly affects the nature of the energy transfer of the binary to the surrounding gas as well as the subsequent evolution. If the impact parameter is large, strong spiral outflows are generated due to the BHs tidally stripping each other's accretion discs. This carries away energy and angular momentum from the binary over a period of time until an equilibrium is reached. The spiral outflows are greatly enhanced for higher AGN disc masses and lead to rapid eccentricity damping of the binary. If the impact parameter is small then the BHs can cross each other's discs, creating more violent and disordered outflows as well as inducing significant accretion drag on the BHs. The latter case results in a CBMD and faster hardening of the binary compared to the former, where the binary after formation, as a fraction of the Hill sphere, has a far larger semi-major axis, prohibiting the formation of a CBMD.
    \item All our binaries objects initially encounter each other with eccentricities greater than unity. Immediately after the initial encounter, its value is still only just shy of this value, implying that initial conditions for binaries formed via this channel having $e\sim0$ is very nonphysical. For our fiducial AGN disc mass/density $e$ remains greater than 0.4 by the end of our runtime which corresponds to thousands of binary orbits for our tighter binaries. This would suggest binaries formed in this pathway would more easily merge secularly via the newly proposed evection induced merger pathway in AGN (e.g \citealt{Munoz2022}, \citealt{Gautham2022}). However, for retrograde binaries we find eccentricity increases due to the gas interactions on far shorter timescales and the binary undergoes a direct merger rather than through GWs over many orbits.
    \item We find accretion is significantly supereddington during capture for our simulation parameters and this dominates the removal of orbital energy of the binary and it's circularisation. Therefore we conclude that accretion should not be neglected in similar future studies. 
    \item The torques of prograde binaries are governed by gravitational interaction with the local gas in the CBMD, while retrograde binaries are governed by the accretion. Both torque sources operate within the same order of magnitude and therefore we encourage future studies to include both effects.
    \item strong bimodality in the binary evolution is identified between prograde and retrograde (w.r.t the CBMD and AGN disc) binaries. Prograde binaries always exhibit eccentricity \textit{damping} due to resonant gas gravitational and accretion torques in favour of the binary motion at apastron. Due to the inverted orbital angular momentum of the retrograde binaries, the resonant torques \textit{excite} their eccentricities. This effect is stronger in both cases for higher AGN gas disc densities where torques from the gas and accretion are stronger and for higher $e$. For prograde binaries formed via close encounters, once the binaries are circularised to $e<0.3$ the torques switch from net positive to net negative and begin to reduce the semi-major axis as opposed to affecting $e$. This is a result of the gravitational drag of quasi-stable streams of gas stretching from the BHs to the cavity being able to form at lower eccentricities.
    \item  For two out of our four retrograde binaries, the increasing torques push apastron to the Hill radius of the binaries as $e\rightarrow1$ and they decouple due to strong perturbations from the SMBH. In our other retrograde models it is demonstrated that the torques can flip the binary orientation from retrograde to prograde as they are not perturbed significantly enough to decouple at the point of flipping. After this transition they then flip to the same aforementioned prograde damping of $e$.
    \item When we consider the disspation of GWs at periastron for our highly eccentric retrograde binaries, we find that the amount of energy dissipated in the last few orbits before decoupling/orbit flipping can exceed the binary orbital energy. This leads to the revelation that our two decoupled retrograde binaries actually undergo merger when GW effects are taken into account. The opportunity for the merger of retrograde falls in a very short window of time as its angular momentum tends to zero and eccentricity to unity. The soul determinant is whether there is a periapsis pass close enough to the moment these quantities reach these values, else the binary can decouple or flip rotation and re-circularise and the pericentre becomes to shallow for GWs to drive inspiral.
    
\end{itemize}

Through our hydrodynamic approach we simulate the full chronology of a merger in the AGN channel under our assumptions, demonstrating that gas aids the formation of binaries and can induce secular torques to harden and merge the binaries, demonstrating through direct simulation the feasibility and efficiency of the AGN merger channel.

\section*{Acknowledgements}
\begin{itemize}
    \item These simulations were performed using the \textit{Hydra} compute cluster at \textit{The University of Oxford}. 
    \item Surface density plots of our simulations were rendered using SPLASH (see \citealt{Price2007}).
    \item This project was supported by funds from the European Research Council (ERC) under the European Union’s Horizon
2020 research and innovation program under grant agreement No
638435 (GalNUC).
    \item  This work was supported by the Science and Technology Facilities Council Grant Number ST/W000903/1.
\end{itemize}

\section*{Data Availability} 

The data underlying this article will be shared on reasonable request to the corresponding author.

\bibliographystyle{mnras}
\bibliography{Paper}

\appendix

\section{Derivative of binary two body specific energy}
\label{app:appendix_derivations}
For our calculation of the relative contributions to the energy change in the binary two body energy we derive a relation involving the acceleration vectors on each BH satellite by taking the time derivative of the specific energy. First we convert two body energy in \eqref{eq:two_body_energy} to the specific energy via dividing by the reduced mass, $\mu$:

\begin{equation}
    \centering
    \frac{E_{\rm bin}}{\mu}=\frac{1}{2}\|\boldsymbol{v}_1-\boldsymbol{v}_2\|^2 - \frac{GM_{\rm{bin}}}{\|\boldsymbol{r}_1-\boldsymbol{r}_2\|}.
\end{equation}

\noindent Here, $M_{\rm{bin}}$ is the total mass of the binary. Then we define the dissipation rate $\varepsilon$ as the time derivative of this quantity, where it is assumed $M=M(t)$, 

\begin{equation}
    \centering
    \varepsilon = \frac{d}{dt}\bigg(\frac{E_{\rm bin}}{\mu}\bigg)=\frac{1}{2}\frac{d\|\boldsymbol{v}_1-\boldsymbol{v}_2\|^2}{dt} - \frac{G\frac{dM}{dt}}{\|\boldsymbol{r}_1-\boldsymbol{r}_2\|} - GM\frac{d}{dt}\bigg(\frac{1}{\|\boldsymbol{r}_1-\boldsymbol{r}_2\|}\bigg).
\end{equation}

\noindent Using the result for a vector $\boldsymbol{v}(t)$ that

\begin{equation}
    \frac{d\|\boldsymbol{v}(t)\|}{dt} = \frac{\boldsymbol{v}(t)\cdot\Dot{\boldsymbol{v}}(t)}{\|\boldsymbol{v}(t)\|} \,\,\text{and}\,\, \frac{d\|\boldsymbol{v}(t)\|^{2}}{dt} = 2\|\boldsymbol{v}(t)\|\frac{d\|v(t)\|}{dt},
\end{equation}

\noindent the derivatives of the first and third terms can be reduced to

\begin{equation}
    \varepsilon = (\boldsymbol{v}_1-\boldsymbol{v}_2)\cdot(\boldsymbol{a}_1-\boldsymbol{a}_2) - \frac{G\dot{M}}{\|\boldsymbol{r}_1-\boldsymbol{r}_2\|} + GM\frac{(\boldsymbol{r}_1-\boldsymbol{r}_2)\cdot(\boldsymbol{v}_1-\boldsymbol{v}_2)}{\|\boldsymbol{r}_1-\boldsymbol{r}_2\|^{3}}.
\end{equation}

\noindent The third term is simply the acceleration of each satellite towards each other dotted with the velocity difference and so can be treated as another acceleration $(\Tilde{\boldsymbol{a}}_1 - \Tilde{\boldsymbol{a}}_2)$ which we denote with a Tilde. Taking the common factor $(\boldsymbol{v}_1-\boldsymbol{v}_2)$ out we can subtract the accelerations between the BHs from their total accelerations:

\begin{equation}
    \varepsilon = (\boldsymbol{v}_1-\boldsymbol{v}_2)\cdot\bigg((\boldsymbol{a}_1-\Tilde{\boldsymbol{a}}_1)-(\boldsymbol{a}_2-\Tilde{\boldsymbol{a}}_2)\bigg) - \frac{G\dot{M}}{\|\boldsymbol{r}_1-\boldsymbol{r}_2\|}.
\end{equation}

\noindent We then define the accelerations due to \textit{external} forces as $\boldsymbol{a}_{\rm 1,ext}\equiv\boldsymbol{a}_1-\Tilde{\boldsymbol{a}}_1$ and $\boldsymbol{a}_{\rm 2,ext}\equiv\boldsymbol{a}_2-\Tilde{\boldsymbol{a}}_2$.

\section{Limiting case: Accretion Capture}
\label{sec:acc_cap}
In Sec. \ref{sec:fiducial_E_diss} we showed that energy dissipation is predominately governed by accretion. \citet{Heppen77} demonstrated that if a satellite accretes enough mass within the Hill sphere of another, it can be remain bound via mass accretion alone. We consider this limiting case in the context of BH binary formation and derive an approximate upper limit criterion for a successful capture, assuming accretion is entirely responsible.
\newline\indent The 2-body specific energy upon the crossing of one BH into the other's Hill sphere is given by
\begin{equation}
    \overline{E_{0}} = \frac{1}{2} v^2 - \frac{\alpha_0}{r_H},
\end{equation}
\noindent with v =  $\|\boldsymbol{v}_1-\boldsymbol{v}_2\|$ as the relative speed upon entering the Hill sphere, $\alpha_0$ the initial gravitational parameter, i.e. $\alpha = G\left( m_0 + m_1 \right)$. After their close (parabolic) encounter, they exit the Hill sphere again, with energy
\begin{equation}
    \overline{E_{1}} = \frac{1}{2} v^2 - \frac{\alpha_1}{r_H},
\end{equation}
\noindent with $\alpha_1$ the new gravitational parameter, which takes into account the accreted mass. As an upper limit, we have also assumed the relative speed of the objects remains unchanged, i.e accretion does not induce any \textit{decelerative} effects. If $\overline{E_{0}} > 0$ and for capture we require $\overline{E}_{1} \le 0$, then a critical accreted mass is given by $\Delta m = \left(\alpha_1 - \alpha_0\right)/G$. But whether this mass can get accreted onto the BHs from the gas within the Hill sphere, depends on the gas density, $\rho$, and the encounter time scale roughly given by $\tau = r_H / v$, as well as the accretion radius (given by Bondi-Hoyle or physical radius). A minimum accretion rate is estimated by
\begin{equation}
    \dot{m}_{\rm{min}} = \frac{\Delta m}{\tau} = \frac{\left(\alpha_1 - \alpha_0\right) v}{G r_H} = \frac{\left(\frac{1}{2}r_H v^2 - \alpha_0\right) v}{G r_H} = \frac{\left(\frac{1}{2} v^2 - \frac{\alpha_0}{r_H}\right) v}{G} = \frac{\overline{E_{0}} v}{G}. 
\end{equation}
\noindent We estimate the accreted mass as $\Delta m = \rho V$, with $V = \sigma r_H$ the volume of accreted material, with $\sigma$ as the accretion cross section. Equating to the minimal accretion rate gives
\begin{equation}
    \frac{\Delta m}{\tau} = \frac{\rho V}{\tau} = \frac{\rho \sigma r_H v}{r_H} = \rho \sigma v \ge \frac{\overline{E_{0}} v}{G}.
\end{equation}
\noindent Simplifying gives
\begin{equation}
    G \rho \sigma \ge \overline{E_0}.
\end{equation}
\noindent We estimate $\sigma$ assuming Bondi-Hoyle (B-H) accretion of material such that $\sigma = \pi r_{\rm{B-H}}^2 \approx 4 \pi \alpha_0^2 / c_s^4$. Putting it together gives
\begin{equation}
    4 \pi G \frac{\rho}{c_s^4} \ge \frac{\overline{E_{0}}}{\alpha_0^2},
\end{equation}
or
\begin{equation}
     \rho \ge \frac{\overline{E_{0}}c_s^4}{4 \pi G \alpha_0^2},
     \label{eq:acc_criterion}
\end{equation}

This criterion in \eqref{eq:acc_criterion} corroborates our findings that capture is easier in higher density environments.


\bsp	
\label{lastpage}
\end{document}